\documentclass[prd,twocolumn,amsmath,amssymb,showpacs,nofootinbib]{revtex4}

\usepackage{graphicx}
\usepackage{dcolumn}
\usepackage{bm}
\usepackage{mathrsfs} 
\usepackage{xspace}

\newcommand{\dzero}{D0\xspace}
\newcommand{\runi}{Run~I\xspace}
\newcommand{\runii}{Run~II\xspace}

\newcommand{\W}{\ensuremath{W}\xspace}
\newcommand{\bquark}{\ensuremath{b}\xspace}

\newcommand{\pevt}{\ensuremath{P_{\rm evt}}\xspace}
\newcommand{\psgn}{\ensuremath{P_{\rm sig}}\xspace}
\newcommand{\pbkg}{\ensuremath{P_{\rm bkg}}\xspace}
\newcommand{\Ntag}{\ensuremath{n_{\rm tag}}\xspace}
\newcommand{\psgnb}{\ensuremath{P_{\rm sig}^{\Ntag}}\xspace}
\newcommand{\pevtb}{\ensuremath{P_{\rm evt}^{\Ntag}}\xspace}
\newcommand{\mtop}{\ensuremath{m_{\rm top}}\xspace}

\newcommand{\mt}{\ensuremath{m_{t}}\xspace}
\newcommand{\ftop}{\ensuremath{f_{\rm top}}\xspace}
\newcommand{\ftopb}{\ensuremath{f_{\rm top}^{\Ntag}}\xspace}
\newcommand{\ftopbest}{\ensuremath{f_{\rm top}^{\rm best}}\xspace}

\newcommand{\Gt}{\ensuremath{\Gamma_{t}}\xspace}
\newcommand{\mW}{\ensuremath{m_{W}}\xspace}
\newcommand{\GW}{\ensuremath{\Gamma_{W}}\xspace}

\newcommand{\subbellnu}{\ensuremath{{b\ell\nu}}\xspace}
\newcommand{\subellnu}{\ensuremath{{\ell\nu}}\xspace}
\newcommand{\subbdubar}{\ensuremath{{\overline{b}d\overline{u}}}\xspace}
\newcommand{\subdubar}{\ensuremath{{d\overline{u}}}\xspace}

\newcommand{\pvecqone}{\ensuremath{\vec{p}_{d}}\xspace}

\newcommand{\pmagqone}{\ensuremath{|\pvecqone|}\xspace}

\newcommand{\mbellnu}{\ensuremath{m_{\subbellnu}}\xspace}
\newcommand{\mellnu}{\ensuremath{m_{\subellnu}}\xspace}
\newcommand{\chatbell}{\ensuremath{{\hat c}_{b\ell}}\xspace}
\newcommand{\mbdubar}{\ensuremath{m_{\subbdubar}}\xspace}
\newcommand{\mdubar}{\ensuremath{m_{\subdubar}}\xspace}
\newcommand{\chatbbard}{\ensuremath{{\hat c}_{\overline{b}d}}\xspace}
\newcommand{\mthad}{\mbdubar} 
\newcommand{\mtlep}{\mbellnu} 
\newcommand{\mtophad}{\mthad}
\newcommand{\mtoplep}{\mtlep}

\newcommand{\mWhad}{\mdubar} 

\newcommand{\qoverptmu}{\ensuremath{\left(q/\pt\right)_\mu}\xspace}
\newcommand{\qoverptmurec}{\ensuremath{\qoverptmu^{\rm rec}}\xspace}
\newcommand{\qoverptmugen}{\ensuremath{\qoverptmu^{\rm gen}}\xspace}

\newcommand{\ptj}{\ensuremath{p_{T,j}}\xspace}
\newcommand{\etj}{\ensuremath{E_{T,j}}\xspace}
\newcommand{\pzbnu}{\ensuremath{p_{z,b\nu}}\xspace}

\newcommand{\et}{\ensuremath{E_{T}}\xspace}
\newcommand{\etmiss}{\ensuremath{E \kern-0.6em\slash_{T}}\xspace}

\newcommand{\pt}{\ensuremath{p_{T}}\xspace}

\newcommand{\qqbar}{\ensuremath{q\bar{q}}\xspace}
\newcommand{\qqbarprime}{\ensuremath{q\overline{q}'}\xspace}
\newcommand{\ppbar}{\ensuremath{p\bar{p}}\xspace}
\newcommand{\ccbar}{\ensuremath{c\bar{c}}\xspace}
\newcommand{\bbbar}{\ensuremath{b\bar{b}}\xspace}
\newcommand{\ttbar}{\ensuremath{t\bar{t}}\xspace}

\newcommand{\wjets}{\ensuremath{W}+{\rm jets}\xspace}
\newcommand{\wjetsshort}{\ensuremath{W}\!+\!{\rm jets}\xspace}

\newcommand{\wjjjj}{\ensuremath{\W jjjj}\xspace}

\newcommand{\wlj}{\wjjjj}  
\newcommand{\whj}{\ensuremath{\W hf}\xspace}
\newcommand{\ejets}{\ensuremath{e}+\rm jets\xspace}
\newcommand{\mujets}{\ensuremath{\mu}+\rm jets\xspace}
\newcommand{\ljets}{\ensuremath{\ell}+\rm jets\xspace}

\newcommand{\madgraph}{{\sc madgraph}\xspace}
\newcommand{\alpgen}{{\sc alpgen}\xspace}
\newcommand{\pythia}{{\sc pythia}\xspace}
\newcommand{\vecbos}{{\sc vecbos}\xspace}
\newcommand{\geant}{{\sc geant}\xspace}
\newcommand{\vegas}{{\sc vegas}\xspace}

\newcommand{\ipb}{\ensuremath{\rm pb^{-1}}\xspace}

\newcommand{\MeV}{\ensuremath{\mathrm{Me\kern-0.1em V}}\xspace}
\newcommand{\GeV}{\ensuremath{\mathrm{Ge\kern-0.1em V}}\xspace}
\newcommand{\GeVc}{\ensuremath{\mathrm{Ge\kern-0.1em V}}\xspace}
\newcommand{\GeVcc}{\ensuremath{\mathrm{Ge\kern-0.1em V}}\xspace}
\newcommand{\TeV}{\ensuremath{\mathrm{Te\kern-0.1em V}}\xspace}

\newcommand{\DeltaR}{\ensuremath{\Delta {\cal R}}\xspace}
\newcommand{\energyoffset}{\ensuremath{E_{\rm off}}\xspace}
\newcommand{\calorimeterresponse}{\ensuremath{R_{\rm cal}}\xspace}
\newcommand{\showeringcorrection}{\ensuremath{C_{\rm cone}}\xspace}

\newcommand{\Eref}[1]{(\ref{#1})}

\newcommand{\MEerrstat}       {\ensuremath{^{+5.0}_{-7.4}}\xspace} 
\newcommand{\MEerrstatenspace}       {\ensuremath{^{\enspace+5.0}_{\enspace-7.4}}\xspace} 
\newcommand{\MEerrstatlong}   {\ensuremath{+5.0~-\hspace{-0.65ex}7.4}\xspace} 
\newcommand{\MEerremstat}     {\ensuremath{^{\enspace+7.3}_{-10.6}}\xspace} 
\newcommand{\MEerrmustat}     {\ensuremath{^{+10.5}_{-10.9}}\xspace} 
\newcommand{\MEerrstatnojes}    {\ensuremath{^{+2.9}_{-3.2}}\xspace} 
\newcommand{\MEerrjes}          {\ensuremath{^{+4.1}_{-6.7}}\xspace} 
\newcommand{\MEjeserrstat}    {\ensuremath{^{+0.052}_{-0.040}}\xspace} 
\newcommand{\MEerrsgnmod} {\ensuremath{\pm 0.34}\xspace} 
\newcommand{\MEerrbkgmod} {\ensuremath{\pm 0.32}\xspace} 
\newcommand{\MEerrpdf}    {\ensuremath{+0.26~-\hspace{-0.65ex}0.40}\xspace} 
\newcommand{\MEerrbjes}   {\ensuremath{\pm 0.71}\xspace} 
\newcommand{\MEerrbclepbr}{\ensuremath{+0.06~-\hspace{-0.65ex}0.07}\xspace} 
\newcommand{\MEerrjespt}  {\ensuremath{\pm 0.25}\xspace} 
\newcommand{\MEerrbresp}  {\ensuremath{+0.87~-\hspace{-0.65ex}0.75}\xspace} 
\newcommand{\MEerrtrg}    {\ensuremath{\pm 0.08}\xspace} 
\newcommand{\MEerrftop}   {\ensuremath{+0.50~-\hspace{-0.65ex}0.17}\xspace} 
\newcommand{\MEerrqcd}    {\ensuremath{\pm 0.67}\xspace} 
\newcommand{\MEerrmccalib}{\ensuremath{\pm 0.17}\xspace} 
\newcommand{\MEerrsyst}   {\ensuremath{^{+1.5}_{-1.4}}\xspace} 
\newcommand{\MEerrsystlong}   {\ensuremath{+1.5~-\hspace{-0.65ex}1.4}\xspace} 
\newcommand{\MEerrtotallong}  {\ensuremath{+5.2~-\hspace{-0.65ex}7.5}\xspace} 
\newcommand{\MEberrstat}       {\ensuremath{^{+4.1}_{-4.5}}\xspace} 
\newcommand{\MEberrstatlong}   {\ensuremath{+4.1~-\hspace{-0.65ex}4.5}\xspace} 
\newcommand{\MEberremstat}     {\ensuremath{^{+4.6}_{-6.3}}\xspace} 
\newcommand{\MEberrmustat}     {\ensuremath{^{+9.4}_{-8.6}}\xspace} 
\newcommand{\MEberrstatnojes}    {\ensuremath{\pm2.5}\xspace} 
\newcommand{\MEberrjes}          {\ensuremath{^{+3.2}_{-3.7}}\xspace} 
\newcommand{\MEbjeserrstat}    {\ensuremath{^{+0.035}_{-0.032}}\xspace} 
\newcommand{\MEberrsgnmod} {\ensuremath{\pm0.46}\xspace} 
\newcommand{\MEberrbkgmod} {\ensuremath{\pm0.40}\xspace} 
\newcommand{\MEberrpdf}    {\ensuremath{+0.16~-\hspace{-0.65ex}0.39}\xspace} 
\newcommand{\MEberrbjes}   {\ensuremath{\pm 0.56}\xspace} 
\newcommand{\MEberrbclepbr}{\ensuremath{\pm0.05}\xspace} 
\newcommand{\MEberrjespt}  {\ensuremath{\pm 0.19}\xspace} 
\newcommand{\MEberrbresp}  {\ensuremath{+0.63~-\hspace{-0.65ex}1.43}\xspace} 
\newcommand{\MEberrtrg}    {\ensuremath{+0.08~-\hspace{-0.65ex}0.13}\xspace} 
\newcommand{\MEberrbtagging}{\ensuremath{\pm 0.24}\xspace}
\newcommand{\MEberrftop}   {\ensuremath{\pm0.15}\xspace} 
\newcommand{\MEberrqcd}    {\ensuremath{\pm0.29}\xspace} 
\newcommand{\MEberrmccalib}{\ensuremath{\pm 0.48}\xspace} 
\newcommand{\MEberrsyst}   {\ensuremath{^{+1.2}_{-1.8}}\xspace} 
\newcommand{\MEberrsystlong}   {\ensuremath{+1.2~-\hspace{-0.65ex}1.8}\xspace} 
\newcommand{\MEberrtotallong}  {\ensuremath{+4.3~-\hspace{-0.65ex}4.9}\xspace} 

\newcommand{\resultstat}{\ensuremath{169.2\MEerrstat\,{\rm (stat.+JES)}}\xspace}
\newcommand{\resultstatenspace}{\ensuremath{169.2\MEerrstatenspace\,{\rm (stat.+JES)}}\xspace}
\newcommand{\resultemstat}{\ensuremath{167.0\MEerremstat\,{\rm (stat.+JES)}}\xspace}
\newcommand{\resultmustat}{\ensuremath{173.0\MEerrmustat\,{\rm (stat.+JES)}}\xspace}
\newcommand{\result}{\ensuremath{\resultstat\,\MEerrsyst\,{\rm (syst.)}}\xspace}

\newcommand{\resultjesstat}{\ensuremath{1.048\MEjeserrstat\,{\rm (stat.)}}\xspace}

\newcommand{\resultftopstat}{\ensuremath{0.33\pm0.06\,{\rm (stat.)}}\xspace}

\newcommand{\resultbstat}{\ensuremath{170.3\MEberrstat\,{\rm (stat.+JES)}}\xspace}
\newcommand{\resultbemstat}{\ensuremath{170.1\MEberremstat\,{\rm (stat.+JES)}}\xspace}
\newcommand{\resultbmustat}{\ensuremath{172.6\MEberrmustat\,{\rm (stat.+JES)}}\xspace}
\newcommand{\resultb}{\ensuremath{\resultbstat\,\MEberrsyst\,{\rm (syst.)}}\xspace}

\newcommand{\resultbjesstat}{\ensuremath{1.027\MEbjeserrstat\,{\rm (stat.)}}\xspace}

\newcommand{\resultbftopstat}{\ensuremath{0.31\pm0.09\,{\rm (stat.)}}\xspace}

\newcommand{\resultbstatzerotag}{\ensuremath{174.4^{+18.5}_{-12.3}\,{\rm (stat.)}}\xspace}
\newcommand{\resultbjesstatzerotag}{\ensuremath{0.986^{+0.084}_{-0.091}\,{\rm (stat.)}}\xspace}
\newcommand{\resultbstatonetag}{\ensuremath{173.1^{\enspace+5.1}_{\enspace-5.2}\,{\rm (stat.)}}\xspace}
\newcommand{\resultbjesstatonetag}{\ensuremath{1.011^{+0.049}_{-0.045}\,{\rm (stat.)}}\xspace}
\newcommand{\resultbstattwotag}{\ensuremath{163.2^{\enspace+6.8}_{\enspace-6.2}\,{\rm (stat.)}}\xspace}
\newcommand{\resultbjesstattwotag}{\ensuremath{1.094^{+0.071}_{-0.066}\,{\rm (stat.)}}\xspace}

\newcommand{\fitemntop}{\ensuremath{40.6~^{+\enspace9.4}_{-\enspace9.1}}\xspace}

\newcommand{\fitemftop}{\ensuremath{\left(47.2\,^{+10.9}_{-10.6}\right)\%}\xspace}

\newcommand{\fitemfqcd}{\ensuremath{\left(17.6~^{+\enspace2.4}_{-\enspace2.2}\right)\%}\xspace}

\newcommand{\fitmuntop}{\ensuremath{25.8~^{+\enspace8.6}_{-\enspace8.1}}\xspace}

\newcommand{\fitmuftop}{\ensuremath{\left(29.0\,^{+\enspace9.7}_{-\enspace9.1}\right)\%}\xspace}

\newcommand{\fitmufqcd}{\ensuremath{\enspace\left(5.1~^{+\enspace0.9}_{-\enspace0.8}\right)\%}\xspace}

\newcommand{\fitntop}{\ensuremath{66.4~^{+12.7}_{-12.2}}\xspace}
\newcommand{\fitftop}{\ensuremath{\left(37.9\,^{+\enspace7.3}_{-\enspace7.0}\right)\%}\xspace}
\newcommand{\fitfqcd}{\ensuremath{\left(11.3\pm 1.2\right)\%}}

\newcommand{\offsetmtoptopo}{\ensuremath{1.375\pm0.085}\xspace}
\newcommand{\offsetatmtoptopo}{\ensuremath{175}\xspace}
\newcommand{\slopemtoptopo}{\ensuremath{1.034\pm0.011}\xspace}
\newcommand{\pullwidthmtoptopo}{\ensuremath{1.06\pm0.01}\xspace}
\newcommand{\offsetjestopo}{\ensuremath{-0.028\pm0.001}\xspace}
\newcommand{\offsetatjestopo}{\ensuremath{1.0}\xspace}
\newcommand{\slopejestopo}{\ensuremath{0.934\pm0.021}\xspace}
\newcommand{\pullwidthjestopo}{\ensuremath{1.09\pm0.01}\xspace}
\newcommand{\offsetmtopbtag}{\ensuremath{1.932\pm0.085}\xspace}
\newcommand{\offsetatmtopbtag}{\ensuremath{175}\xspace}
\newcommand{\slopemtopbtag}{\ensuremath{1.018\pm0.011}\xspace}
\newcommand{\pullwidthmtopbtag}{\ensuremath{1.11\pm0.01}\xspace}
\newcommand{\offsetjesbtag}{\ensuremath{-0.028\pm0.001}\xspace}
\newcommand{\offsetatjesbtag}{\ensuremath{1.0}\xspace}
\newcommand{\slopejesbtag}{\ensuremath{0.945\pm0.021}\xspace}
\newcommand{\pullwidthjesbtag}{\ensuremath{1.09\pm0.01}\xspace}

\newcommand{\beginwidetext}{\begin{widetext}}
\newcommand{\myendwidetext}{\end{widetext}}
\newcommand{\smallskiponlyintwocol}{\smallskip}
\newcommand{\newlineonlyintwocol}{\\}
\newcommand{\nonumberonlyintwocol}{\nonumber}

\newcommand{\tableDATAkinematiccuts}{
\begin{table}[htbp]
\begin{center}
\begin{tabular}{@{\,}l@{\quad}r@{\quad}l@{\,}}
\hline
\hline
\smallskip
 charged lepton            & \begin{tabular}{@{}r@{}}$\et\!>\!20\,\GeV$\\$\pt\!>\!20\,\GeV$\end{tabular}
                           & \begin{tabular}{@{}l@{}}$|\eta|\!<\!1.1$ (electrons)\\ $|\eta|\!<\!2.0$ (muons)\end{tabular}\\
\smallskip
 exactly 4 jets            & $\et\!>\!20\,\GeV$     & $|\eta|\!<\!2.5$ \\
 missing transverse energy\hspace{-1ex} & $\etmiss\!>\!20\,\GeV$ & \\
\hline
\hline
\end{tabular}
\caption{A summary of the kinematic event selection.  In addition,
  quality and isolation criteria are applied.}
\label{kinematiccuts.table}
\end{center}
\end{table}
}

\newcommand{\eqDATAdeltaphicutmujets}{
\begin{eqnarray}
  \nonumber
    \Delta\phi\left(\mu,\,\etmiss\right) \ >
  & &
    0.1\pi \left( 1 - \frac{\etmiss}{50\,\GeV} \right)
    \ \ {\rm and} 
  \\
  \label{deltaphicut_mujets.eqn}
    \Delta\phi\left(\mu,\,\etmiss\right) \ <
  & \pi \, - \! &
    0.2\pi \left( 1 - \frac{\etmiss}{30\,\GeV} \right)
\end{eqnarray}
}

\newcommand{\eqDATAepssgnbkg}{
\begin{eqnarray}
  \nonumber
    \epsilon^{\Ntag}_{\rm sig} 
  & \!\! = \!\! & 
    \left\langle
    \frac{1}{2}
    \left(   \epsilon^{\Ntag}_{t\overline{t}} (\{\alpha_i\}=bbud)
           + \epsilon^{\Ntag}_{t\overline{t}} (\{\alpha_i\}=bbcs) \right) 
    \right\rangle
    \ ,
  \\
  \label{eq:epssgnbkg}
    \epsilon^{\Ntag}_{\rm bkg} 
  & \!\! = \!\! & 
    \left\langle
    \sum_{\Phi} f_{\Phi} \epsilon^{\Ntag}_{\Phi} 
    \right\rangle
    \ ,
\end{eqnarray}
}

\newcommand{\eqMEpevt}{
\begin{eqnarray}
  \label{eq:MEpevt}
    P\!_{evt}\!\left(x;\mtop,JES,\ftop\right) = 
                     \ftop\! & \!\!\!\! &\! \psgn\! \left(x;\mtop,JES\right) \nonumber \\
     +\! \left(1\!-\!\ftop\right)\! & \!\!\!\! &\! \pbkg\! \left(x;JES\right) 
  \ .  
\end{eqnarray}
}

\newcommand{\eqMEdsigmahs}{
\begin{eqnarray}
  \label{eq:MEdsigmahs}
    {\rm d}\sigma(\qqbar\!\to\!\ttbar\!\to\! y;\mtop)
   = 
    \frac{(2 \pi)^{4}\! \left|\mathscr{M}\left(\qqbar\!\to\!\ttbar\!\to\! y\right)\right|^{2}}
         {q_1 q_2 s}
    {\rm d}\Phi_{6}
   \ . \nonumberonlyintwocol \newlineonlyintwocol
\end{eqnarray}
}

\newcommand{\eqMEdPpp}{
\begin{eqnarray}
  \label{eq:MEdPpp}
    {\rm d}\sigma(\ppbar\to\ttbar\to y;\,\mtop) && \\
    = 
    \int\limits_{q_1,q_2} \mathop{\sum_{q_1,\,q_2}^{}}_{\rm flavors} &&
    {\rm d}q_1 {\rm d}q_2\ f(q_{1})\ f(q_{2}) \nonumber \\ 
    && {\rm d}\sigma(\qqbar\to\ttbar\to y;\,\mtop)
  \ , \nonumber
\end{eqnarray}
}

\newcommand{\eqMEdsigmapp}{
\begin{eqnarray}
  \label{eq:MEdsigmapp}
    &&{\rm d}\sigma (\ppbar\to\ttbar\to x;\,\mtop,JES) \\
    &&=
    \int\limits_{y} {\rm d}\sigma (\ppbar\to\ttbar\to y;\,\mtop)  
    W(x,y;\,JES) \nonumber 
  \ .
\end{eqnarray}
}

\newcommand{\eqMEsigmaobs}{
\begin{eqnarray}
  \label{eq:MEsigmaobs}
    \!\!\!\!\!\!\!\!
    && \sigma_{\rm obs}(\ppbar\to\ttbar;\,\mtop,JES) \\
    \!\!\!\!\!\!\!\!
    && =
    \int\limits_{x,y} \!
        {\rm d}\sigma(\ppbar\to\ttbar\to y;\,\mtop) 
    W(x,y;\,JES)
    f_{\rm acc}(x)
  \ , \nonumber
\end{eqnarray}
}

\newcommand{\eqMEpsgn}{
\begin{eqnarray}
  \label{eq:MEpsgn}
    &&\psgn(x;\,\mtop,JES) \\
  && = 
    \frac{{\rm d}\sigma(\ppbar\to\ttbar\to x;\,\mtop,JES)}
         {\sigma_{\rm obs}(\ppbar\to\ttbar;\,\mtop,JES)}
  \nonumber \\
  && =  
    \frac{1}{\sigma_{\rm obs}(\ppbar\to\ttbar;\,\mtop,JES)}
  \nonumber \\
  && 
    \quad \times
    \int\limits_{q_1,q_2,y} 
    \mathop{\sum_{q_1,\,q_2}^{}}_{\rm flavors}
    {\rm d}q_1 {\rm d}q_2 f(q_{1}) f(q_{2}) 
  \nonumber \\
  && \ \quad \quad \qquad \qquad \ \frac{(2 \pi)^{4} \left|\mathscr{M}(\qqbar\to\ttbar\to y)\right|^{2}}
         {q_1 q_2 s}
    \,{\rm d}\Phi_{6} 
  \nonumber \\
  &&  \ \quad \quad \qquad \qquad \ W(x,y;\,JES)
  \ . \nonumber
\end{eqnarray}
}

\newcommand{\eqMEpbkg}{
\begin{eqnarray}
  \label{eq:MEpbkg}
  \!\!\!\!\!\! && \pbkg(x;\,JES)
  \\
  \!\!\!\!\!\! && =
    \frac{1}{\sigma_{\rm obs}(\ppbar\!\to\!\wjetsshort;JES)}
  \nonumber \\
  \!\!\!\!\!\! &&
    \quad \times\!
    \int\limits_{q_1,q_2,y} \!
    \mathop{\sum_{q_1,\,q_2}^{}}_{\rm flavors}\!
    {\rm d}q_1 {\rm d}q_2\, f(q_{1})\, f(q_{2})\,
  \nonumber \\
  \!\!\!\!\!\! && \quad \quad \qquad \qquad \ \frac{(2 \pi)^{4} \left|\mathscr{M}(\qqbar\to\wjetsshort\to y)\right|^{2}}
         {q_1 q_2 s}\,
    {\rm d}\Phi_{6}\,
  \nonumber \\
  \!\!\!\!\!\! && \quad \quad \qquad \qquad \ W(x,y;JES)
  \ , \nonumber
\end{eqnarray}
}

\newcommand{\eqMElhoodfnc}{
\begin{eqnarray}
  \label{eq:MElhood-fnc}
  &&
    L(x_1,..,x_N;\,\mtop,JES,\ftop) \\
  \nonumber
  &&
  = 
    \prod_{i=1}^{N}\pevt(x_i;\,\mtop,JES,\ftop)
  \, .
\end{eqnarray}
}

\newcommand{\eqMElhoodnoftop}{
\begin{eqnarray}
\label{eq:MElhood-noftop}
  &&  L\left(x_1,..,x_N;\,\mtop,JES,\ftopbest(\mtop,JES)\right)
  \\
  && = \prod_{i=1}^{N} 
      P_{\rm evt}\left(x_i;\,\mtop,JES,\ftopbest(\mtop,JES)\right)
  \nonumber
\end{eqnarray}
}

\newcommand{\eqMEtfdefinitiontopo}{
\begin{eqnarray}
  \label{eq:tfdefinition-topo}
  \!\!\!\! &&
    W(x,y;\,JES)
  \\
  \nonumber
  \!\!\!\! &&
  = 
    W_{\mu}\left( \qoverptmurec,\qoverptmugen \right)
  \\
  \nonumber
  \!\!\!\! &&
  \phantom{=}
    \times
    \frac{1}{24} 
    \sum_{i=1}^{24}
    \delta({\rm angles}) 
    \prod_{j=1}^{4} 
    W_{\rm jet}(E_{{\rm jet}\,j},\ E_{{\rm quark}\,k};\ JES)
  \ ,
\end{eqnarray}
}

\newcommand{\eqMEtfdefinitionbtag}{
\begin{eqnarray}
  \label{eq:tfdefinition-btag}
  \!\!\!\! &&
    W(x,y;\,JES)
  \\
  \nonumber
  \!\!\!\! &&
  = 
    W_{\mu}\left( \qoverptmurec,\qoverptmugen \right)
  \\
  \nonumber
  \!\!\!\! &&
  \phantom{=}
    \times
    \frac{\displaystyle\sum_{i=1}^{24}
          w_i \
          \delta({\rm angles})
	  \prod_{j=1}^{4} 
	  W_{\rm jet}(E_{{\rm jet}\,j},\ E_{{\rm quark}\,k};\ JES)}
         {\displaystyle\sum_{i=1}^{24} w_i}
\end{eqnarray}
}

\newcommand{\eqMEtf}{
\begin{eqnarray}
\label{eq:MEtf}
\!\!\!\!\!\!&&
W_{\rm jet}(E_{j},E_{q};JES\!=\!1)
=
\frac{1}{\sqrt{2\pi}(p_2+p_3p_5)} \times \\
\!\!\!\!\!\!&&
\nonumber
\Bigglb[ \exp\!{\left(\!-\frac{[(E_{j}\!-\!E_{q})\!-\!p_1]^2}{2\,p_2^2}
\right)} 
\!+\! p_3\exp\!{\left(\!-\frac{\![(E_{j}\!-\!E_{q})\!-\!p_4]^2}{2\,p_5^2}\!\right)}\! \Biggrb] \,. \nonumber
\end{eqnarray}
}

\newcommand{\eqMEtfmu}{
\begin{eqnarray}
  \label{eq:MEtfmu}
  && W_{\mu}\left( \qoverptmurec,\qoverptmugen \right) \\
  && =\
    \frac{1}{\sqrt{2\pi}\sigma}\, 
    \exp\!\left[ -\frac{1}{2} \left( \frac{ \qoverptmurec - \qoverptmugen }
                                          { \sigma }
                            \right)^2
        \right]
  \ , \nonumber
\end{eqnarray}
}

\newcommand{\eqMEsignalMEFFbar}{
\begin{eqnarray}
\label{eq:MEsignalME_F}
  F
& = &
  \frac{g_w^4}{4}
  \left( \frac{   \mbellnu^2 - \mellnu^2 }
              {   \left( \mbellnu^2 - \mt^2 \right)^2 
                + \left( \mt \Gt \right)^2 } \right) \nonumber \\
  & & \times
  \left( \frac{   \mbellnu^2 \left( 1 - \chatbell^2 \right)
                + \mellnu^2 \left( 1 + \chatbell \right)^2 }
              {   \left( \mellnu^2 - \mW^2 \right)^2
                + \left( \mW \GW \right)^2 } \right)
\ , \\ \nonumber \\
\label{eq:MEsignalME_Fbar}
  \overline{F}
& = &
  \frac{g_w^4}{4}
  \left( \frac{   \mbdubar^2 - \mdubar^2 }
              {   \left( \mbdubar^2 - \mt^2 \right)^2 
                + \left( \mt \Gt \right)^2 } \right) \nonumber \\
  & & \times
  \left( \frac{   \mbdubar^2 \left( 1 - \chatbbard^2 \right)
                + \mdubar^2 \left( 1 + \chatbbard \right)^2 }
              {   \left( \mdubar^2 - \mW^2 \right)^2
                + \left( \mW \GW \right)^2 } \right)
\end{eqnarray}
}

\newcommand{\eqMEsignalMEFbarsymm}{
\begin{eqnarray}
\label{eq:MEsignalME_Fbarsymm}
  \overline{F}
 & = & 
  \frac{g_w^4}{4}
  \left( \frac{   \mbdubar^2 - \mdubar^2 }
              {   \left( \mbdubar^2 - \mt^2 \right)^2 
                + \left( \mt \Gt \right)^2 } \right) \nonumber \\
  & & \times              
  \left( \frac{   \mbdubar^2 \left( 1 - \chatbbard^2 \right)
                + \mdubar^2 \left( 1 + \chatbbard^2 \right) }
              {   \left( \mdubar^2 - \mW^2 \right)^2
                + \left( \mW \GW \right)^2 } \right)
\end{eqnarray}
}

\newcommand{\captionparamstwo}{$b$ quark transfer function parameters for jets without a
                               muon (top) and for jets containing a muon (bottom) ($a_i$ in GeV).}

\newcommand{\eqMTOPFITTOPOoffsetpullcorrection}{
\begin{eqnarray}
  \nonumber
      \mtop 
    & = &
        \displaystyle
        \frac{ \mtop^{\rm fit} - o_{\mtop} - 175\,\GeV }
             { s_{\mtop} }
        + 175\,\GeV
    \ ,
  \\
  \nonumber
      \Delta\mtop
    & = &
      w_{\mtop} \left(\Delta\mtop\right)^{\rm fit}
    \ ,
  \vspace{2ex}\\
  \nonumber
      JES
    & = &
        \displaystyle
        \frac{ JES^{\,\rm fit} - o_{JES} - 1 }
             { s_{JES} }
        + 1
    \ ,\ {\rm and}
  \\
  \label{MTOPFITTOPOoffsetpullcorrection.eqn}
      \Delta JES
    & = &
      w_{JES} \left(\Delta JES\right)^{\rm fit}
    \ .
\end{eqnarray}
}

\newcommand{\figMEcalibwidth}{0.25}

\newcommand{\eqMElhoodfncbtag}{
\begin{eqnarray}
  \label{eq:MElhood-fncb}
  &&
    L(x_1,..,x_N;\,\mtop,JES,\ftop) 
  \\
  \nonumber
  &&
  = 
    \prod_{\Ntag=0,1,\geq2} \prod_{i=1}^{N^{\Ntag}} 
    \pevtb(x_i;\,\mtop,JES,\ftop^{\Ntag})
  \ ,
\end{eqnarray}
}

\newcommand{\eqMEpevtbtag}{
\begin{eqnarray}
    \pevtb\!\left(x;\,\mtop,JES,\ftop\right) 
   = \ftopb \!& \!\!\!\! &\! \psgnb\left(x;\,\mtop,JES\right)  
  \nonumber 
  \\
  \label{eq:MEpevtb}
    + (1\!-\!\ftopb) \!& \!\!\!\! &\! \pbkg\left(x;\,JES\right) 
  \, .
\end{eqnarray}
}

\newcommand{\eqMElhoodnoftopbtag}{
\begin{eqnarray}
\label{eq:MElhood-noftopb}
  &&
    \!\!\!\!\! L\left(x_1,..,x_N;\,\mtop,JES,\ftopbest(\mtop,JES)\right) 
  \\
  \nonumber
  &&
  \!\!\!\!\! = \!\!\!\!\!\!\!\!\! 
    \prod_{\Ntag=0,1,\geq2} \!\! \prod_{i=1}^{N^{\Ntag}} 
    \!\!\!\!\!\pevtb\!\!\left(\!x_i;\mtop,JES,{\ftopbest}^{\Ntag}(\mtop,JES)\right)
\end{eqnarray}
}

\newcommand{\figbtagwidth}{0.45}
\newcommand{\figbtagwidthsmall}{0.25}

\begin{document}
\hspace{5.2in}\mbox{FERMILAB-PUB-06/353-E}

\title{Measurement of the top quark mass in
the lepton+jets final state with the matrix element method}

%
\author{                                                                      
V.M.~Abazov,$^{35}$                                                           
B.~Abbott,$^{75}$                                                             
M.~Abolins,$^{65}$                                                            
B.S.~Acharya,$^{28}$                                                          
M.~Adams,$^{51}$                                                              
T.~Adams,$^{49}$                                                              
M.~Agelou,$^{17}$                                                             
E.~Aguilo,$^{5}$                                                              
S.H.~Ahn,$^{30}$                                                              
M.~Ahsan,$^{59}$
% ...
\\                                                                            
\vskip 0.30cm                                                                 
\centerline{(D\O\ Collaboration)}                                             
\vskip 0.30cm                                                                 
}                                                                             
\affiliation{                                                                 
\centerline{$^{1}$Universidad de Buenos Aires, Buenos Aires, Argentina}       
\centerline{$^{2}$LAFEX, Centro Brasileiro de Pesquisas F{\'\i}sicas,         
                  Rio de Janeiro, Brazil}                                     
\centerline{$^{3}$Universidade do Estado do Rio de Janeiro,                   
                  Rio de Janeiro, Brazil}                                     
\centerline{$^{4}$Instituto de F\'{\i}sica Te\'orica, Universidade            
                  Estadual Paulista, S\~ao Paulo, Brazil}                     
\centerline{$^{5}$University of Alberta, Edmonton, Alberta, Canada,           
                  Simon Fraser University, Burnaby, British Columbia, Canada,}
\centerline{York University, Toronto, Ontario, Canada, and                    
                  McGill University, Montreal, Quebec, Canada}                
\centerline{$^{6}$University of Science and Technology of China, Hefei,       
                  People's Republic of China}                                 
\centerline{$^{7}$Universidad de los Andes, Bogot\'{a}, Colombia}             
\centerline{$^{8}$Center for Particle Physics, Charles University,            
                  Prague, Czech Republic}                                     
\centerline{$^{9}$Czech Technical University, Prague, Czech Republic}         
\centerline{$^{10}$Center for Particle Physics, Institute of Physics,         
                   Academy of Sciences of the Czech Republic,                 
                   Prague, Czech Republic}                                    
\centerline{$^{11}$Universidad San Francisco de Quito, Quito, Ecuador}        
\centerline{$^{12}$Laboratoire de Physique Corpusculaire, IN2P3-CNRS,         
                   Universit\'e Blaise Pascal, Clermont-Ferrand, France}      
\centerline{$^{13}$Laboratoire de Physique Subatomique et de Cosmologie,      
                   IN2P3-CNRS, Universite de Grenoble 1, Grenoble, France}    
\centerline{$^{14}$CPPM, IN2P3-CNRS, Universit\'e de la M\'editerran\'ee,     
                   Marseille, France}                                         
\centerline{$^{15}$IN2P3-CNRS, Laboratoire de l'Acc\'el\'erateur              
                   Lin\'eaire, Orsay, France}                                 
\centerline{$^{16}$LPNHE, IN2P3-CNRS, Universit\'es Paris VI and VII,         
                   Paris, France}                                             
\centerline{$^{17}$DAPNIA/Service de Physique des Particules, CEA, Saclay,    
                   France}                                                    
\centerline{$^{18}$IPHC, IN2P3-CNRS, Universit\'e Louis Pasteur, Strasbourg,  
                    France, and Universit\'e de Haute Alsace,                 
                    Mulhouse, France}                                         
\centerline{$^{19}$Institut de Physique Nucl\'eaire de Lyon, IN2P3-CNRS,      
                   Universit\'e Claude Bernard, Villeurbanne, France}         
\centerline{$^{20}$III. Physikalisches Institut A, RWTH Aachen,               
                   Aachen, Germany}                                           
\centerline{$^{21}$Physikalisches Institut, Universit{\"a}t Bonn,             
                   Bonn, Germany}                                             
\centerline{$^{22}$Physikalisches Institut, Universit{\"a}t Freiburg,         
                   Freiburg, Germany}                                         
\centerline{$^{23}$Institut f{\"u}r Physik, Universit{\"a}t Mainz,            
                   Mainz, Germany}                                            
\centerline{$^{24}$Ludwig-Maximilians-Universit{\"a}t M{\"u}nchen,            
                   M{\"u}nchen, Germany}                                      
\centerline{$^{25}$Fachbereich Physik, University of Wuppertal,               
                   Wuppertal, Germany}                                        
\centerline{$^{26}$Panjab University, Chandigarh, India}                      
\centerline{$^{27}$Delhi University, Delhi, India}                            
\centerline{$^{28}$Tata Institute of Fundamental Research, Mumbai, India}     
\centerline{$^{29}$University College Dublin, Dublin, Ireland}                
\centerline{$^{30}$Korea Detector Laboratory, Korea University,               
                   Seoul, Korea}                                              
\centerline{$^{31}$SungKyunKwan University, Suwon, Korea}                     
\centerline{$^{32}$CINVESTAV, Mexico City, Mexico}                            
\centerline{$^{33}$FOM-Institute NIKHEF and University of                     
                   Amsterdam/NIKHEF, Amsterdam, The Netherlands}              
\centerline{$^{34}$Radboud University Nijmegen/NIKHEF, Nijmegen, The          
                  Netherlands}                                                
\centerline{$^{35}$Joint Institute for Nuclear Research, Dubna, Russia}       
\centerline{$^{36}$Institute for Theoretical and Experimental Physics,        
                   Moscow, Russia}                                            
\centerline{$^{37}$Moscow State University, Moscow, Russia}                   
\centerline{$^{38}$Institute for High Energy Physics, Protvino, Russia}       
\centerline{$^{39}$Petersburg Nuclear Physics Institute,                      
                   St. Petersburg, Russia}                                    
\centerline{$^{40}$Lund University, Lund, Sweden, Royal Institute of          
                   Technology and Stockholm University, Stockholm,            
                   Sweden, and}                                               
\centerline{Uppsala University, Uppsala, Sweden}                              
\centerline{$^{41}$Physik Institut der Universit{\"a}t Z{\"u}rich,            
                   Z{\"u}rich, Switzerland}                                   
\centerline{$^{42}$Lancaster University, Lancaster, United Kingdom}           
\centerline{$^{43}$Imperial College, London, United Kingdom}                  
\centerline{$^{44}$University of Manchester, Manchester, United Kingdom}      
\centerline{$^{45}$University of Arizona, Tucson, Arizona 85721, USA}         
\centerline{$^{46}$Lawrence Berkeley National Laboratory and University of    
                   California, Berkeley, California 94720, USA}               
\centerline{$^{47}$California State University, Fresno, California 93740, USA}
\centerline{$^{48}$University of California, Riverside, California 92521, USA}
\centerline{$^{49}$Florida State University, Tallahassee, Florida 32306, USA} 
\centerline{$^{50}$Fermi National Accelerator Laboratory,                     
            Batavia, Illinois 60510, USA}                                     
\centerline{$^{51}$University of Illinois at Chicago,                         
            Chicago, Illinois 60607, USA}                                     
\centerline{$^{52}$Northern Illinois University, DeKalb, Illinois 60115, USA} 
\centerline{$^{53}$Northwestern University, Evanston, Illinois 60208, USA}    
\centerline{$^{54}$Indiana University, Bloomington, Indiana 47405, USA}       
\centerline{$^{55}$University of Notre Dame, Notre Dame, Indiana 46556, USA}  
\centerline{$^{56}$Purdue University Calumet, Hammond, Indiana 46323, USA}    
\centerline{$^{57}$Iowa State University, Ames, Iowa 50011, USA}              
\centerline{$^{58}$University of Kansas, Lawrence, Kansas 66045, USA}         
\centerline{$^{59}$Kansas State University, Manhattan, Kansas 66506, USA}     
\centerline{$^{60}$Louisiana Tech University, Ruston, Louisiana 71272, USA}   
\centerline{$^{61}$University of Maryland, College Park, Maryland 20742, USA} 
\centerline{$^{62}$Boston University, Boston, Massachusetts 02215, USA}       
\centerline{$^{63}$Northeastern University, Boston, Massachusetts 02115, USA} 
\centerline{$^{64}$University of Michigan, Ann Arbor, Michigan 48109, USA}    
\centerline{$^{65}$Michigan State University,                                 
            East Lansing, Michigan 48824, USA}                                
\centerline{$^{66}$University of Mississippi,                                 
            University, Mississippi 38677, USA}                               
\centerline{$^{67}$University of Nebraska, Lincoln, Nebraska 68588, USA}      
\centerline{$^{68}$Princeton University, Princeton, New Jersey 08544, USA}    
\centerline{$^{69}$State University of New York, Buffalo, New York 14260, USA}
\centerline{$^{70}$Columbia University, New York, New York 10027, USA}        
\centerline{$^{71}$University of Rochester, Rochester, New York 14627, USA}   
\centerline{$^{72}$State University of New York,                              
            Stony Brook, New York 11794, USA}                                 
\centerline{$^{73}$Brookhaven National Laboratory, Upton, New York 11973, USA}
\centerline{$^{74}$Langston University, Langston, Oklahoma 73050, USA}        
\centerline{$^{75}$University of Oklahoma, Norman, Oklahoma 73019, USA}       
\centerline{$^{76}$Oklahoma State University, Stillwater, Oklahoma 74078, USA}
\centerline{$^{77}$Brown University, Providence, Rhode Island 02912, USA}     
\centerline{$^{78}$University of Texas, Arlington, Texas 76019, USA}          
\centerline{$^{79}$Southern Methodist University, Dallas, Texas 75275, USA}   
\centerline{$^{80}$Rice University, Houston, Texas 77005, USA}                
\centerline{$^{81}$University of Virginia, Charlottesville,                   
            Virginia 22901, USA}                                              
\centerline{$^{82}$University of Washington, Seattle, Washington 98195, USA}  
}                                                                             

\date{September 27, 2006}
           
\begin{abstract}

We present a measurement of the top quark mass with the Matrix Element
method in the lepton+jets final state.
As the energy scale for calorimeter jets 
represents the dominant source of systematic
uncertainty, the Matrix Element likelihood is extended by an
additional parameter, which is defined as a global multiplicative 
factor applied to the standard energy scale.
The top quark mass is obtained from a fit that
yields the combined statistical and systematic jet energy scale
uncertainty.
Using a data set of $370\,\ipb$ taken with the \dzero
experiment at \runii of the Fermilab Tevatron Collider, the mass of the
top quark is measured using topological information to be:
\begin{equation}
  \hspace{2pt}
  \mtop^{\ljets}(\mbox{\rm topo})    = \result\,\GeVcc  \nonumber \ ,
\end{equation}
and when information about identified $b$ jets is included:
\begin{equation}
  \mtop^{\ljets}(\mbox{\rm $b$-tag}) = \resultb\,\GeVcc \nonumber \ .
\end{equation}
The measurements yield a jet energy scale consistent with the
reference scale.

\end{abstract}
\pacs{14.65.Ha, 12.15.Ff} 

\maketitle


\section{Introduction}
\label{sec:intro}
The origin of mass of the elementary fermions of the standard model
is one of the central questions in particle physics.
Of the six known quark flavors, the top quark is unique in that its mass
can be measured to the percent-level with the current data
of the Fermilab Tevatron Collider.
There is particular interest in a precision measurement of the top quark
mass because of its dominant contribution in loop corrections to electroweak
observables such as the $\rho$ parameter.
Within the standard model, a precise determination of the top quark
mass in combination with existing electroweak data can place significant 
constraints on the mass of the Higgs boson~\cite{bib:Zbible}.

To date, proton-antiproton collisions at the Fermilab Tevatron Collider provide
the only possibility to produce top quarks.
During \runi of the Tevatron in the 1990s, at a proton-antiproton
center-of-mass energy of $\sqrt{s}=1.8\,\TeV$, the top quark was
discovered by the CDF and
\dzero~\cite{bib:topdiscovery} experiments, and its mass was
measured~\cite{bib:topmassruni}. 
Since the beginning of \runii in 2002,
the Tevatron is running with an increased
luminosity at a center-of-mass energy of $\sqrt{s}=1.96\,\TeV$.
The CDF experiment has recently published a measurement of the top
quark mass using \runii data~\cite{bib:topmasscdfrunii}.
The first \dzero measurement of the top quark mass at \runii is
described in this paper.

Top quarks are produced in proton-antiproton collisions either in
pairs (production of \ttbar pairs via the strong interaction) or 
singly (via the electroweak interaction).
Only the first process has been observed so far and is used to measure
the top quark mass.
In the standard model, the top quark essentially always decays to a
\bquark~quark and a real \W boson.
The topology of a \ttbar event is therefore determined by the subsequent
\W boson decays.
The so-called lepton+jets topology, where one \W boson decays to an
electron or muon and the corresponding neutrino while the other decays
hadronically, allows the most precise experimental measurement of the
top quark mass.
These events are characterized by an energetic, isolated electron or
muon (charge conjugate modes are implicitly included
throughout this paper), missing transverse energy relative to
the beamline from the
neutrino, and four energetic jets.

This paper describes a measurement of the top quark mass with the
\dzero detector, using lepton+jets events from 370\,\ipb of data
collected during \runii of the Fermilab Tevatron Collider.
To make maximal use of kinematic information, the events selected
are analyzed with the Matrix Element method.
This method was developed by \dzero for the \runi measurement of the
top quark mass~\cite{bib:nature} and led to the single most precise
measurement during \runi.
For each event, a probability is calculated as a function of the top
quark mass that this event has arisen from \ttbar production.
A similar probability is computed for the main background process, 
which is the production of a leptonically decaying \W boson produced
in association with jets.
The detector resolution is taken into account in the calculation of
these probabilities.
The top quark mass is then extracted from the joint probability
calculated for all selected events.
To reduce the sensitivity to the energy scale of the jets measured in
the calorimeter, the Matrix Element method has been extended so 
this scale can be determined simultaneously with the top quark mass 
from the same event sample~\cite{bib:philipp,bib:topmasscdfrunii}.

The paper is organized as follows.
Section~\ref{sec:dzero} gives a brief overview of the \dzero detector.
The event reconstruction, selection, and simulation are discussed in
Section~\ref{sec:data}.
A detailed description of the Matrix Element method is given in
Section~\ref{sec:ME}.
The top quark mass fit is described in Sections~\ref{sec:MEfit}
and~\ref{sec:MEfitbtag} for the analyses before and after the use
of $b$-tagging information, and 
Section~\ref{sec:systuncs} lists the systematic uncertainties.
Section~\ref{sec:conclusions} summarizes the results.

\section{The \dzero Detector}
\label{sec:dzero}
We use a right-handed coordinate system whose origin is at the 
center of the detector, with the proton beam defining
the positive $z$ direction.
The \dzero detector consists of a magnetic central-tracking system, 
comprising a silicon microstrip tracker (SMT) and a central fiber 
tracker (CFT), both located within a 2~T superconducting solenoidal 
magnet~\cite{run2det}. The SMT has $\approx 800,000$ individual strips, 
with typical pitch of $50-80$ $\mu$m, and a design optimized for 
tracking and vertexing capability at pseudorapidities of $|\eta|<2.5$. 
The system has a six-barrel longitudinal structure, each with a set 
of four layers arranged axially around the beam pipe, and interspersed 
with 16 radial disks. The CFT has eight thin coaxial barrels, each 
supporting two doublets of overlapping scintillating fibers of 0.835~mm 
diameter, one doublet being parallel to the collision axis, and the 
other alternating by $\pm 3^{\circ}$ relative to the axis. Light signals 
are transferred via clear fibers to solid-state photon counters (VLPC) 
that have $\approx 80$\% quantum efficiency.

Central and forward preshower detectors located just outside of the
superconducting coil (in front of the calorimetry) are constructed of
several layers of extruded triangular scintillator strips that are read 
out using wavelength-shifting fibers and VLPCs. The next layer of 
detection involves three liquid-argon/uranium calorimeters: a central 
section (CC) covering $|\eta|$ up to $\approx 1.1$, and two end
calorimeters (EC) that extend coverage to $|\eta|\approx 4.2$, all 
housed in separate cryostats~\cite{run1det}. In addition to the preshower 
detectors, scintillators between the CC and EC cryostats provide 
sampling of developing showers at $1.1<|\eta|<1.4$.

A muon system~\cite{run2muon} is located beyond the calorimetry and consists of a 
layer of tracking detectors and scintillation trigger counters 
before 1.8~T iron toroids, followed by two similar layers after
the toroids. Tracking at $|\eta|<1$ relies on 10~cm wide drift
tubes~\cite{run1det}, while 1~cm mini-drift tubes are used at
$1<|\eta|<2$.

Luminosity is measured using plastic scintillator arrays located in front 
of the EC cryostats, covering $2.7 < |\eta| < 4.4$. 
Trigger and data acquisition systems are designed to accommodate 
the high luminosities of Run II. Based on preliminary information from 
tracking, calorimetry, and muon systems, the output of the first level 
of the trigger is used to limit the rate for accepted events to 
$\approx$ 2~kHz. At the next trigger stage, with more refined 
information, the rate is reduced further to $\approx$ 1~kHz. These
first two levels of triggering rely mainly on hardware and firmware.
The third and final level of the trigger, with access to all the event 
information, uses software algorithms and a computing farm, and reduces 
the output rate to $\approx$ 50~Hz, which is written to tape.

\section{Event Reconstruction, Selection, and Simulation}
\label{sec:data}
This paper describes the analysis of 370\,\ipb of data taken between
April 2002 and August 2004.
Events considered for the analysis must initially pass trigger conditions 
requiring the presence of an electron or muon and a jet.
In the \ejets trigger, an electron with transverse momentum
$\pt>15\,\GeV$ within $|\eta|<1.1$ is required.  
In addition, a jet 
reconstructed using a cone algorithm with radius
$\DeltaR \equiv \sqrt{(\Delta\eta)^2 + (\Delta\phi)^2} = 0.5$ is
required, with a minimum jet transverse energy,
$\et$, of 15 or 20\,\GeV depending on the
period of data taking.
In the \mujets trigger, a muon detected outside the toroidal
magnet (corresponding to an effective minimum momentum of around
3\,\GeV) is required, along with a jet with transverse energy 
$\et$ of at least 20 or 25\,\GeV, 
depending on the data taking period.

The offline reconstruction and selection of the events is 
described in detail in the following sections.
The kinematic selection criteria are also summarized in
Table~\ref{kinematiccuts.table}.
A total of 
86 \ejets
and
89 \mujets
events are selected.

\tableDATAkinematiccuts

\subsection{Charged Lepton Selection}
\label{sec:leptonsel}
Candidates for a charged lepton from \W decay
are required to have a transverse momentum $\pt$ of at
least 20\,\GeV and must be within $|\eta|<1.1$ for electrons and
$|\eta|<2.0$ for muons.
In addition, charged leptons have to pass quality and isolation
criteria described below.

Electrons must deposit at least 90\% of their energy in the
electromagnetic calorimeter within a cone of radius $\DeltaR = 0.2$
around the shower axis.
The transverse and longitudinal shower shapes must be consistent with
those expected for an electron, based on Monte Carlo simulation,
with efficiencies corrected for observed differences between data
and Monte Carlo.
A good spatial match of the reconstructed track in the tracking
system and the shower position in the calorimeter is required.
Electrons must be isolated, i.e.,\ the energy in the
hollow cone $0.2 < \DeltaR < 0.4$ around the shower axis 
must not exceed 15\% of the electron energy.
Finally, a likelihood is formed by combining the above variables with 
information about the impact parameter of the matched track relative to
the primary interaction vertex, the number of 
tracks in a cone with radius $\DeltaR=0.05$ 
around the electron candidate, the $\pt$ of tracks (excluding the 
track matched to the electron) within a 
cone with radius $\DeltaR=0.4$, and the number of strips in the 
central preshower detector associated with the electron.
The value of this likelihood is required
to be consistent with expectations for high-$\pt$ isolated electrons.

For each muon, a match of muon track segments inside and outside the toroid is
required.
The timing information from scintillator hits associated with the muon
must be consistent with that of a particle produced in the \ppbar
collision, thereby rejecting cosmic rays.
A track reconstructed in the tracking system and pointing to the event
vertex is required to be matched to the track in the muon system.
The muon must be separated from jets, satisfying $\DeltaR(\mu,\,{\rm jet})>0.5$
for all jets in the event.
Finally, the muon must pass an isolation criterion based on the energy of
calorimeter clusters and tracks around the muon: The calorimeter
transverse energy in the hollow cone $0.1 < \DeltaR < 0.4$ around the muon
direction is required to be less than 8\% of
the muon transverse momentum, and the sum of transverse momenta of all
other tracks within a cone of radius $\DeltaR=0.5$ around the muon
direction must be smaller than 6\% of the muon $\pt$.

\subsection{Jet Reconstruction and Selection}
\label{sec:jetsel}
Jets are defined using 
a 
cone algorithm
with radius
$\DeltaR = 0.5$.  
They are required to have pseudorapidity $|\eta|<2.5$.
Calorimeter cells with negative energy or with energy below four times
the width of the average electronics noise are suppressed (unless they 
neighbor a cell of high positive energy, where the threshold is
lowered by a factor of two) in order to improve the calorimeter 
performance.
In the reconstruction, jets are considered only if they have a
minimum raw energy of 8\,\GeV.
Jets must then pass the following quality requirements:
\begin{itemize}
\item
the energy reconstructed in the electromagnetic part of the
calorimeter must be between 5\% and 95\% of
the total jet energy;
\item
the fraction of energy in the outer hadronic calorimeter must be below
40\%;
\item
the energy ratio of the most and second most energetic calorimeter
cells in the jet must be below 10;
\item
the most energetic calorimeter cell must not contain more than 90\% of
the jet energy;
\item
the jet is required to be confirmed
by the independent trigger readout; and
\item
jets within $\DeltaR<0.5$ of an isolated electromagnetic object 
(electron or photon) with $\et>15\,\GeV$ reconstructed in the calorimeter
are rejected.  The electromagnetic objects used here are obtained with
a selection similar to the electron selection described in
Section~\ref{sec:leptonsel},
but without the requirement of a track match or a cut on the
likelihood.
\end{itemize}

The analysis is restricted to events with exactly four
jets; these four jets must each have $\et>20\,\GeV$ after jet 
energy scale correction, which is described below.
The motivation for the requirement of four jets is that for each event, 
a signal probability \psgn is calculated using a leading-order matrix
element for \ttbar production, as described below in Section~\ref{sec:MEpsgn}.
Decays in which additional radiation is emitted as well as \ttbar
pairs produced in association with other jets are not modeled in the
probability.

\subsection{Jet Energy Scale}
\label{sec:jes}
The measured energy $E_{\rm jet}^{\rm reco}$ of a reconstructed jet is given
by the sum of energies deposited in the calorimeter cells associated
with the jet by the cone algorithm.
The energy $E_{\rm jet}^{\rm corr}$ of the jet before
interaction with the calorimeter is obtained from the reconstructed
jet energy as
\begin{equation}
\label{eq:jetcorr}
E_{\rm jet}^{\rm corr} = \frac{E_{\rm jet}^{\rm reco} - \energyoffset}
                      {\calorimeterresponse \showeringcorrection} \ .
\end{equation}
The corrections are applied to account for several effects:
\begin{itemize}
\item
  {\bf\boldmath Energy Offset $\energyoffset$:} Energy in the clustered cells which is
  due to noise, the underlying event, multiple interactions, energy
  pile-up, and uranium noise lead to a global offset of jet
  energies. This offset
  $\energyoffset$ is determined from energy densities in minimum bias
  events.
\item 
  {\bf\boldmath Showering Corrections $\showeringcorrection$:} A
  fraction of the jet energy is
  deposited outside of the finite-size jet cone. 
  Jet energy density profiles are analyzed to obtain the corresponding
  correction $\showeringcorrection$.
\item 
  {\bf\boldmath Calorimeter Response $\calorimeterresponse$:} Jets 
  consist of different
  particles (mostly photons, pions, kaons, \mbox{(anti-)}protons, and
  neutrons), to which the calorimeter response differs.  
  Furthermore, the energy reponse of the calorimeter
  is slightly nonlinear. The response 
  $\calorimeterresponse$ is determined from 
  $\gamma$+jets events by requiring transverse momentum balance. The
  photon energy scale is assumed to be identical to the electron scale and is
  measured independently using $Z\to ee$ events.
\end{itemize}
Note that $E_{\rm jet}^{\rm corr}$ is not the parton energy:
the parton may radiate additional quarks or gluons before
hadronization, which may or may not end up in the jet cone.
The relation between the jet and parton energies is parameterized with 
a transfer function, see Section~\ref{sec:MEtransferfcn}.
The jet energy scale is determined separately for data and Monte 
Carlo jets.
The scale depends both on the energy of the jet and on the pseudorapidity.
All jet energies in data and Monte Carlo events are corrected
according to the appropriate jet energy scale, and these corrections
are propagated to the missing transverse energy, see
Section~\ref{sec:reco-met}.

The uncertainty on the jet energy scale was the dominating systematic
uncertainty on most previous measurements of the top quark mass.
To reduce this systematic uncertainty, information from the jets arising 
from the hadronic \W decay can be used to determine an overall jet energy 
scale factor, $JES$, simultaneously with the top quark mass.
A value of $JES\neq1$ means that all jet energies need to be scaled by
a factor of $JES$ relative to the reference scale described above.

\subsection{Missing Transverse Energy}
\label{sec:reco-met}
Neutrinos can
only be identified indirectly by the imbalance of the event in the
transverse plane.
This imbalance is reconstructed from the vector sum of all calorimeter
cells with significant energy (cf.\ Section~\ref{sec:jetsel}).
The missing transverse energy is corrected for the energy scale of
jets and for muons in the event.
Only events with \mbox{$\etmiss>20\,\GeV$} are considered.

In addition, 
a cut on the difference $\Delta\phi$ between the azimuthal angles
of the lepton momentum and the missing transverse energy vector is 
imposed to reject events in which the transverse energy imbalance
originates from a poor lepton energy measurement.
This requirement depends on the scalar value of \etmiss and is
\begin{equation}
  \label{deltaphicut_ejets.eqn}
      \Delta\phi\left(e,\,\etmiss\right)   
      > 
      0.7\pi - \frac{0.045}{\GeV}\etmiss
\end{equation}
in the \ejets channel and
\eqDATAdeltaphicutmujets
in the \mujets channel.

\subsection{{\boldmath$b$} Jet Identification}
\label{sec:btagging}

A $\ttbar$ event contains two $b$ jets, while jets produced in
association with $W$ bosons predominantly originate from light quarks or
gluons. 
The signal to background ratio is therefore significantly
enhanced when requiring that one or more of the jets be identified as
$b$ jets ($b$-tagged).
\dzero developed a lifetime based $b$-tagging algorithm referred to
as SVT~\cite{bib:btagxsect}.
The algorithm starts by identifying  tracks with significant
impact parameter with respect to the primary vertex.  
Only tracks that are displaced by more than two standard
deviations are considered. 
The algorithm then requires that
these tracks form a secondary vertex displaced by more than seven standard
deviations from the primary vertex. 
For each track participating in secondary 
vertex reconstruction its impact parameter 
must have a positive projection onto the jet axis. 
Tracks with a negative projection appear to
originate from behind the primary vertex, which is a sign of a
mismeasurement. 
Tracks with negative impact parameter are used to quantify
the mistagging probability.

The performance of the SVT algorithm is extensively tested on data. 
The $b$-tagging
efficiency is verified on a dijet data sample whose $b$ jet content is
enhanced by
requiring that one of the jets be associated with a muon. 
The distribution of the transverse momentum of the muon 
relative to the associated jet axis is used to extract the
fraction of $b$ jets before and after tagging.  
The probability to tag a
light quark jet (mistag rate) is inferred from the 
rate of secondary vertices with negative impact parameter,
corrected for the contribution of heavy flavor jets to such tags and
the presence of long-lived particles in light quark jets. 
Both corrections are derived from Monte Carlo simulation. 
Both the $b$-jet tagging efficiency $\epsilon_{\rm jet}(b)$ and the light-jet
tagging rate $\epsilon_{\rm jet}(u,\ d,\ s,\ g)$ 
are parameterized as functions of the transverse jet energy and pseudorapidity. 
The efficiency $\epsilon_{\rm jet}(c)$ to tag a $c$ quark jet is
estimated based on the Monte Carlo prediction for the $b$ to $c$-jet
tagging efficiency ratio. 
These parameterizations are used to predict 
the probability for a jet of a certain flavor to be tagged.

\subsection{Simulation}
\label{sec:simulation}
Large samples of Monte Carlo simulated events are used
to determine the detector resolution, to calibrate the
method, and to cross-check the results for the top quark mass obtained
in the data.
The \alpgen~\cite{bib:alpgen} event generator is used for both signal 
and background simulation. 
The hadronization and fragmentation process is simulated using 
\pythia~\cite{bib:pythia}. 
Signal \ttbar events are simulated for top 
quark mass values of 160, 170, 175, 180, and 190~\GeV.
The main background is from $\W$+4~jets events.

All simulated events are passed through a detailed simulation of the
detector response and are then subjected to the same selection
criteria as the data.
The probability that a simulated event would have passed the trigger
conditions is calculated, taking into account the relative integrated
luminosities for which the various trigger conditions were in use.
This probability is typically between 0.9 and 1 and is accounted for 
when simulated events are used in this analysis.

Background from QCD multijet processes has not been generated in the 
simulation; instead, events that pass a selection with reversed
isolation cuts for the charged lepton have been used to model this
background.

\subsection{Sample composition}
\label{sec:sample}

Even though the
Matrix Element method yields the \ttbar content \ftop of the selected 
data sample together with the top quark mass and jet energy scale,
an independent estimate of the sample purity is obtained using a topological 
likelihood discriminant, as described in this section.
This result for the sample composition does not directly enter
in the top quark mass fit; it is used only
\begin{itemize}
\item
to obtain
the relative normalization of the signal and background probabilities
as described in Section~\ref{sec:MEpbkg} ---
this allows fitting the sample purity without large corrections to the 
result --- and
\item
to choose the sample purity in ensemble tests 
(cf.\ Sections~\ref{sec:MEvalidation}, \ref{sec:MEcalibration},
and~\ref{sec:MEcalibrationb}) according to the sample composition in
the data --- in order to
compute the expected fit uncertainties.
\end{itemize}

\begin{figure}
\begin{center}
\includegraphics[width=0.45\textwidth]{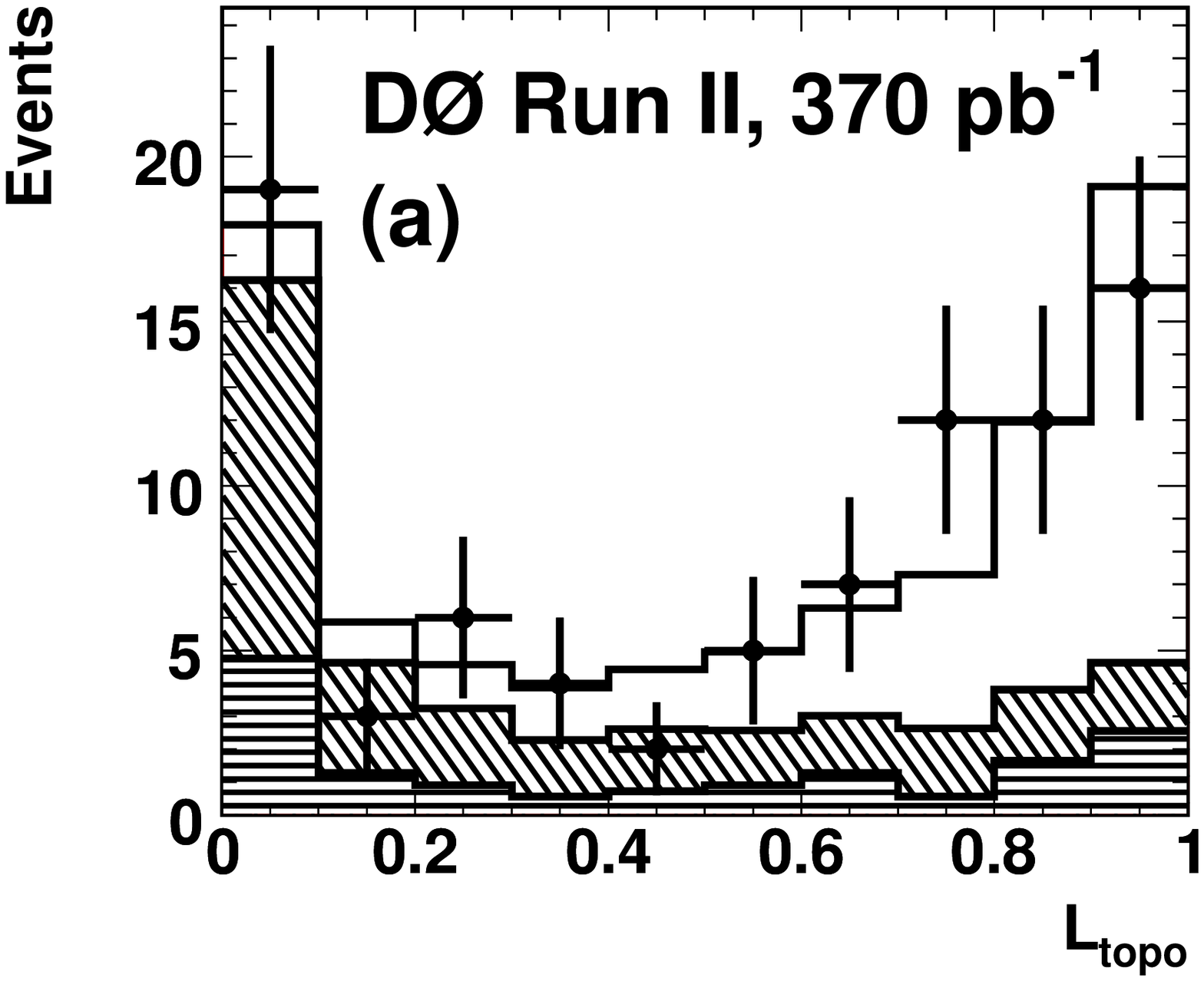}
\includegraphics[width=0.45\textwidth]{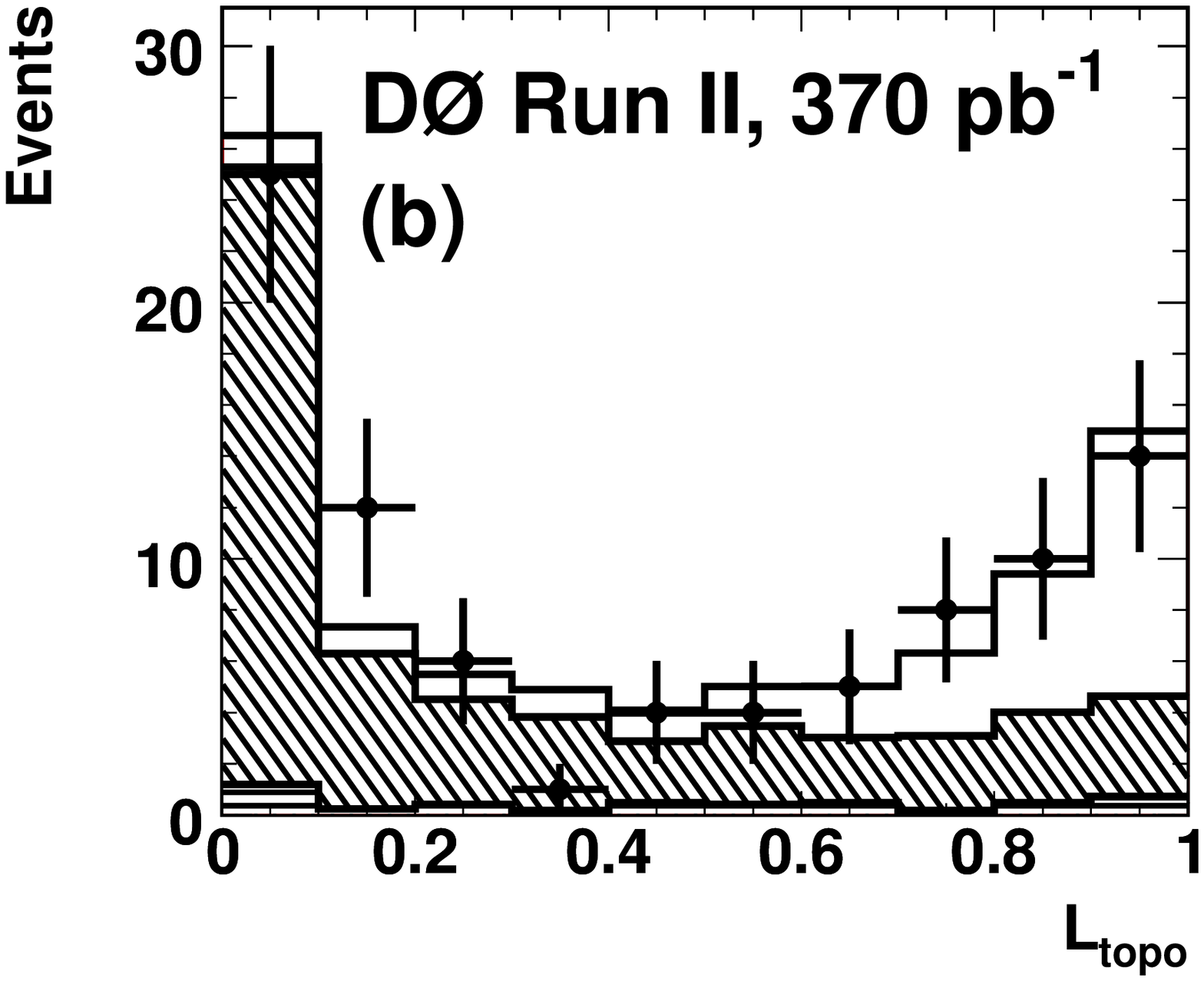}
\caption{Distribution of the topological likelihood for the $370\,\ipb$ \dzero
\runii data sample.  The distribution for
\ejets events is shown in plot (a) and for \mujets events in
plot (b).  The points with error bars indicate the data, and the 
fitted fractions of \ttbar events (open area), \wjets events (diagonally
hatched), and QCD multijet events (horizontally hatched area) 
are superimposed.}
\label{fig:MEtopolhd}
\end{center}
\end{figure}

The relative contributions of \ttbar, \wjets, and QCD multijet events
to the selected data sample are determined before $b$ tagging is applied.
A likelihood discriminant based on topological variables 
is calculated for every selected event.
The technique is the same as described in~\cite{bib:topoxsect}.
A fit to the observed
distribution yields the fractions of \ttbar, \wjets, and multijet  
events in the data sample, separately for \ejets and \mujets events.
The fits are shown in Figure~\ref{fig:MEtopolhd}, and the results are
summarized in Table~\ref{tab:MEcomposition}.
Note that because of differences between the selection criteria
of \ejets and \mujets events (most notably, the $|\eta|$ requirement,
but also the criteria for selecting isolated leptons), 
the numbers of selected \ttbar, \wjets, and QCD events in the two
channels are not expected to be equal.

\begin{table}
\begin{center}
\begin{tabular}{c@{\quad}c@{\quad}c@{\quad}c@{\quad}c}
\hline
\hline
&&&&\vspace{-2ex}\\
channel & $N_{\rm evts}$ & $N_{\rm top}^{\rm topo}$  
        & $f_{\rm top}^{\rm topo}$  
        & $f_{\rm QCD}^{\rm topo}$ \\
&&&&\vspace{-2ex}\\
\hline
&&&&\vspace{-2ex}\\
\medskip			      			      
\ejets  & $\enspace86$ & \fitemntop & \fitemftop & \fitemfqcd \\
\mujets & $\enspace89$ & \fitmuntop & \fitmuftop & \fitmufqcd \\
&&&&\vspace{-2ex}\\				   				   
\hline						   						   
&&&&\vspace{-2ex}\\				   				   
\ljets  &        $175$ & \fitntop   & \fitftop   & \fitfqcd   \\
&&&&\vspace{-2ex}\\
\hline
\hline
\end{tabular}
\caption{Composition of the \ejets, \mujets, and \ljets data samples,
estimated with the topological likelihood technique.
The fractions are constrained as 
$
    f_{\rm top}^{\rm topo}
  + f_{W+\rm jets}^{\rm topo}
  + f_{\rm QCD}^{\rm topo}
  = 1
$.}
\label{tab:MEcomposition}                                                       
\end{center}
\end{table}

The relative contributions from background 
events with a $W$ boson and four jets
with different flavor composition are estimated using the \alpgen
generator.
The fractions $f_\Phi$ of each of the six flavor configurations
$\Phi=jjjj$, $\bbbar jj$, $\ccbar jj$, $(\bbbar)jjj$, $(\ccbar)jjj$, 
and $cjjj$ 
are listed in Table~\ref{tab:HFfract}~\cite{bib:btagxsect}.
The symbol $j$ denotes a
light jet not containing a charm or bottom quark, and
the symbols $(\bbbar)$ and $(\ccbar)$ refer 
to situations where two heavy flavor quarks end up in the same jet.
These fractions are obtained without a $b$-tagging requirement;
in the $b$-tagging analysis where the event sample is divided into
three classes according to the number of $b$-tagged jets per event
(cf.\ Section~\ref{sec:MEfitbtag-fit}), the individual numbers for each of the
three separate classes are significantly different.

\begin{table}
\begin{center}
\begin{tabular}{lrcl}
\hline
\hline
Contribution            & \multicolumn{3}{r}{$W$ + $\geq 4\ {\rm jets}$}  \\ 
\hline
$Wb\bar{b}jj$		& $( 2.72$ & $\pm$ & $0.11 )\,\%$ \\ 
$Wc\bar{c}jj$		& $( 4.31$ & $\pm$ & $0.20 )\,\%$ \\ 
$W(b\bar{b})jjj$	& $( 2.70$ & $\pm$ & $0.15 )\,\%$ \\ 
$W(c\bar{c})jjj$	& $( 4.69$ & $\pm$ & $0.36 )\,\%$ \\ 
$Wcjjj$			& $( 4.88$ & $\pm$ & $0.17 )\,\%$ \\ 
$Wjjjj$                 & $( 80.71$& $\pm$ & $0.43 )\,\%$ \\ 
\hline
\hline
\end{tabular}
\caption{Fractions $f_{\Phi}$ of different flavor subprocesses 
contributing to the \wjets sample.}
\label{tab:HFfract}
\end{center}
\end{table}

\section{The Matrix Element Method}
\label{sec:ME}
In this section, the measurement of the top quark mass using the Matrix
Element method is described.
The method is similar to the one of~\cite{bib:nature};
however, the calculation of the signal probability has been revised,
the normalization of the background probability is determined
differently, and the
method now allows a simultaneous measurement of the top quark mass and
the jet energy scale~\cite{bib:philipp}.

An overview of the Matrix Element method is given in
Section~\ref{sec:MEmethod}.
Section~\ref{sec:MEtransferfcn} describes how $b$-tagging information
is used in the analysis and discusses the parameterization of the
detector response.
Details on the computation of the probabilities \psgn and \pbkg are given in 
Sections~\ref{sec:MEpsgn} and~\ref{sec:MEpbkg}.

\subsection{The Event Probability}
\label{sec:MEmethod}
To make maximal use of the kinematic information on the top quark mass
contained in the event sample (or each individual event category
in the case of the analysis using $b$-tagging information), 
for each selected event 
a probability \pevt that this event is observed is calculated 
as a function of the assumed top quark mass and jet energy scale.
The probabilities from all events are then combined to
obtain the sample probability as a function of assumed mass and jet
energy scale, and the top quark mass measurement is extracted from
this sample probability.
To make the probability calculation tractable, simplifying assumptions
in the description of the physics processes and the detector response
are introduced as described in this section.
Before applying it to the data, the measurement technique is 
calibrated using fully simulated events in the \dzero detector, and 
the assumptions in the description of the physics processes are
accounted for by systematic uncertainties.

It is assumed that the physics processes that can lead to the observed
event do not interfere.
The probability \pevt then in principle has to be composed from
probabilities for all these processes as
\begin{equation}
  \pevt = \sum_{{\rm processes}\ i} f_i P_i
  \ ,
\end{equation}
where $P_i$ is the probability for a given process $i$ and $f_i$
denotes the fraction of events from that process in the event sample.
In this analysis, \pevt is composed from probabilities for
two processes, \ttbar production and \wjets events, as
\eqMEpevt
Here, $x$ denotes the kinematic variables of the event, \ftop is the
signal fraction of the event sample, and \psgn and \pbkg are the 
probabilities for \ttbar and \wjets production, respectively.
The largest background contribution is from \wjets events.
Therefore, \pbkg is taken to be the probability for \wjets production.
Contributions from QCD multijet events are not treated explicitly 
and are considered as a systematic uncertainty.
The signal probability \psgn accounts for both possible flavor 
compositions in the hadronic \W decay in \ljets \ttbar events, 
$\W \to u\overline{d}'$ and $\W \to c\overline{s}'$:
\begin{equation}
  \psgn = \frac{1}{2}\left(   \psgn^{\W \to u\overline{d}'}
                            + \psgn^{\W \to c\overline{s}'} \right)
  \ .
\end{equation}
Because the event kinematics are the same, both final states are
treated simultaneously in the probability calculation.

To evaluate the \ttbar probability, all configurations of
\ttbar decay products that could have led to the observed
event $x$ are considered.
This includes different hadronic \W decays as discussed above and
all possible configurations $y$ of the final state particles' four-momenta.
The probability density for given partonic
final state four-momenta $y$ to be produced in the hard scattering process is 
proportional to the differential cross section ${\rm d}\sigma$ of the 
corresponding process, given by
\eqMEdsigmahs
The symbol $\mathscr{M}$ denotes the matrix element for the process 
$\qqbar\to\ttbar\to b (\ell\nu) \overline{b} (\qqbarprime)$,
$s$ is the center-of-mass energy squared, $q_1$ and $q_2$ are the
momentum fractions of the colliding partons (which are assumed to be 
massless) within the colliding proton and antiproton, and
${\rm d}\Phi_{6}$ is an element of six-body phase space.
Here the symbol $\qqbarprime$ stands for $u\overline{d}'$ or $c\overline{s}'$.

To obtain the differential cross section 
${\rm d}\sigma(\ppbar\to\ttbar\to y;\,\mtop)$
in \ppbar collisions, the differential cross section
from equation \Eref{eq:MEdsigmahs} is convoluted 
with the parton density functions (PDF) for all possible flavor
compositions of the colliding quark and antiquark,
\eqMEdPpp
where $f(q)$ denotes the probability density to find a parton of
given flavor and momentum fraction $q$ in the proton or antiproton.

The finite detector resolution is taken into account via a
convolution with
a transfer function $W(x,y;\,JES)$ that describes the probability to
reconstruct a partonic final state $y$ as $x$ in the detector.
The differential cross section to observe a given reconstructed \ttbar
event then becomes
\eqMEdsigmapp
Because \psgn describes \ljets \ttbar events with both
$\W \to u\overline{d}'$ and $\W \to c\overline{s}'$ decays,
the transfer function depends
on the quark flavors produced in the hadronic $W$ decay
when $b$-tagging information is used.
This is further discussed in Section~\ref{sec:MEtransferfcn}.

Only events that are inside the 
detector acceptance and pass the trigger conditions and offline
event selection are used in the measurement.
The corresponding overall detector efficiency depends both on \mtop
and on the jet energy scale.
This is taken into account in the cross section $\sigma_{\rm obs}$ of
\ttbar events observed in the detector:
\eqMEsigmaobs
where $f_{\rm acc}=1$ for selected events and $f_{\rm acc}=0$ otherwise.

The differential probability to observe a \ttbar event as $x$ in the detector
is then given by
\eqMEpsgn
The parametrization of the matrix element and the computation of 
\psgn are described in Section~\ref{sec:MEpsgn}.

Similarly, the differential background probability is computed as
\eqMEpbkg
where the matrix element and the total observed cross section for
the process $\ppbar\to\wjets$ have been used accordingly.  
Since the matrix element for \wjets production does not
depend on \mtop, \pbkg is independent of \mtop; however, \pbkg in
principle does
depend on the jet energy scale through the transfer function.
Details about the \pbkg calculation can be found in 
Section~\ref{sec:MEpbkg}.

To extract the top quark mass from a set of $N$ measured
events $x_1,..,x_N$, a likelihood function is built from the
individual event probabilities calculated according to
Equation~(\ref{eq:MEpevt}) as
\eqMElhoodfnc
For every assumed pair of values $(\mtop,\ JES)$, the 
value \ftopbest that maximizes the likelihood is determined.
The top quark mass and jet energy scale are then obtained by maximizing
the likelihood
\eqMElhoodnoftop
with respect to \mtop and $JES$, taking the correlation 
between both parameters into account.

\subsection{Description of the Detector Response}
\label{sec:MEtransferfcn}
The transfer function $W(x,y;\,JES)$ relates the characteristics $y$
of the final state partons to the 
measurements $x$ in the detector.
The symbol $x$ denotes measurements of the jet and charged lepton
energies or momenta and directions as well as $b$-tagging information
for the jets.
A parameterization of the detector resolution is used in the
probability calculation because the full \geant-based simulation would
be too slow.
The full simulation is however used to generate the simulated events
with which the method is calibrated.

The transfer function is assumed to factorize into contributions from 
each measured final state particle.
The angles of all measured \ttbar decay products as well as the energy of
electrons are assumed to be well-measured; in other words, the transfer 
functions for these quantities are given by $\delta$-distributions.
This allows reducing the dimensionality
of the integration over 6-particle phase space as described in 
Sections~\ref{sec:MEpsgn} and~\ref{sec:MEpbkg}.
Consequently, contributions to the integral only arise if the directions
of the quark momenta in the final state agree with the measured jet
directions.
In addition to the energy resolution, 
one has to take into account the fact that the jets in the
detector cannot be assigned unambiguously
to a specific parton from the \ttbar decay.
Consequently, all 24 permutations of jet-quark
assignments are considered.

In this section, the general form of the transfer function 
in the topological and $b$-tagging analyses is first discussed,
followed by a description of
the jet energy and muon transverse momentum resolutions.

\subsubsection{Transfer Function in the Topological Analysis}
\label{sec:tf-topo}
If no $b$-tagging information is used, the transfer function 
$W(x,y;\,JES)$ is given by
\eqMEtfdefinitiontopo
where $W_\mu$ and $W_{\rm jet}$ stand for factors describing the 
muon transverse momentum and jet energy resolutions, respectively.  
The sum is over the 24 different assignments of jets $j$ to 
partons $k$.
The factor $\delta({\rm angles})$ denotes the $\delta$
distributions that ensure that assumed and reconstructed particle
directions are identical, as discussed above.
For \ejets events, the factor for the muon transverse
momentum resolution is replaced with another $\delta$-distribution.
The neutrino is not measured in the detector and does not enter the 
transfer function.
The jet transfer functions for light quark and charm jets are 
taken to be identical, and in the calculation of the background 
probability, all jets are assumed to be described by the light
quark transfer function.
Because the matrix elements for $W \to u\overline{d}'$ and 
$W \to c\overline{s}'$ decays are equal, and the different 
flavor contributions to the \wjets process are all parameterized by the 
\wjets matrix element without heavy flavor quarks in the final state,
no distinction between different processes
is necessary in the topological analysis.

\subsubsection{Transfer Function in the $b$-tagging Analysis}
\label{sec:tf-btag}
In the topological analysis, the information from the reconstructed 
jet energies determines the relative weight of different
jet-parton assignments for a given partonic final state.
The inclusion of $b$-tagging information allows an improved
identification of the correct permutation.
This additional information enters the probability calculation by
weighting different permutations $i$ of jet-parton assignments
with weights $w_i$ according to which jets, if any,
are $b$-tagged.  
This allows to give those permutations a larger weight that assign
tagged jets to $b$ quarks and untagged
ones to light quarks.
The transfer function is thus
\eqMEtfdefinitionbtag
(with a $\delta$-distribution instead of the factor $W_\mu$ in the
case of \ejets events).

The weight $w_i$ for a permutation $i$ is parameterized as a product
of individual weights $w_i(j)$ for each jet.  
The latter are a function of
the jet flavor hypothesis $\alpha_k$ and the jet transverse energy
$\etj$ and pseudorapidity $\eta_j$.  
For tagged jets, $w_i(j)$ is equal to the per-jet 
tagging efficiency $\epsilon_{\rm jet}(\alpha_k;\ \etj,\,\eta_j)$
where $\alpha_k$ labels the three possible parton assignments to the jet:
(a) $b$ quarks, (b) $c$ quarks, and (c) light quarks or gluons.  
For untagged jets, the $w_i(j)$ factors are equal to 
$1-\epsilon_{\rm jet}(\alpha_k;\ \etj,\,\eta_j)$.  
If an event does not contain any $b$-tagged jet, all the weights
$w_i(j)$ are set to 1.0.

To compute the signal probability of events containing $b$-tagged jets, 
assumptions on the jet flavors are made for the calculation of the $w_i$
such that hadronic $W$ decays to 
$u\overline{d}'$ and $c\overline{s}'$ final states
need not be distinguished in the 
matrix element, allowing for a reduction of the computation time.
If an event contains exactly one $b$-tagged jet, the quarks from the 
hadronic $W$ decay are both assumed to be light quarks ($u$, $d$, or $s$).
This is justified since the tagging efficiencies for $b$ jets are 
much larger than those for other flavors, and there are two $b$ jets
per event.
For events with two or more $b$-tagged jets, a charm jet from the 
hadronic $W$ decay is tagged in a non-negligible fraction of cases.
Consequently, the quarks from the hadronic $W$ decay are assumed to be charm
quarks if the corresponding jet has been tagged, and light quarks
otherwise.

\begin{figure}
\begin{center}
\includegraphics[width=\figMEcalibwidth\textwidth]{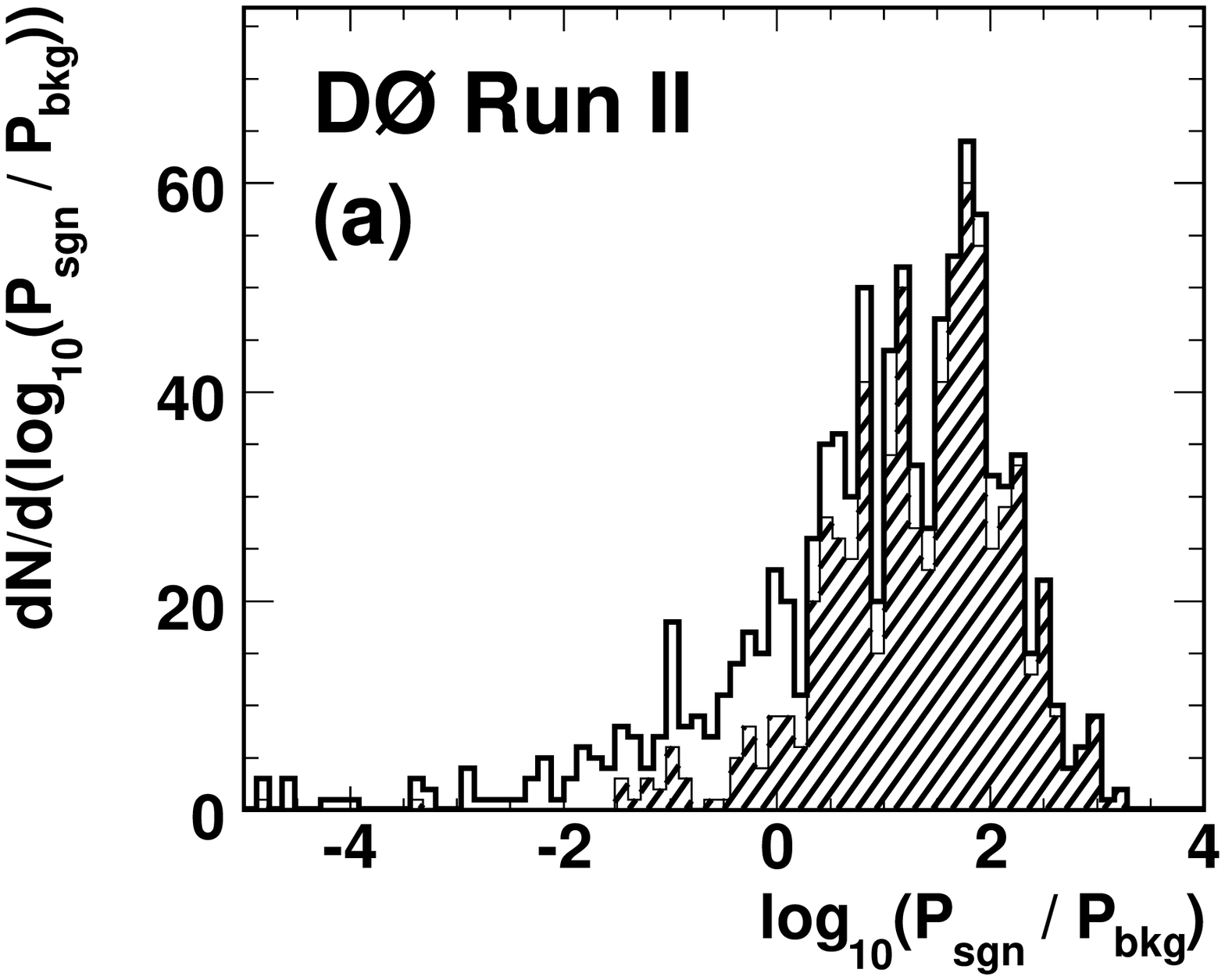}
\hspace{-0.03\textwidth}
\includegraphics[width=\figMEcalibwidth\textwidth]{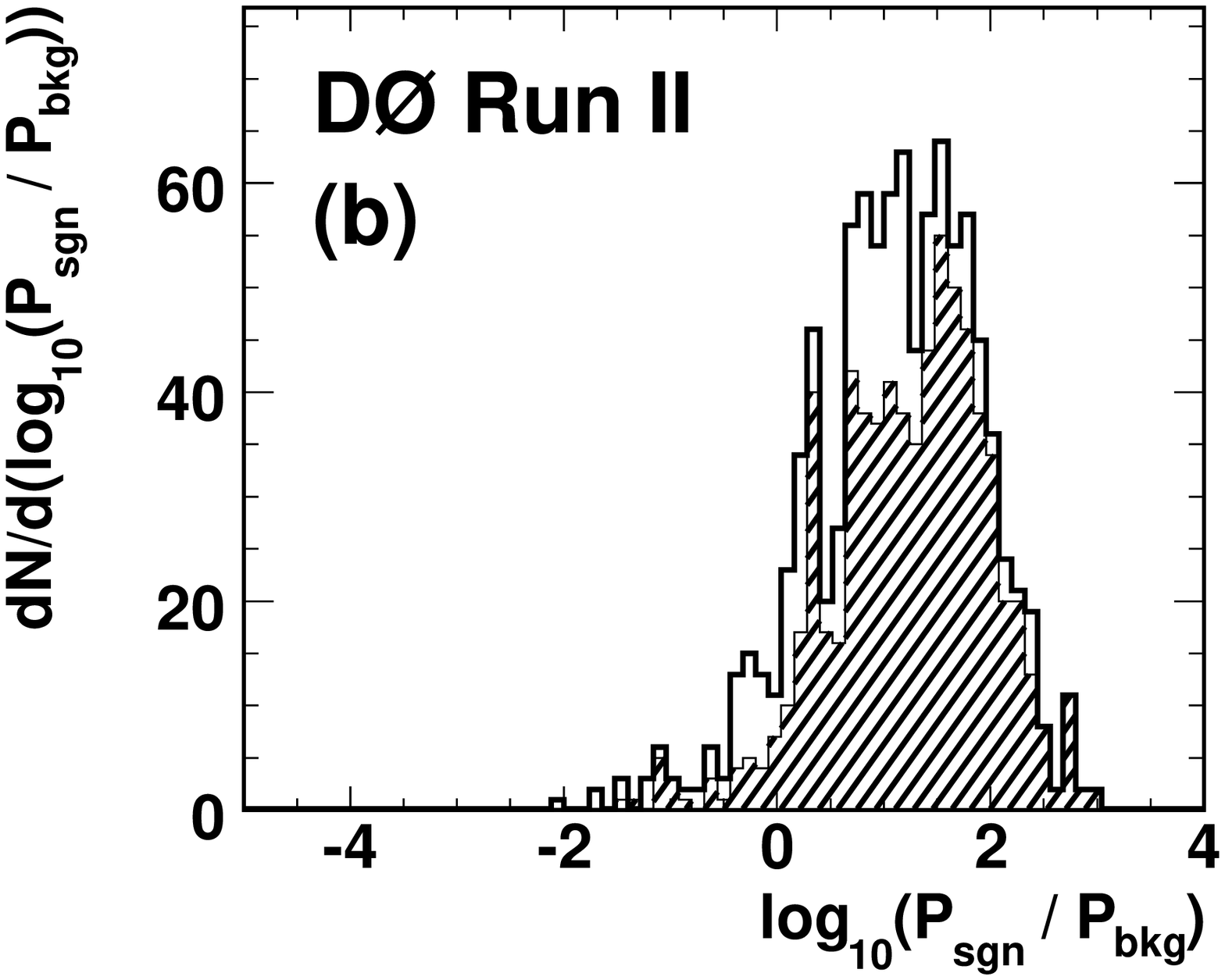}
\caption{Monte Carlo study of the effect of charm-jet tagging on the
  signal to background probability ratio in the $b$-tagging
  analysis, for \ttbar events generated with $\mtop=175\,\GeV$ 
  that contain two $b$-tagged jets.  
  The \psgn values are calculated for the assumption $\mtop=175\,\GeV$.
  (a) Only the two jet-parton assignments in which tagged jets
  are assigned to $b$ quarks are considered. 
  (b) All weighted jet parton-assignments enter the probability calculation.
  In both plots, the hatched histogram corresponds to those cases
  where the two $b$-tagged jets are correctly assigned to $b$ quarks,
  which happens $84\%$ of the time in the double tag sample.}
\label{fig:nopruning}
\end{center}
\end{figure}

The need to include jet-parton assignments with a tagged charm jet
in the probability calculation can be seen by comparing the signal and
background probabilities.
Figure~\ref{fig:nopruning}(a) shows the ratio of signal to background
probabilities calculated in a large sample of simulated 
\ttbar events with two $b$-tagged jets
when only the two jet-parton assignments in which tagged jets
are assigned to $b$ quarks are considered in the signal probability
calculation. 
The hatched histogram shows the correct assignments only, whereas the 
open histogram shows all combinations, including the ones in which a 
charm quark from the \W decay was tagged.
Figure~\ref{fig:nopruning}(b) shows the same ratio when all combinations are
included with their corresponding weight as discussed above.
The tail for low signal to background probability ratios in
Fig.~\ref{fig:nopruning}(a) arises because the correct jet-parton
assignment is not included in the calculation in events where one of
the tagged jets comes from a charm quark.
It clearly shows the need to include these assignments in the signal
probability calculation.

Because the different 
flavor contributions to the \wjets process are parameterized by the 
\wjets matrix element without heavy flavor quarks in the final state,
the weights $w_i$ for the 
background probability are all equal even if $b$-tagged jets are
present.
Therefore, the background probability calculated for the topological
analysis is used in the $b$-tagging analysis without modifications.

\subsubsection{Parameterization of the Jet Energy Resolution}
\label{sec:MEejetresolution}
The transfer function for calorimeter jets,
$W_{\rm jet}(E_{j},E_{q};JES)$,
yields the probability for a measurement $E_{j}$ in the
detector if the true quark energy is $E_{q}$.
For the case $JES=1$, it is parameterized as
\eqMEtf
The parameters $p_i$ are themselves functions of the quark energy, and are
parameterized as linear functions of the quark energy so that
\begin{equation}
p_i = a_i + E_{q}\cdot b_i
\ ,
\end{equation}
with $a_3$ set to 0.

\begin{figure}[htb]
\includegraphics[width=0.45\textwidth]{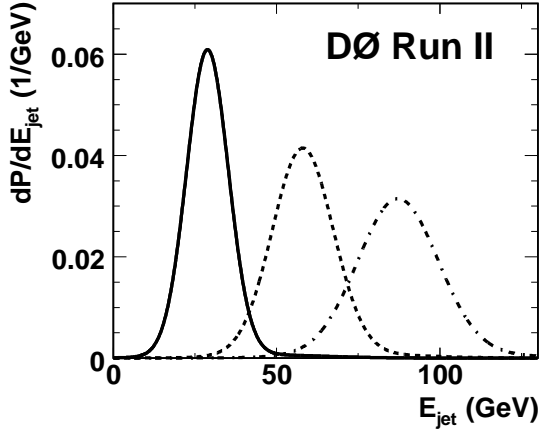}
\caption{Jet transfer functions for light quark jets,
$0.0<\left|\eta\right|<0.5$, for parton energies $E_{p}=30\,{\rm GeV}$
(solid), $60\,{\rm GeV}$ (dashed), and $90\,{\rm GeV}$ (dash-dotted
curve).  The parametrization corresponds to the reference jet
energy scale, $JES=1.0$.}
\label{fig:transfer}
\end{figure}

The parameters $a_i$ and $b_i$ are determined from simulated events,
after all jet energy corrections have been applied.
The parton and jet energies are fed to an unbinned likelihood fit that
minimizes the $\chi^2$ of the fit to Equation~(\ref{eq:MEtf}) with respect
to $a_i$ and $b_i$.
A different set of parameters is derived for each of four $\eta$ regions:
$|\eta|<0.5$, $0.5<|\eta|<1.0$, $1.0<|\eta|<1.5$, and
$1.5<|\eta|<2.5$, and for three different quark varieties:
light quarks ($u$, $d$, $s$, $c$), $b$ quarks with a soft muon tag
in the associated jet, and all other $b$ quarks.
A total of $120$ parameters describe the transfer function for all
jets, and are given in
Tables~\ref{table:params1} and~\ref{table:params2}.
The transfer function for light quarks in the region $|\eta|<0.5$ is 
shown in Fig.~\ref{fig:transfer}.

For $JES\neq1$, the jet transfer function is modified as follows:
\begin{equation}
  \label{eq:MEtfjes}
    W_{\rm jet}(E_{j},E_{q}; JES) 
  =
    \frac{W_{\rm jet}(\frac{E_{j}}{JES},E_{q};1)}{JES} \ .
\end{equation}

\begin{table}[htbp]
\begin{center}
\begin{tabular}{lr@{$\times$}lr@{$\times$}lr@{$\times$}lr@{$\times$}l}
\hline
\hline
 & \multicolumn{8}{c}{$|\eta|$ region} \\
par & \multicolumn{2}{c}{$<0.5$} & \multicolumn{2}{c}{$0.5-1.0$} & \multicolumn{2}{c}{$1.0-1.5$} & \multicolumn{2}{c}{$>1.5$} \\
 \hline
$a_1$ &$-3.00$ & $10^{-1}$ &$ 7.30$ & $10^{-1}$ &$ 4.00$ & $10^{ 0}$ &$ 1.01$ & $10^{ 1}$ \\
$b_1$ &$-2.80$ & $10^{-2}$ &$-5.20$ & $10^{-2}$ &$-1.08$ & $10^{-1}$ &$-1.16$ & $10^{-1}$ \\
$a_2$ &$ 3.47$ & $10^{ 0}$ &$ 2.05$ & $10^{ 0}$ &$ 2.65$ & $10^{ 0}$ &$ 5.54$ & $10^{ 0}$ \\
$b_2$ &$ 9.70$ & $10^{-2}$ &$ 1.44$ & $10^{-1}$ &$ 1.51$ & $10^{-1}$ &$ 1.22$ & $10^{-1}$ \\
$a_3$ & \multicolumn{2}{l}{\phantom{$-$}0.0} & \multicolumn{2}{l}{\phantom{$-$}0.0} & \multicolumn{2}{l}{\phantom{$-$}0.0} & \multicolumn{2}{l}{\phantom{$-$}0.0} \\
$b_3$ &$ 3.73$ & $10^{-4}$ &$ 3.98$ & $10^{-4}$ &$ 7.74$ & $10^{-4}$ &$ 1.06$ & $10^{-3}$ \\
$a_4$ &$ 1.81$ & $10^{ 1}$ &$ 2.23$ & $10^{ 1}$ &$ 1.71$ & $10^{ 1}$ &$ 3.77$ & $10^{ 1}$ \\
$b_4$ &$-1.70$ & $10^{-1}$ &$-1.57$ & $10^{-1}$ &$ 3.09$ & $10^{-2}$ &$-1.54$ & $10^{-1}$ \\
$a_5$ &$ 1.71$ & $10^{ 1}$ &$ 1.98$ & $10^{ 1}$ &$ 2.00$ & $10^{ 1}$ &$ 2.91$ & $10^{ 1}$ \\
$b_5$ &$ 9.70$ & $10^{-2}$ &$ 8.04$ & $10^{-2}$ &$ 5.61$ & $10^{-2}$ &$-4.45$ & $10^{-2}$ \\
\hline
\hline
\end{tabular}
\caption{Light quark transfer function parameters ($a_i$ in GeV).}
\label{table:params1}
\end{center}
\end{table}
\begin{table}[htbp]
\begin{center}
\begin{tabular}{lr@{$\times$}lr@{$\times$}lr@{$\times$}lr@{$\times$}l}
\hline
\hline
 & \multicolumn{8}{c}{$|\eta|$ region} \\
par & \multicolumn{2}{c}{$<0.5$} & \multicolumn{2}{c}{$0.5-1.0$} & \multicolumn{2}{c}{$1.0-1.5$} & \multicolumn{2}{c}{$>1.5$} \\
 \hline
$a_1$ &$-5.08$ & $10^{ 0}$ &$-2.38$ & $10^{ 0}$ &$ 0.68$ & $10^{-1}$ &$ 3.30$ & $10^{ 0}$ \\
$b_1$ &$ 2.40$ & $10^{-3}$ &$-6.50$ & $10^{-2}$ &$-1.24$ & $10^{-1}$ &$-3.37$ & $10^{-1}$ \\
$a_2$ &$ 3.80$ & $10^{ 0}$ &$ 2.40$ & $10^{ 0}$ &$ 9.10$ & $10^{-1}$ &$ 1.32$ & $10^{ 1}$ \\
$b_2$ &$ 8.70$ & $10^{-2}$ &$ 1.55$ & $10^{-1}$ &$ 1.81$ & $10^{-1}$ &$ 1.32$ & $10^{-1}$ \\
$a_3$ & \multicolumn{2}{l}{\phantom{$-$}0.0} & \multicolumn{2}{l}{\phantom{$-$}0.0} & \multicolumn{2}{l}{\phantom{$-$}0.0} & \multicolumn{2}{l}{\phantom{$-$}0.0} \\
$b_3$ &$ 2.12$ & $10^{-3}$ &$ 3.49$ & $10^{-4}$ &$ 7.46$ & $10^{-4}$ &$ 4.06$ & $10^{-2}$ \\
$a_4$ &$ 2.23$ & $10^{-1}$ &$ 2.62$ & $10^{ 1}$ &$ 1.17$ & $10^{ 1}$ &$-1.90$ & $10^{ 0}$ \\
$b_4$ &$-1.81$ & $10^{-1}$ &$-4.07$ & $10^{-1}$ &$-7.50$ & $10^{-2}$ &$-5.09$ & $10^{-2}$ \\
$a_5$ &$ 1.12$ & $10^{ 1}$ &$ 2.01$ & $10^{ 1}$ &$ 1.80$ & $10^{ 1}$ &$ 3.42$ & $10^{ 0}$ \\
$b_5$ &$ 1.12$ & $10^{-1}$ &$ 1.22$ & $10^{-1}$ &$ 7.50$ & $10^{-2}$ &$ 1.34$ & $10^{-1}$ \\
\hline
\hline
\smallskiponlyintwocol
\end{tabular}
\begin{tabular}{lr@{$\times$}lr@{$\times$}lr@{$\times$}lr@{$\times$}l}
\hline
\hline
 & \multicolumn{8}{c}{$|\eta|$ region} \\
par & \multicolumn{2}{c}{$<0.5$} & \multicolumn{2}{c}{$0.5-1.0$} & \multicolumn{2}{c}{$1.0-1.5$} & \multicolumn{2}{c}{$>1.5$} \\
\hline
$a_1$ &$ 1.10$ & $10^{ 1}$ &$ 4.97$ & $10^{ 0}$ &$ 1.29$ & $10^{ 1}$ &$ 1.36$ & $10^{ 1}$ \\
$b_1$ &$-1.33$ & $10^{-1}$ &$ 5.30$ & $10^{-3}$ &$-1.65$ & $10^{-1}$ &$-1.32$ & $10^{-1}$ \\
$a_2$ &$ 2.99$ & $10^{ 0}$ &$ 3.85$ & $10^{ 0}$ &$ 4.02$ & $10^{ 0}$ &$ 5.42$ & $10^{ 0}$ \\
$b_2$ &$ 1.18$ & $10^{-1}$ &$ 4.00$ & $10^{-2}$ &$ 1.25$ & $10^{-1}$ &$ 1.18$ & $10^{-1}$ \\
$a_3$ & \multicolumn{2}{l}{\phantom{$-$}0.0} & \multicolumn{2}{l}{\phantom{$-$}0.0} & \multicolumn{2}{l}{\phantom{$-$}0.0} & \multicolumn{2}{l}{\phantom{$-$}0.0} \\
$b_3$ &$ 3.02$ & $10^{-4}$ &$ 1.14$ & $10^{-2}$ &$ 4.30$ & $10^{-4}$ &$ 2.42$ & $10^{-4}$ \\
$a_4$ &$ 4.53$ & $10^{ 1}$ &$ 1.33$ & $10^{ 1}$ &$ 4.51$ & $10^{ 1}$ &$ 7.18$ & $10^{ 1}$ \\
$b_4$ &$-4.54$ & $10^{-1}$ &$-1.91$ & $10^{-1}$ &$-2.15$ & $10^{-1}$ &$-1.24$ & $10^{-1}$ \\
$a_5$ &$ 1.58$ & $10^{ 1}$ &$ 5.60$ & $10^{ 0}$ &$ 1.39$ & $10^{ 1}$ &$ 1.64$ & $10^{ 1}$ \\
$b_5$ &$ 2.25$ & $10^{-1}$ &$ 1.35$ & $10^{-1}$ &$ 1.42$ & $10^{-1}$ &$ 3.40$ & $10^{-2}$ \\
\hline
\hline
\end{tabular}
\caption{\captionparamstwo}
\label{table:params2}
\end{center}
\end{table}

\subsubsection{Parameterization of the Muon Momentum Resolution}
\label{sec:MEptinvmuresolution}
To describe the resolution of the central tracking chamber, 
the resolution of the charge divided by the transverse momentum of a
particle is considered as a function of pseudorapidity.
The muon transfer function is parameterized as
\eqMEtfmu
where $q$ denotes the charge and $\pt$ the transverse momentum of a
generated (gen) muon or its reconstructed (rec) track.
The resolution
\begin{equation}
  \label{eq:MEsigmamu}
    \sigma
  =
    \left\{
    \begin{array}{cc}
        \sigma_0
      &
        {\rm for}\ |\eta|\le\eta_0
      \vspace{2ex}\\
        \sqrt{   \sigma_0^2 
               + \left[ c \left( |\eta| - \eta_0 \right) \right]^2 
             }
       &
        {\rm for}\ |\eta|>\eta_0
    \end{array}
    \right.
\end{equation}
is obtained from muon tracks in simulated events
with the following values:
\begin{eqnarray}
  \nonumber
    \sigma_0
  & = &
    2.760 \times 10^{-3} / {\rm GeV}
  \\
  \label{eq:MEmutfpars}
    c
  & = &
    5.93 \times 10^{-3} / {\rm GeV}
  \\
  \nonumber
    \eta_0
  & = &
    1.277
  \ .
\end{eqnarray}

The muon charge is not used in the calculation of \psgn and \pbkg;
however, for muons with large transverse momentum it is important to
take the possibility of charge misidentification into account in the
transfer function.

\subsection{Calculation of the Signal Probability {\boldmath\psgn}}
\label{sec:MEpsgn}
The leading order matrix element for the process $\qqbar\to\ttbar$ is
taken to compute \psgn.  
Neglecting spin correlations, the matrix element is given
by~\cite{bib:MAHLON} 
\begin{equation}
\label{eq:MEsignalME}
  |\mathscr{M}|^2 = \frac{g_s^4}{9} F \overline{F} \left( 2 - \beta^2
  s_{qt}^2 \right) ,
\end{equation}
where $g_s^2/(4\pi)=\alpha_s$ is the strong coupling constant, 
$\beta$ is the velocity
of the top quarks in the \ttbar rest frame, and $s_{qt}$ denotes the
sine of the angle between the incoming parton and the outgoing top
quark in the \ttbar rest frame.
If the top quark decay products include the leptonically decaying $W$,
while the antitop decay includes the hadronically decaying $W$, one
has
\eqMEsignalMEFFbar
(for the other case, replace $b \leftrightarrow \overline{b}$, $\ell
\leftrightarrow d$, and $\nu \leftrightarrow \overline{u}$).
Here, $g_w$ denotes the weak charge ($G_{\rm F}/\sqrt{2}=g_w^2/8 m_W^2$),
\mt and \mW are the masses of the top quark (which is to be
measured) and the $W$ boson, and \Gt and \GW are their widths. 
Invariant top and $W$ masses in a particular event are denoted by
$m_{xyz}$ and $m_{yz}$, respectively, where $x$, $y$, and $z$ are the
decay products.
The cosine of the angle between particles $x$ and $y$ in the $W$ rest
frame is denoted by ${\hat c}_{xy}$.
Here and in the following, the symbols $d$ and $\overline{u}$ stand
for all possible decay products in a hadronic \W decay.
The top quark width is calculated as a function of the top quark 
mass according to~\cite{bib:PDG}.

The correct association of reconstructed jets with the final state
quarks in Equations~(\ref{eq:MEsignalME_F}) and (\ref{eq:MEsignalME_Fbar})
is not known.
Therefore, the transfer function takes into account all 24 jet-parton
assignments as described in Section~\ref{sec:MEtransferfcn}.
However, in the case of the signal probability, 
the mean value of the two assignments with the 4-momenta of the
quarks from the hadronic $W$ decay interchanged is computed explicitly
by using the symmetrized formula
\eqMEsignalMEFbarsymm
instead of~\Eref{eq:MEsignalME_Fbar}, where only the terms containing 
$\chatbbard$ are affected. Consequently, only a summation 
over 12 different jet-quark assignments remains to be evaluated.

The computation of the signal probability \psgn 
involves an integral over the momenta of the colliding partons and
over 6-body phase space to cover all
possible partonic final states, cf.~Equation~(\ref{eq:MEpsgn}).
The number of dimensions of the integration is reduced by the 
following conditions:
\begin{itemize}
\item
  The transverse momentum of the colliding partons is assumed to be
  zero.  Conservation 
  of 4-momentum then implies zero transverse momentum of the
  \ttbar system because the leading order matrix element is used to 
  describe \ttbar production.  Also, the $z$ momentum and energy of
  the \ttbar system are known from the momenta of the colliding partons.
\item
  The directions of the quarks and the charged lepton in the final
  state are assumed to be exactly measured. 
\item
  The energy of electrons from $W$ decay is assumed to be perfectly
  measured. The corresponding statement is not
  necessarily true for high momentum muons, and an integration over
  the muon momentum is performed.
\end{itemize}

After these considerations, an integration over the quark momenta,
the charged lepton momentum (\mujets events only), and the
longitudinal component of the neutrino momentum remains to be calculated.
This calculation is performed numerically with the Monte Carlo program
\vegas~\cite{bib:VEGAS1,bib:VEGAS2}.
The algorithm works
most efficiently if the one-dimensional projections of the integrand
onto the individual integration variables have well-localized peaks.
The Breit-Wigner peaks of the integrand 
corresponding to the two top quark and two $W$
boson decays in the \ttbar matrix element are more localized than the
peaks from the jet transfer functions, suggesting that the masses are better
integration variables leading to faster convergence.
The computation of the parton kinematics from the integration
variables must however be performed in each integration step,
and this task simplifies to solving a quadratic equation when
choosing \pzbnu as an integration variable
instead of the mass of the leptonically decaying $W$
(both solutions of the quadratic equation are considered when determining
\psgn).
Therefore,
the following integration variables are chosen for the
computation of \psgn:
\begin{itemize}
\item
  the magnitude \pmagqone of the momentum of one of the quarks from
  the hadronic $W$ decay, with $0 \le \pmagqone \le 500\ \GeVc$,
\item
  the squared mass $\mWhad^2$ of the hadronically decaying $W$, $0 \le
  \mWhad^2 \le (400\ \GeV)^2$,
\item
  the squared mass $\mtophad^2$ of the top quark with the hadronic $W$
  decay, $0 \le \mtophad^2 \le (500\ \GeV)^2$,
\item
  the squared mass $\mtoplep^2$ of the top quark with the leptonic $W$
  decay, $0 \le \mtoplep^2 \le (500\ \GeV)^2$,
\item
  the $z$ component \pzbnu of the sum of the momenta of the
  $b$ quark and neutrino from the top quark with the leptonic $W$
  decay, $-500\ \GeV \le \pzbnu \le +500\ \GeV$, and
\item
  the muon charge divided by the muon transverse momentum (in
  the \mujets channel only), $-1/(100\ \MeV) \le \qoverptmu \le
  +1/(100\ \MeV)$.
\end{itemize}

Thus, for each point in the (\pmagqone, $\mWhad^2$, $\mtophad^2$,
$\mtoplep^2$, \pzbnu [, \qoverptmu]) integration space the
following computation is performed for each of the 12 possible
jet-parton assignments (where the symmetrized form of the matrix
element according to Equation~(\ref{eq:MEsignalME_Fbarsymm}) is used):
\begin{enumerate}
\item 
  The 4-momenta of the \ttbar decay products are calculated from the
  values of the integration variables, the measured jet and lepton
  angles, and the electron energy (in the \ejets case).
\item 
  The matrix element is evaluated according to
  Equations~(\ref{eq:MEsignalME}),~(\ref{eq:MEsignalME_F}),
  and~(\ref{eq:MEsignalME_Fbarsymm}).
\item
  The parton distribution functions are evaluated.  For consistency
  with the leading-order matrix element, we use the
  CTEQ5L~\cite{bib:CTEQ5L} parton distribution functions, summing over
  all possible quark flavors.
\item 
  The probabilities to observe the measured jet energies and muon
  transverse momentum given the energies and momentum computed in the
  first step are evaluated using transfer functions.
\item 
  The Jacobian determinant for the transformation from momenta in
  Cartesian coordinates to the (\pmagqone, $\mWhad^2$, $\mtophad^2$,
  $\mtoplep^2$, \pzbnu [, \qoverptmu]) integration space is
  included. 
\end{enumerate}
The precision of the \psgn calculation varies from typically
$2\,\%$ to a maximum of $10\,\%$.

\begin{figure}
\begin{center}
\includegraphics[width=0.45\textwidth]{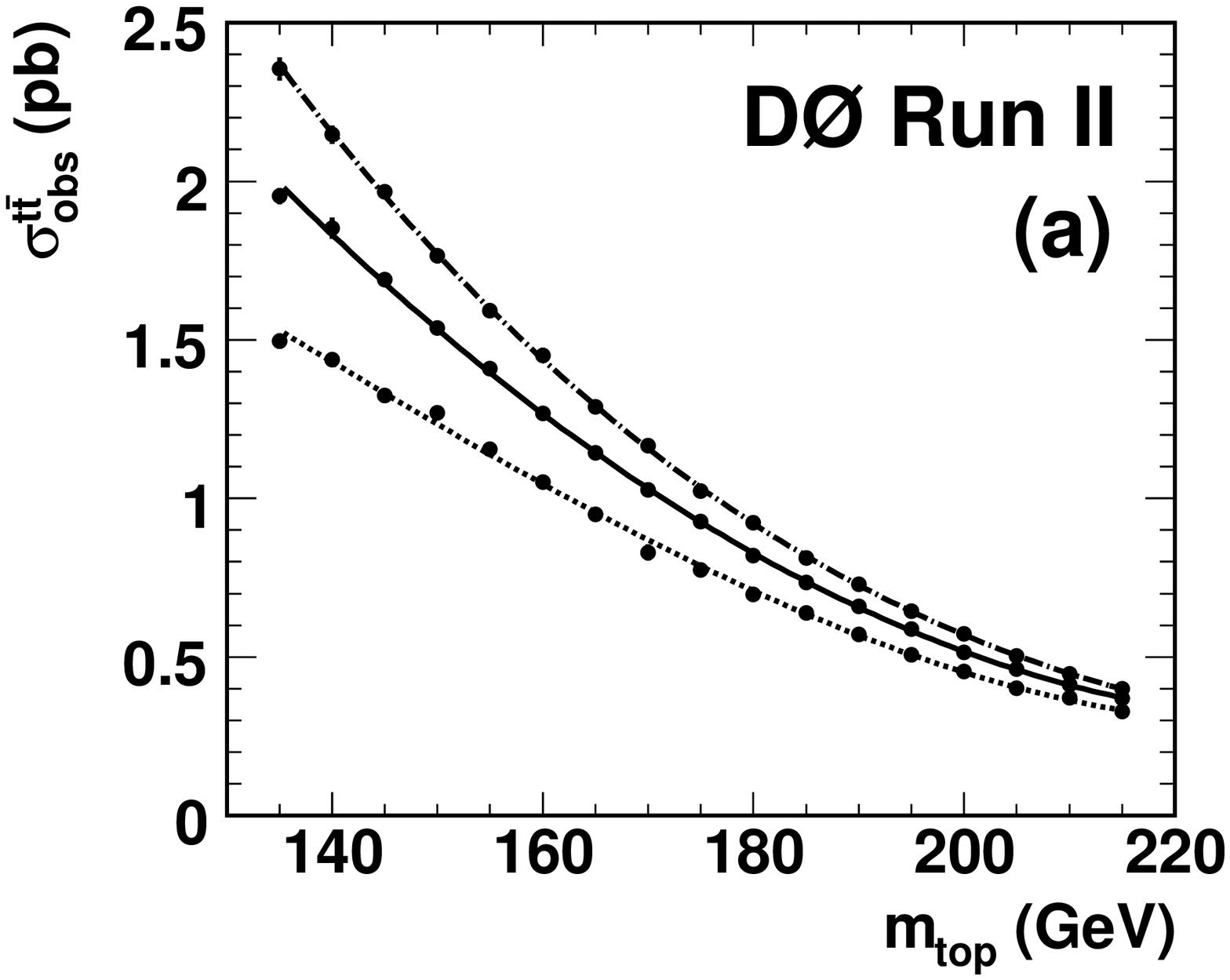}
\includegraphics[width=0.45\textwidth]{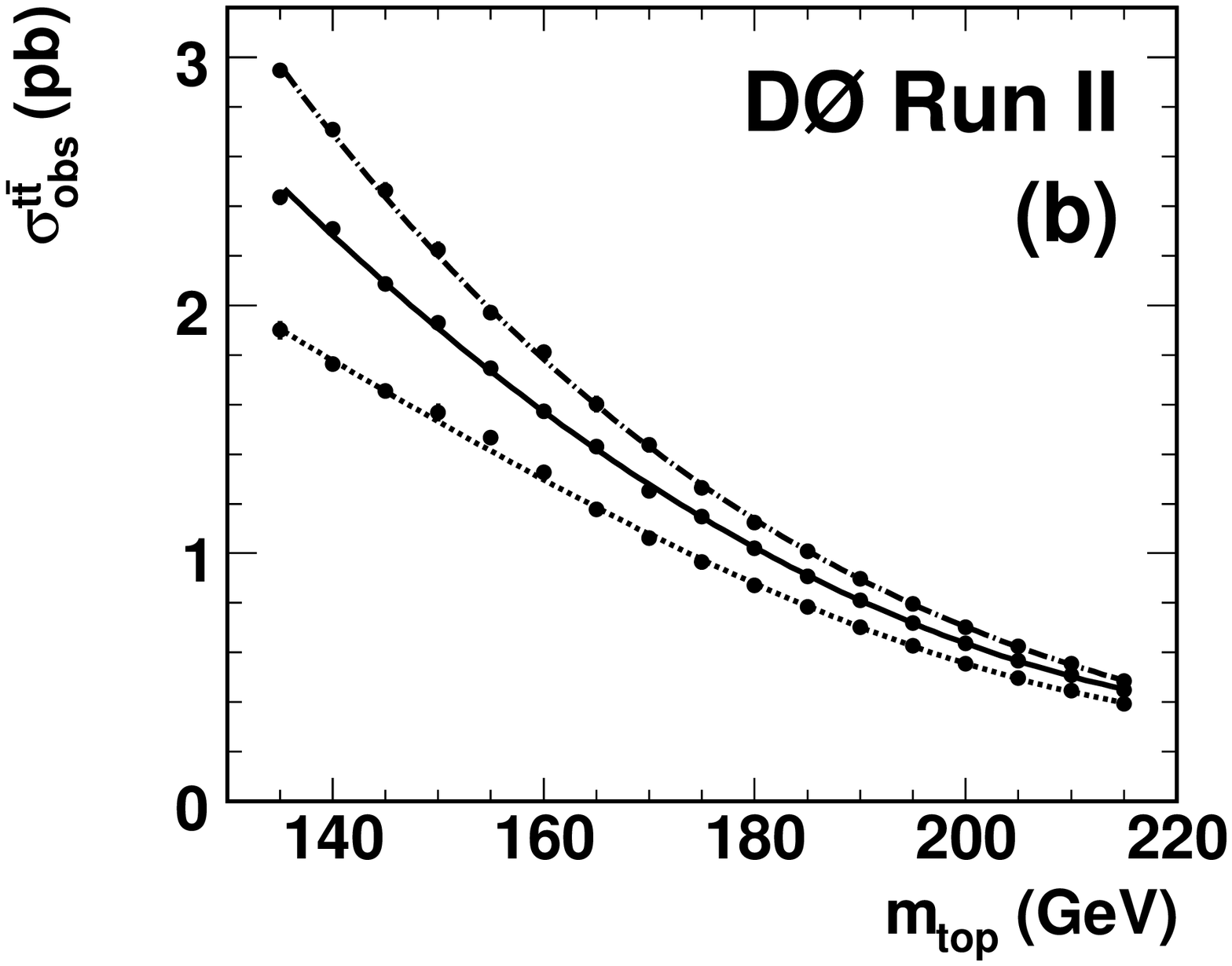}
\caption{Observed \ttbar cross section computed with the leading-order
  matrix element for (a) \ejets and (b) \mujets
  events as a function of the top quark mass \mtop for different
  choices of the $JES$ scale factor: $JES=1.12$ (dash-dotted), 
  $JES=1.0$ (solid), and $JES=0.88$ (dotted lines).}
\label{fig:MEnrmpsgn}
\end{center}
\end{figure}

To normalize the signal probability, the integral 
$\int {\rm d}\sigma(\ppbar\to\ttbar\to x;\,\mtop,JES) f_{\rm acc}(x)$
over 16-dimen\-sional phase space has been computed
as a function of \mtop
and $JES$.
The detector acceptance and efficiency is taken into account
as outlined in Equation~(\ref{eq:MEsigmaobs}).
The results are shown in Fig.~\ref{fig:MEnrmpsgn} for \ejets and
\mujets events as a function of \mtop for various choices of the
$JES$ scale factor.

\subsection{Calculation of the Background Probability {\boldmath\pbkg}}
\label{sec:MEpbkg}
To calculate \pbkg, the jet
directions and the charged electron or muon
are taken as well-measured.
The integral over the quark energies in Equation~(\ref{eq:MEpbkg}) is
performed by generating Monte Carlo events with parton energies
distributed according to the jet transfer function.
In these Monte Carlo events, the neutrino transverse momentum is 
given by the condition that the transverse momentum of the \wjets
system be zero, while the invariant mass of the charged lepton and
neutrino is assumed to be equal to the \W mass to obtain the 
neutrino $z$ momentum (both solutions are considered).
The \vecbos~\cite{bib:VECBOS} parameterization of the matrix element is
used.
The mean result from all 24 possible assignments of jets to quarks in the 
matrix element is calculated.
A minimum of 10 Monte Carlo events is generated for each measured
event $x$, and the relative spread of the
resulting \pbkg values is evaluated as the standard deviation divided
by the mean.
If the relative spread is larger than 10\%, another 10
Monte Carlo events are evaluated, and this procedure is repeated
until a 10\% relative uncertainty is 
reached or a maximum of 100 Monte Carlo events has been considered.

To normalize the background probability density, 
$\sigma_{\rm obs}(\ppbar\to\wjets;\,JES)$ is chosen such
that the total signal fraction \ftop in the analysis without $b$ tagging
is reproduced in the fit to 
simulated event samples containing \ttbar and \wjets events.
This makes use of the fact that \ftop is underestimated in the fit if
the background probabilities are too large and vice versa.

\begin{figure}
\begin{center}
\includegraphics[width=0.45\textwidth]{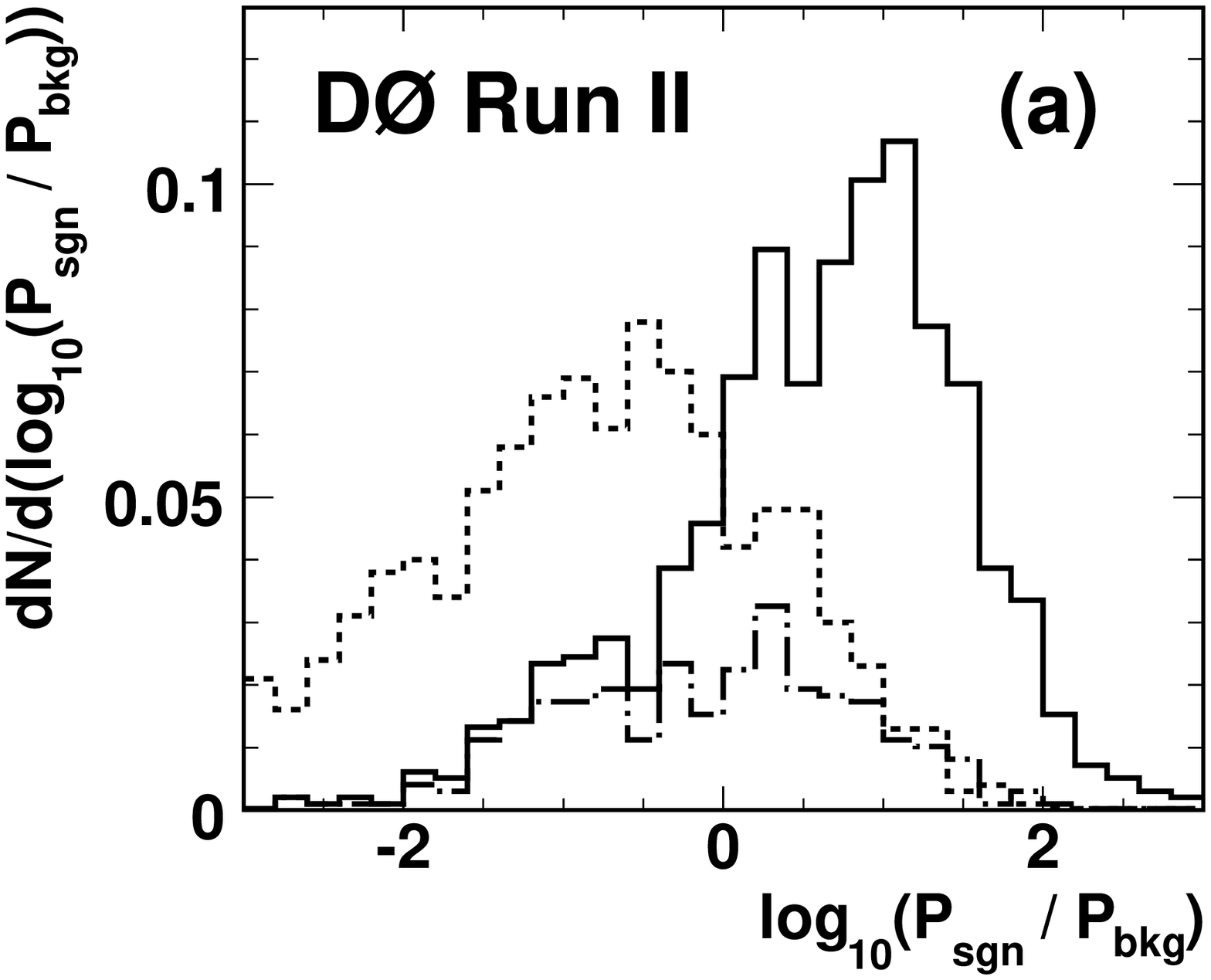}
\includegraphics[width=0.45\textwidth]{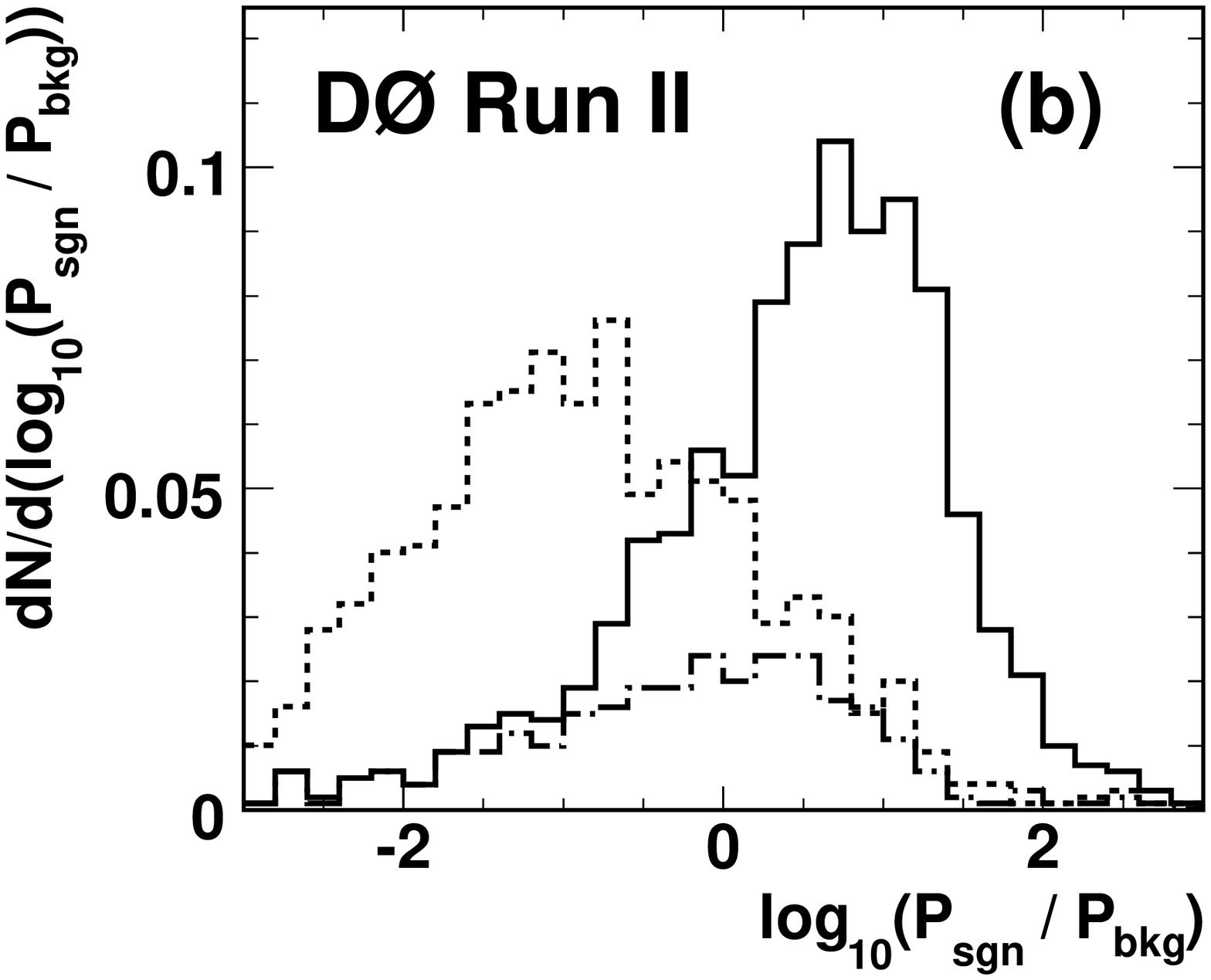}
\caption{Distributions of $\log_{10}(\psgn/\pbkg)$ for \ttbar events with 
  $\mtop=175\,\GeV$ (solid) and \wjets events (dashed lines) for
  (a) \ejets events and (b) \mujets events.
  The \psgn values are calculated for the assumption $\mtop=175\,\GeV$.
  The distributions for signal and background events are
  normalized individually.  The distributions for those \ttbar events that fail
  the requirement of jets matched to partons are shown
  separately (dash-dotted lines).}
\label{fig:MEpsgnandpbkg}
\end{center}
\end{figure}

In the simulation, about
$20-30\,\%$ of \ttbar events have jets and partons that
cannot be unambiguously matched, i.e.,\ at least one of the four
reconstructed jets cannot be matched to a
parton from the \ttbar decay within $\DeltaR<0.5$.
These events yield poor top mass information and degrade the uncertainty
estimate of the likelihood fit.
Figure~\ref{fig:MEpsgnandpbkg} illustrates that
jet-parton matched \ttbar events tend to have a higher signal than background
probability density, which is how the mass fit identifies them as
signal-like.
There is no such separation for signal events in which one or more
jets cannot be matched to a parton, so that these events contribute much 
less mass information to the final likelihood.
This observation is consistent with the fact that a leading-order
matrix element is used to describe \ttbar events.
Therefore, only jet-parton matched events are used to calibrate the
\pbkg normalization.
On average, the \ftop fit will consequently yield the fraction
of jet-parton matched (leading-order) \ttbar events in the event sample.
The quoted \ftop values are corrected for this effect.

The \pbkg normalization is determined as follows:
\begin{itemize}

\item 
A large ensemble of simulated \ttbar and \wjets events is composed with 
the signal fraction as determined by the topological 
likelihood fit described in Section~\ref{sec:sample}.
\item 
The top mass likelihood fit described in Section~\ref{sec:MEfit} is
applied to the sample and the \pbkg normalization is adjusted 
iteratively until the fit result yields the true signal fraction.
\item 
The normalization of \pbkg cannot depend on the top quark mass.
Therefore, the above steps are applied to \ttbar Monte Carlo samples
with different generated masses. The mean of all results is
taken as the \pbkg normalization.
\end{itemize}
This procedure is applied separately for \ejets and \mujets events.
Note that the topological likelihood discriminant is only used to determine the
normalization of the background probability and the sample 
composition for ensemble tests used to calibrate the procedure. 
The topological likelihood discriminant does not otherwise enter 
the top quark mass fit.

\section{Top Quark Mass Measurement Using Topological Information}
\label{sec:MEfit}

\subsection{Top Quark Mass Fit}
\label{sec:MEfit-fit}
The top quark mass and overall jet energy scale $JES$ are determined
as optimal values of the 
likelihood for the sample of selected events, which depends on the \psgn 
and \pbkg values.
For each measured event, \psgn is calculated for
various values of \mtop in steps of $2.5\,\GeVcc$ and various values
of $JES$ in steps of $0.01$.
It has been found that it is not necessary to compute the background
probability for different values of the jet energy scale.
Therefore, all \pbkg values are computed for $JES=1$ only.
Both \psgn and \pbkg are normalized as described in
Sections~\ref{sec:MEpsgn} and~\ref{sec:MEpbkg}, using separate constants for
\ejets and \mujets events.
The top quark mass measurements on the \ejets, \mujets, and combined
\ljets event samples are in each case derived from the likelihood
of the event sample, given by Equation~(\ref{eq:MElhood-fnc}), in the 
way described below.

For given values of \mtop and $JES$, each event probability 
$\pevt=\ftop\psgn+(1-\ftop)\pbkg$ depends on the signal fraction \ftop
of the sample,
and consequently, the value of the likelihood for the event sample 
is a function of \ftop.
For each (\mtop,$JES$) parameter pair, the best
\ftop parameter value is determined, and the likelihood value
corresponding to this value is used in further computations.
The overall result quoted for the fitted signal fraction \ftop is
derived from the
value obtained at the point in the grid of (\mtop,$JES$) assumptions 
with the maximum
likelihood value for the event sample.
The uncertainty on \ftop is computed by varying \ftop at fixed \mtop and
$JES$ until $\Delta(-\ln L)=+\frac{1}{2}$.
This uncertainty does not account for correlations between \ftop, \mtop, and $JES$.

The result for the top quark mass is obtained from a projection of the 
two-dimensional grid of likelihood values onto the \mtop axis.
In this projection, the correlation between \mtop and the $JES$
parameter is taken into account.
The probability for a given \mtop hypothesis is obtained as the integral
over the likelihood as a function of $JES$, using linear interpolation
between the grid points and
Gaussian extrapolation to account for the tails for $JES$ values outside
the range considered in the grid.

The probabilities as a function of assumed top mass are converted
to $-\ln L$ values.
These $-\ln L$ points are then fitted with a fourth order polynomial in the 
region defined by the condition $\Delta\ln L<3$ around the best value.
The $-\ln L$ points on either side of the $\Delta\ln L<3$ region are each
fitted with a parabola, and Gaussian extrapolation is used to describe
the tails outside the range of \mtop hypotheses considered.
The \mtop value that maximizes the fitted probability is taken to be
the measured value of the top quark mass.
The lower and upper uncertainties on the top mass are defined 
such that 68\% of the total probability integral is enclosed by the 
corresponding top mass values, with equal probabilities at both
limits of the 68\% confidence level region.

The same projection and fitting procedure is applied to determine the 
value of the $JES$ parameter.

\subsection{Validation of the Method}
\label{sec:MEvalidation}
The method is first validated using parton-level simulated \ttbar
and \wjets events.
These have been generated with leading-order event generators
(\madgraph~\cite{bib:MADGRAPH} for \ttbar events, \alpgen
for \wjets events), i.e.,\ no
initial or final state radiation is included.
The jet energies in these events are smeared according to 
the transfer functions described in Section~\ref{sec:MEtransferfcn}
(the treatment of the muon transverse momentum integration has
been checked with additional ensemble tests not described here).

Ensembles are composed with $75$ events, $40\,\%$
of which are \ttbar signal events.
A total of $1000$ events for top masses of $160$, $170$, $175$, $180$, and
$190\,\GeVcc$ each are used, along with $1000$ \wjets events.
In addition, samples with $\mtop=175\,\GeVcc$ with all jet energies
scaled by $0.95$ and $1.05$ are prepared in order to validate the
$JES$ fit result.
All events are required to pass the kinematic selection
criteria listed in Table~\ref{kinematiccuts.table}.  The final state
jets and the charged lepton must be separated according to
$\DeltaR(j,j')>0.5$ and $\DeltaR(\ell,j)>0.5$.
The signal normalization is obtained according to this selection, see
Section~\ref{sec:MEpsgn}.
\mtop and $JES$ are obtained for each ensemble as described in
Section~\ref{sec:MEfit-fit}.
The results of this test 
show that the fitted top mass and jet energy scale are 
unbiased within statistical uncertainties of $300\,\MeV$
and $0.003$, respectively.
Furthermore, the fitted \mtop value does not depend on the input 
$JES$ value used
in the ensemble generation, and similarly, the fitted $JES$ value is
independent of the true input top mass.

To test that the uncertainties obtained from the fit describe the actual measurement
uncertainty, the deviation of the fitted top mass 
from the true value is divided by the fitted measurement uncertainty.
The upper (lower) measurement uncertainty is taken if the fitted value
is below (above) the true value.
This definition is chosen to account for the possibility of asymmetric
uncertainties.
This distribution of deviations normalized by the measurement
uncertainty is
fitted with a Gaussian, and its width, commonly referred to as 
``pull width,'' is in agreement with $1.0$.
This is also the case for the jet energy scale measurement, for which 
same test has been performed.

The events used in the test outlined in this section have been generated with 
the same simplified model that is used in the probability calculation
for the description of the production of signal and background events
and the detector response, cf.\ Section~\ref{sec:ME}.
As it cannot be assumed that this simplified model correctly
reproduces every aspect of the data, the method for measuring
the top quark mass has been calibrated with Monte Carlo events that
have been generated with the full \dzero simulation.
Any deviations observed in this calibration step are taken into
account in the final result.
The calibration is described in the following section.

\subsection{Calibration of the Method}
\label{sec:MEcalibration}
The default \dzero Monte Carlo events, generated as described in 
Section~\ref{sec:data} and passed through the full simulation of
the \dzero detector, are found to describe the data well.
They are therefore used to derive the final calibration of the fitting
procedure.
\ttbar samples with top quark masses of $160$, $170$, $175$,
$180$, and $190~\GeVcc$ and a \wjets sample are used.
In addition, samples with $\mtop=175\,\GeVcc$ where all jet energies are
scaled by $0.92$, $0.96$, $1.04$, and $1.08$ are prepared in order to
calibrate the $JES$ fit.
For each sample and each lepton channel (\ejets and \mujets), \psgn
and \pbkg are calculated for $1000$ events which pass the event
selection.
Ensembles are drawn from these event pools, with an ensemble composition
as measured for the data sample.
Each probability is normalized according to the flavor of the
isolated lepton (see Sections~\ref{sec:MEpsgn} and
\ref{sec:MEpbkg}).
The QCD contribution is not added during the calibration but treated
as a systematic uncertainty (cf.\ Section~\ref{sec:systuncs}).

\begin{figure}
\begin{center}
\includegraphics[width=\figMEcalibwidth\textwidth]{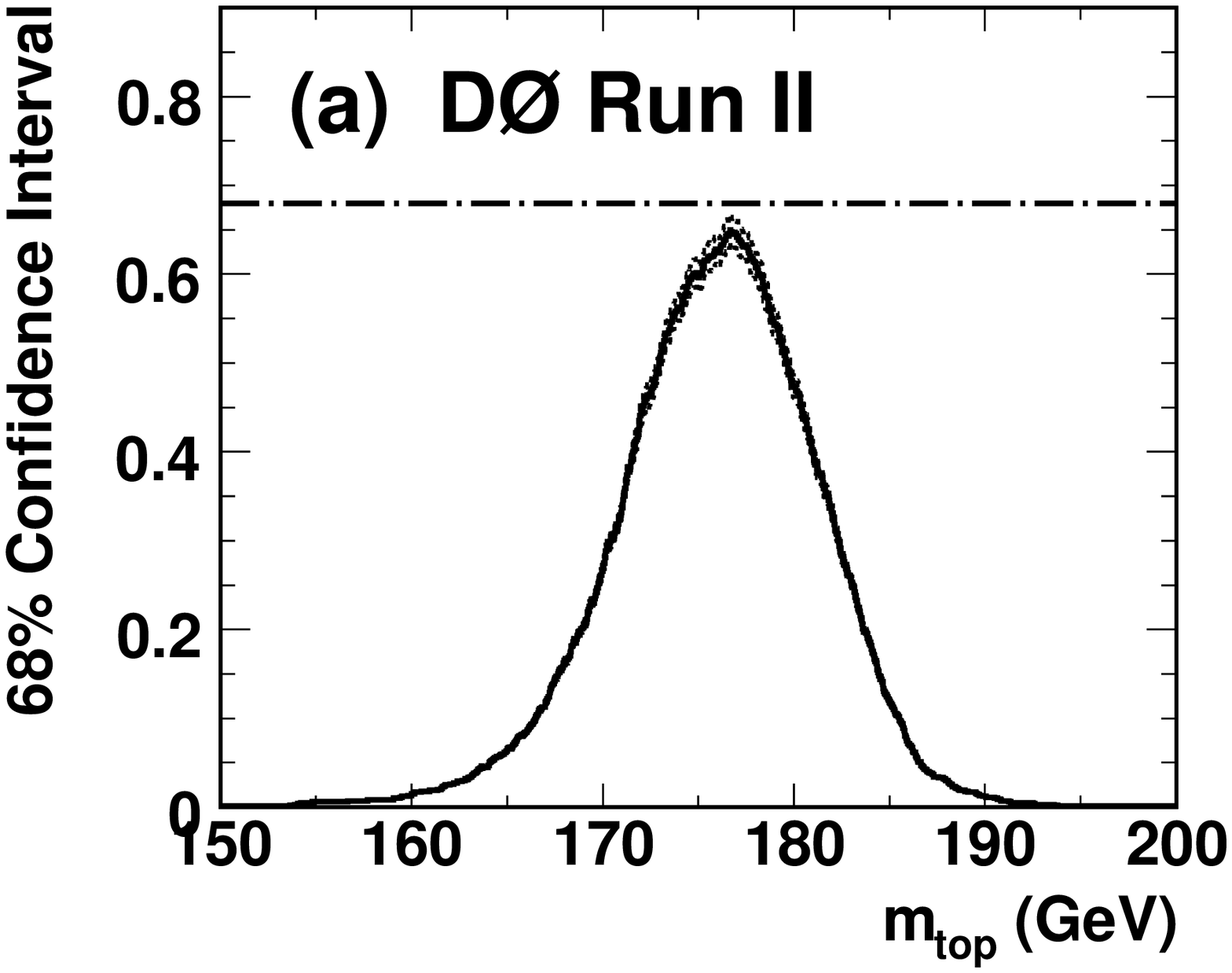}
\hspace{-0.03\textwidth}
\includegraphics[width=\figMEcalibwidth\textwidth]{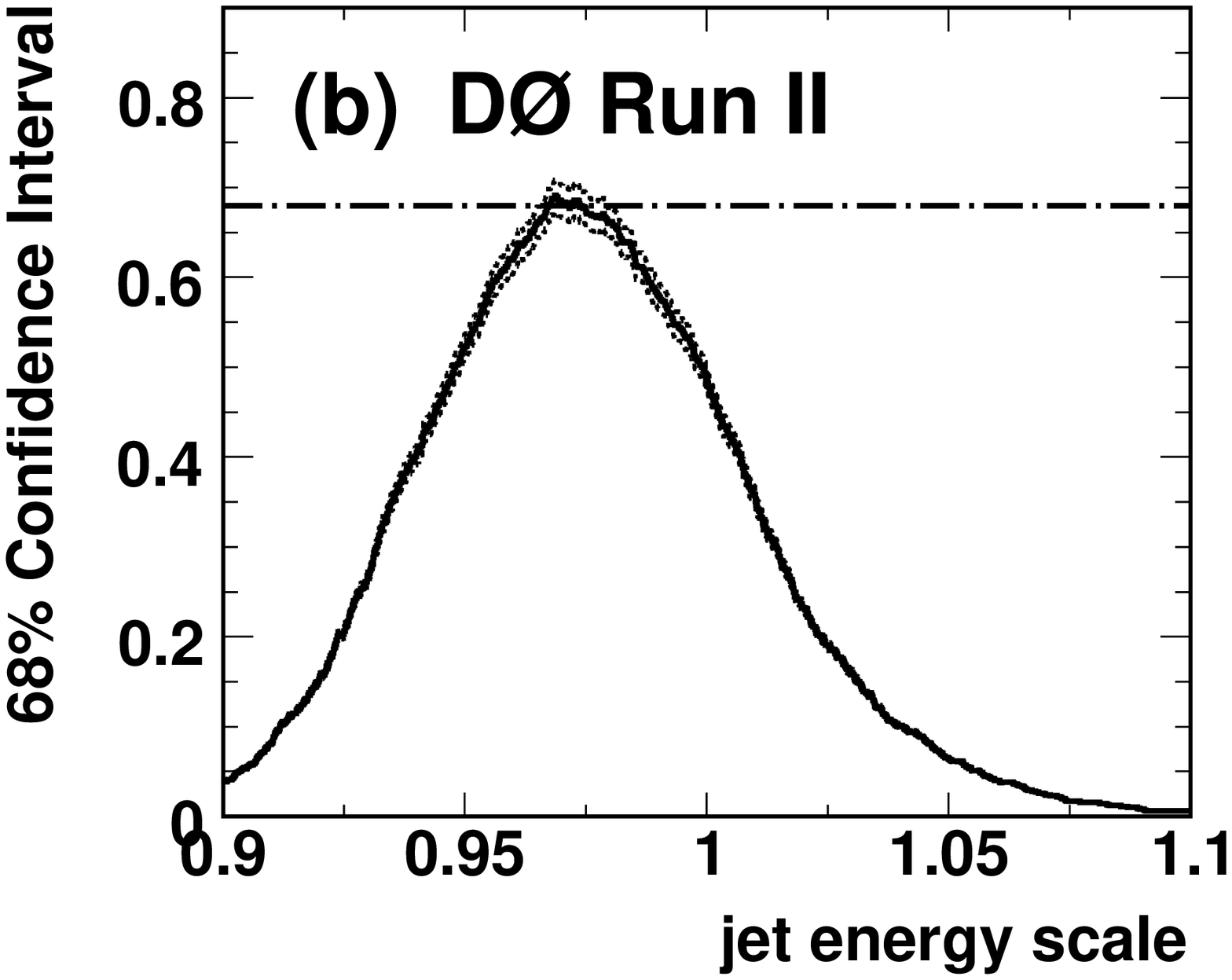}
\caption{Calibration of the Matrix Element mass fitting 
procedure for the topological analysis, using ensembles with a
top quark mass of 175\,\GeV and $JES=1.0$.  The 68\% confidence interval
distributions for (a) the measured top quark mass and (b) the jet energy
scale is given by the solid, the upper and lower error
bands by the dashed histograms.  A value of 0.68 as indicated by
the dash-dotted line would mean that the corresponding \mtop ($JES$) value 
is included in the fitted $68\%$ confidence interval in 68\% of the
ensembles.}
\label{fig:CL-topo}
\end{center}
\end{figure}

\begin{figure}
\begin{center}
\includegraphics[width=\figMEcalibwidth\textwidth]{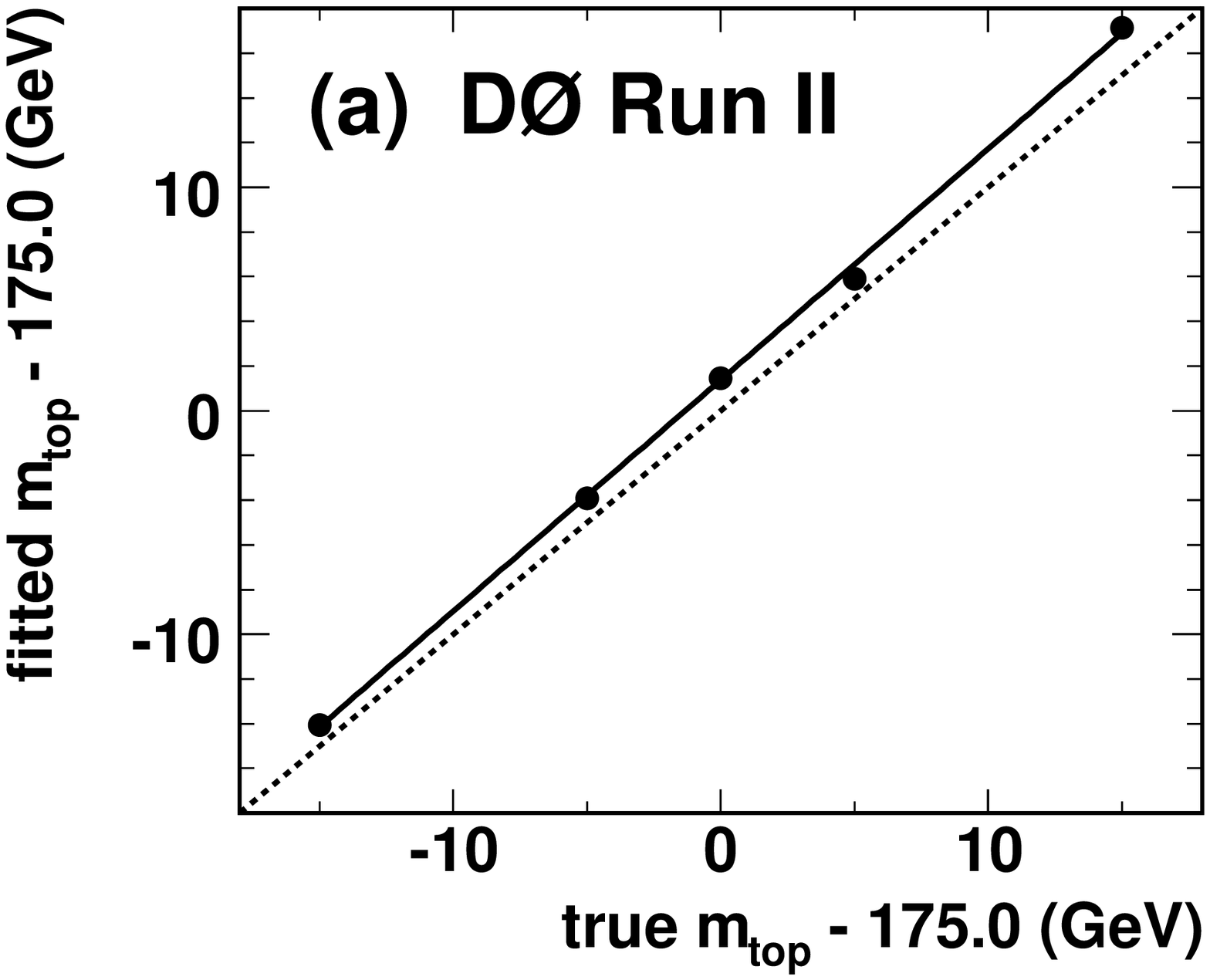}
\hspace{-0.03\textwidth}
\includegraphics[width=\figMEcalibwidth\textwidth]{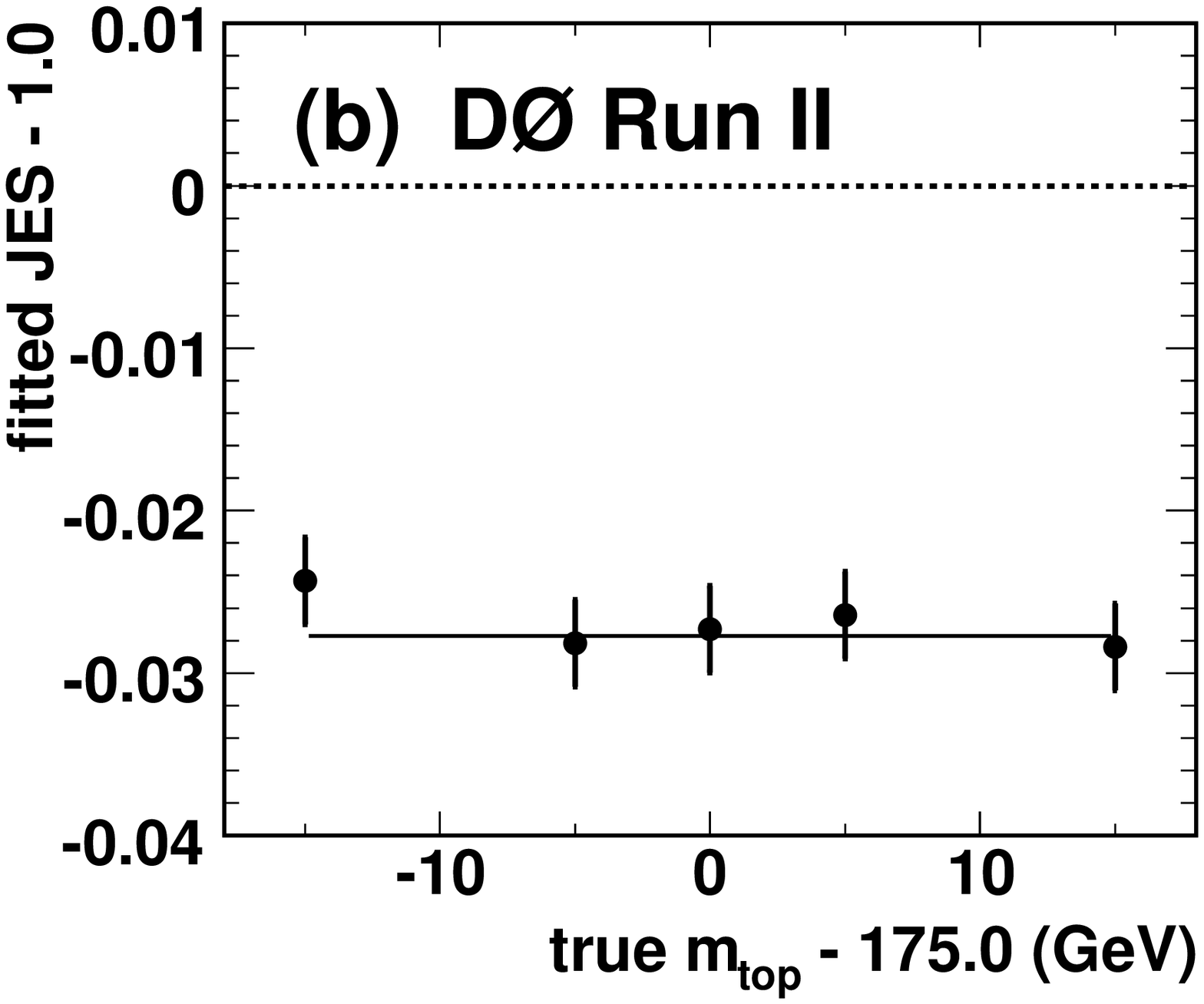}
\includegraphics[width=\figMEcalibwidth\textwidth]{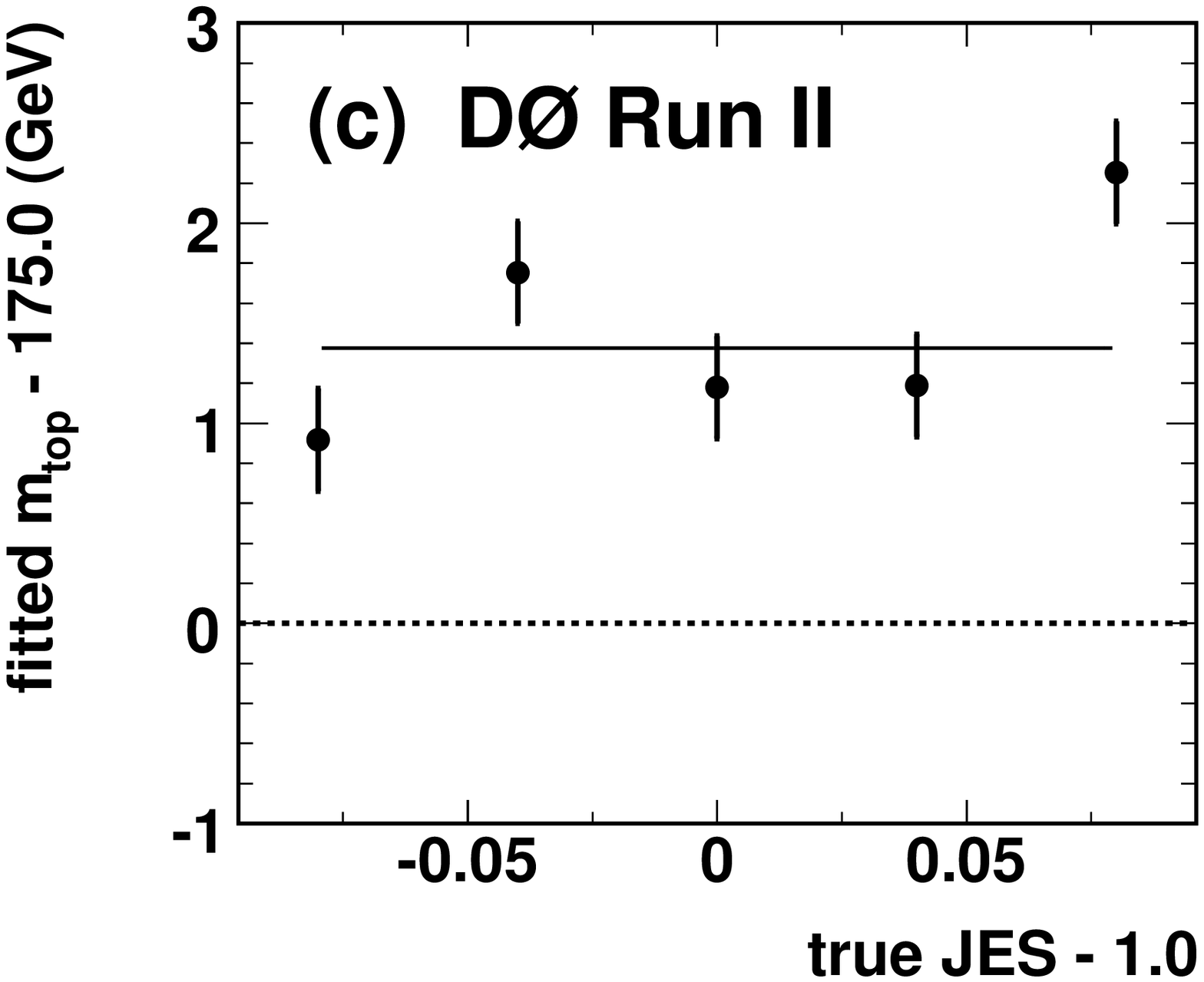}
\hspace{-0.03\textwidth}
\includegraphics[width=\figMEcalibwidth\textwidth]{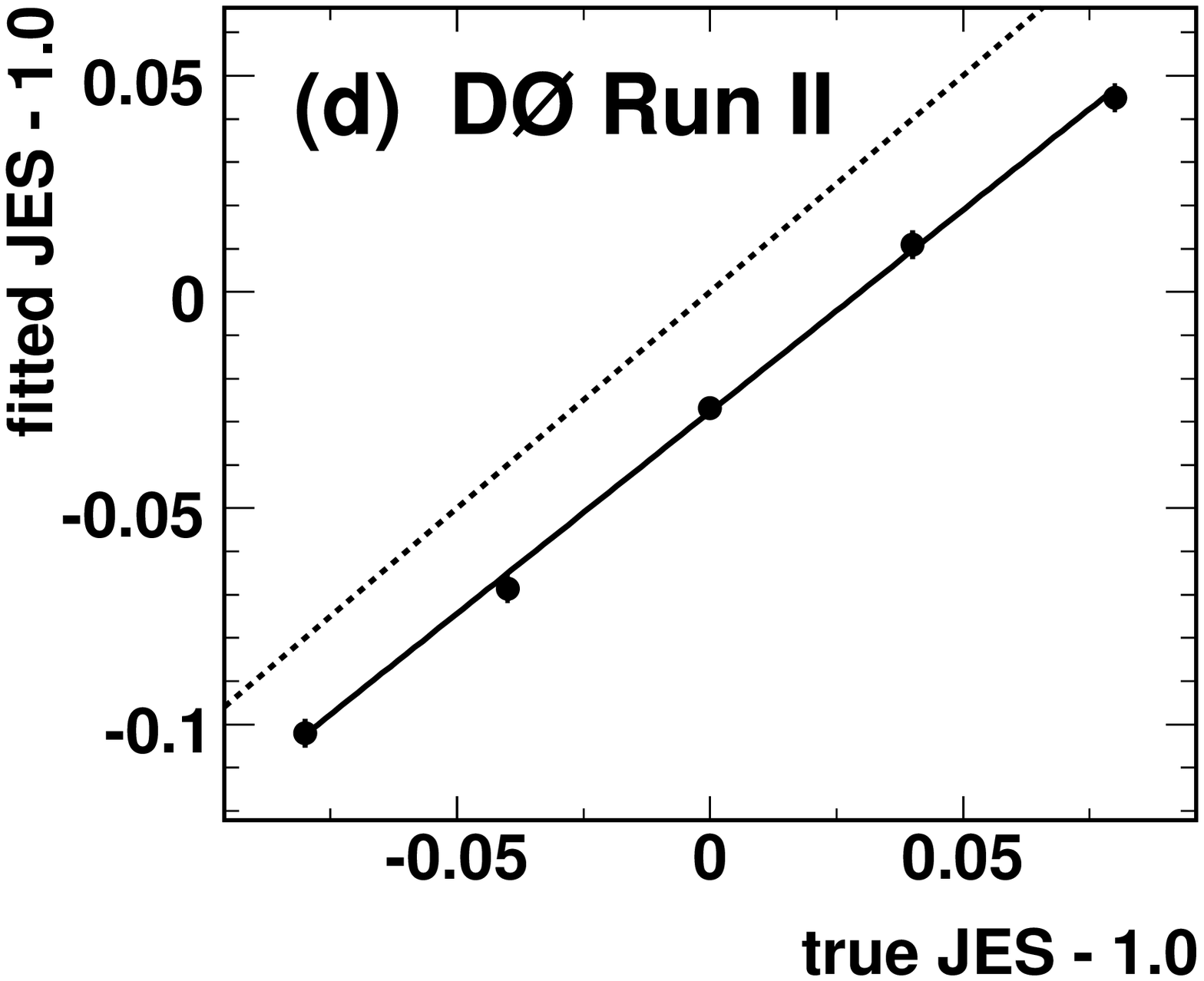}
\caption{Calibration of the Matrix Element mass fitting 
procedure for the topological analysis.  The upper plots show the 
reconstructed top mass (a) and the measured jet energy scale
(b) as a function of the input top mass.
The two lower plots show the reconstructed top mass (c) and the 
measured jet energy scale (d) as a function of the
input jet energy scale.
The solid lines show the results of linear fits to the points,
which are used to calibrate the measurement technique.
The dashed lines would be obtained for equal 
fitted and true values of \mtop and $JES$.}
\label{fig:MEcalib-topo}
\end{center}
\end{figure}

\begin{figure}
\begin{center}
\includegraphics[width=\figMEcalibwidth\textwidth]{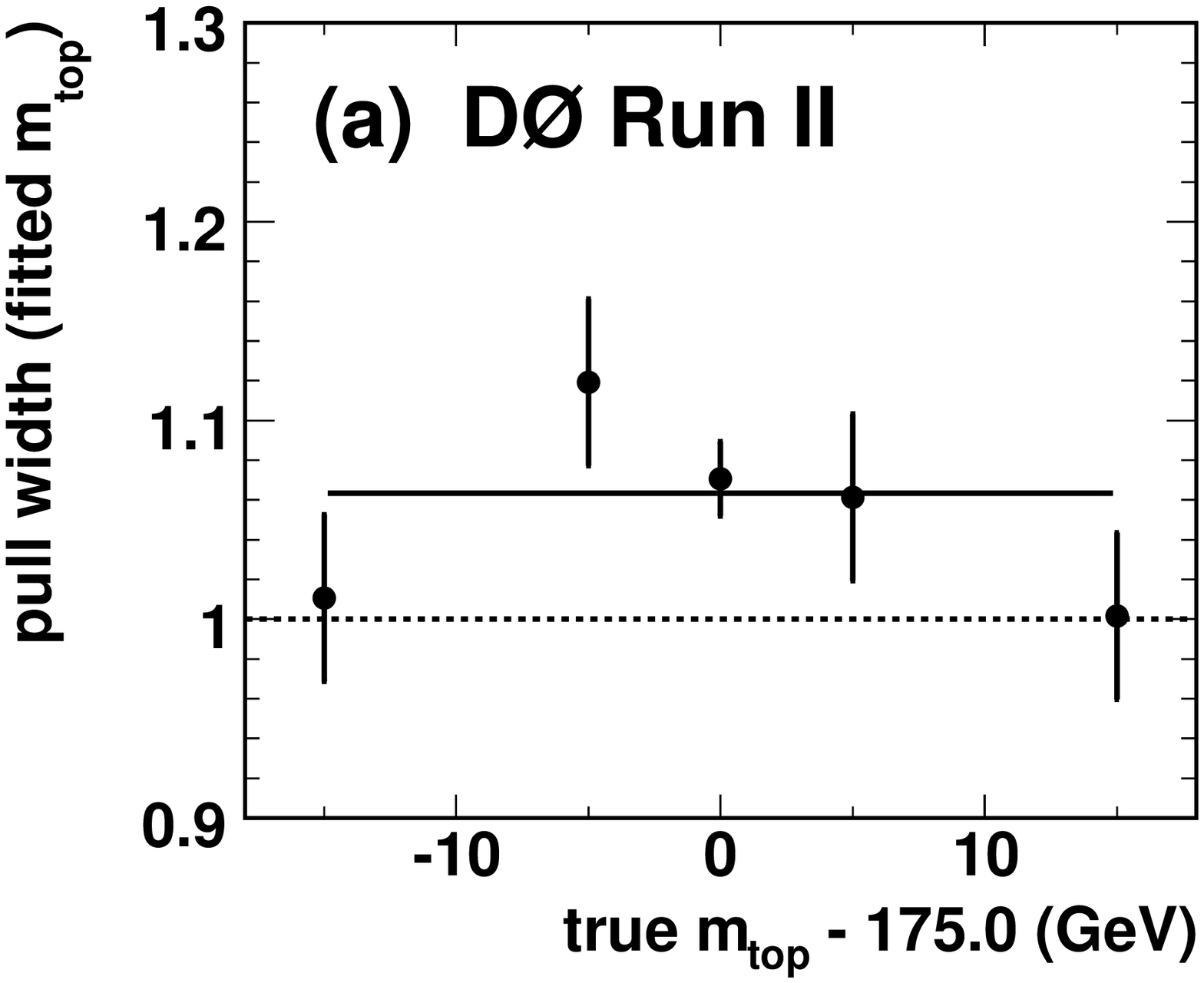}
\hspace{-0.03\textwidth}
\includegraphics[width=\figMEcalibwidth\textwidth]{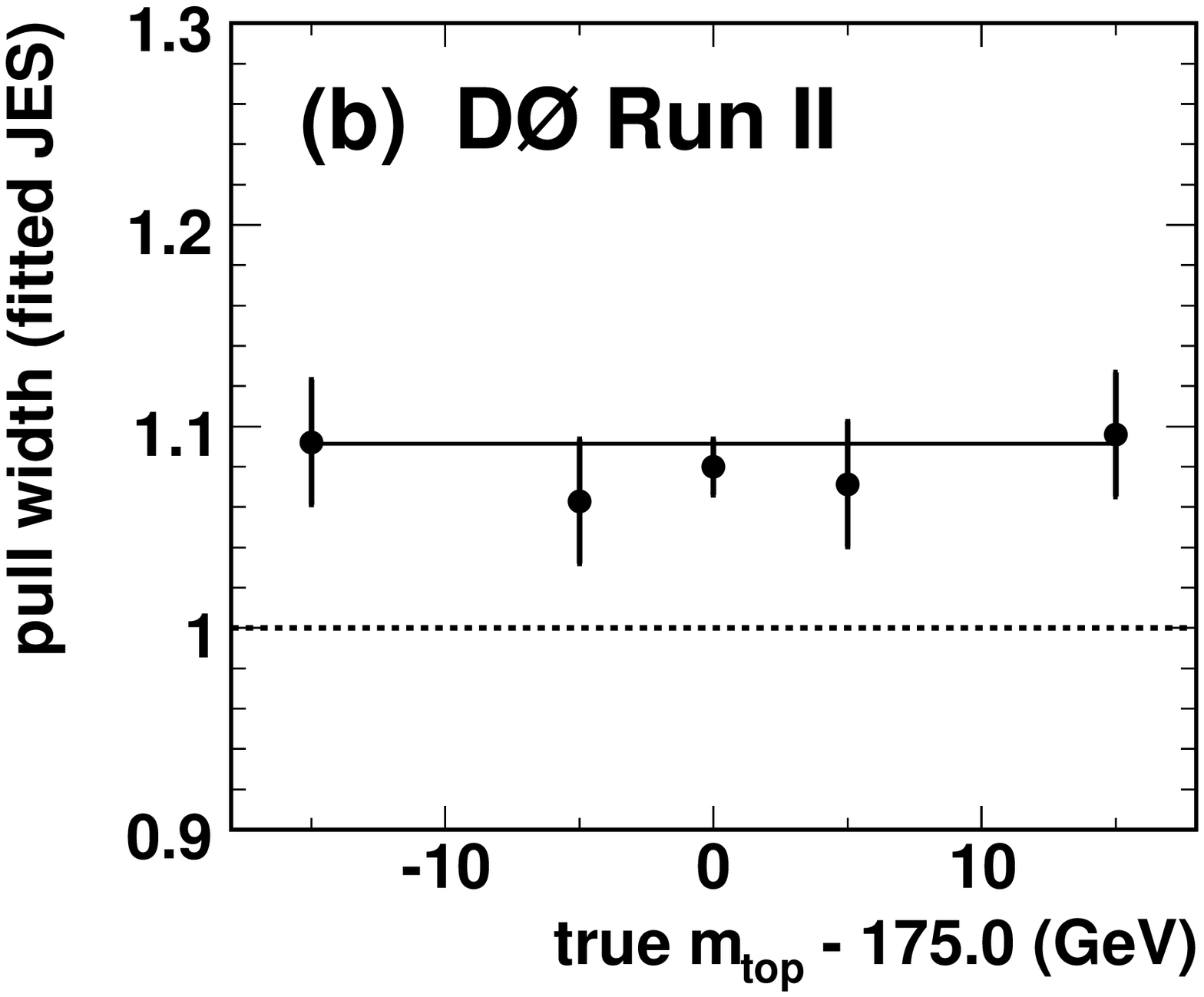}
\includegraphics[width=\figMEcalibwidth\textwidth]{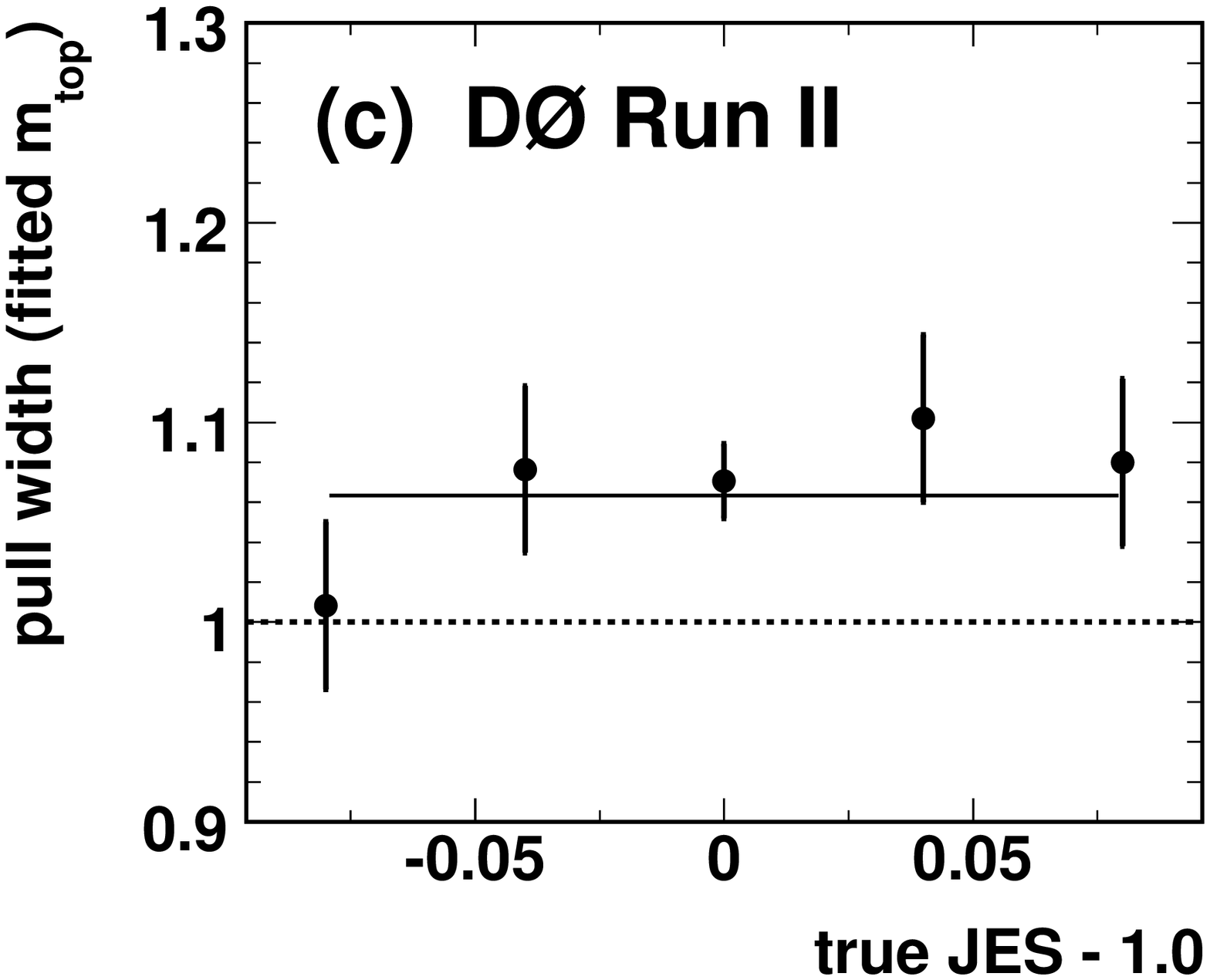}
\hspace{-0.03\textwidth}
\includegraphics[width=\figMEcalibwidth\textwidth]{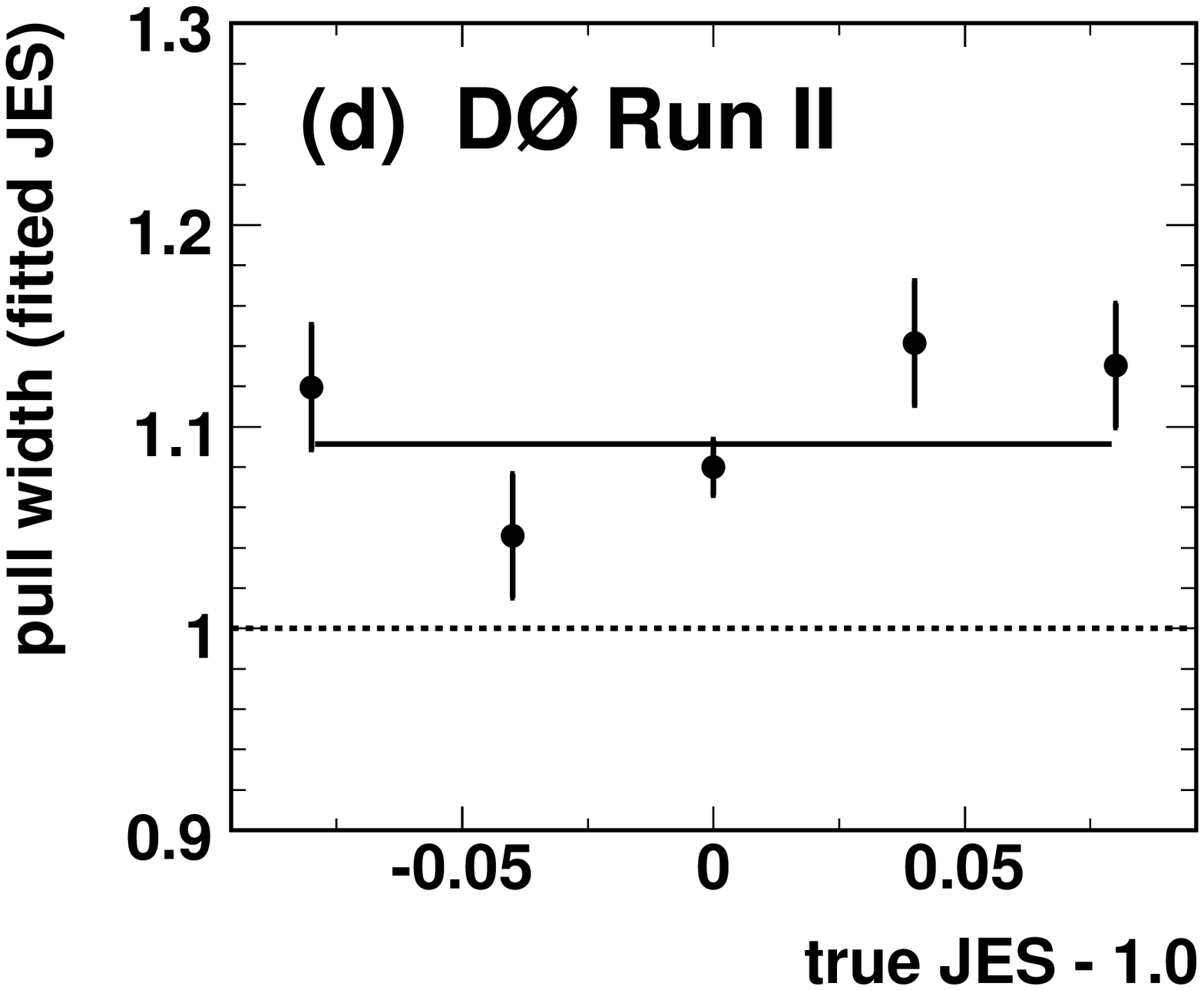}
\caption{Calibration of the Matrix Element mass fitting 
procedure for the topological analysis.  The upper plots show the 
widths of the pull distributions for the top mass
(a) and jet energy scale (b)
as a function of the input top mass.
The two lower plots show the widths of the pull distributions for the top mass
(c) and jet energy scale (d) as a function of the
input jet energy scale.
The solid lines show the mean pull width, while
the dashed lines indicate a pull width of 1.0.}
\label{fig:MEpullcalib-topo}
\end{center}
\end{figure}

In Fig.~\ref{fig:CL-topo}, $68\%$ confidence interval distributions 
are shown for ensembles with $\mtop=175\,\GeV$ and $JES=1.0$.
For many pseudo-experiments, the true \mtop ($JES$) value is expected 
to be within the fitted uncertainties in 68\% of the pseudo-experiments
(corresponding to a value of the conficence interval distribution of 
$0.68$ for this value), while other \mtop ($JES$) values should be 
less likely to be within the uncertainties.
When the error interval resulting from the integration of $68\%$ of
the likelihood distribution does not include the input top mass
($JES$) value $68\%$ of the time, the uncertainty is inflated to correspond
to an integration over a larger interval.
The calibration results for the combined fit to the \ejets and \mujets
ensembles are shown in
Figs.~\ref{fig:MEcalib-topo} and \ref{fig:MEpullcalib-topo}.
The fit results are corrected for the offsets $o$ and slopes $s$,
and for the deviations of the pull width $w$ from 1.0 given in 
Table~\ref{tab:calib-topo} to obtain the final results and their
statistical uncertainties as follows:
\eqMTOPFITTOPOoffsetpullcorrection

\begin{table}
\begin{center}
\begin{tabular}{c@{\quad}r@{\quad}c@{\quad}c}
\hline
\hline
  &
    \multicolumn{1}{c}{offset $o$}
  &
    slope $s$
  &
    pull width $w$
\\
\hline
    \mtop
  &
        \offsetmtoptopo GeV
  &
    \slopemtoptopo
  &
    \pullwidthmtoptopo
\\
    $JES$
  &
        \offsetjestopo \phantom{GeV}
  &
    \slopejestopo
  &
    \pullwidthjestopo
\\
\hline
\hline
\end{tabular}
\caption{Calibration of the Matrix Element mass fitting 
procedure for the topological analysis.  The offsets are 
quoted for a true top quark mass of \offsetatmtoptopo~GeV
and a true jet energy scale of \offsetatjestopo, respectively.
Only statistical uncertainties are quoted in this table.}
\label{tab:calib-topo}
\end{center}
\end{table}

\subsection{Result}
\label{sec:MEresult}
\begin{figure}
\begin{center}
\includegraphics[width=0.45\textwidth]{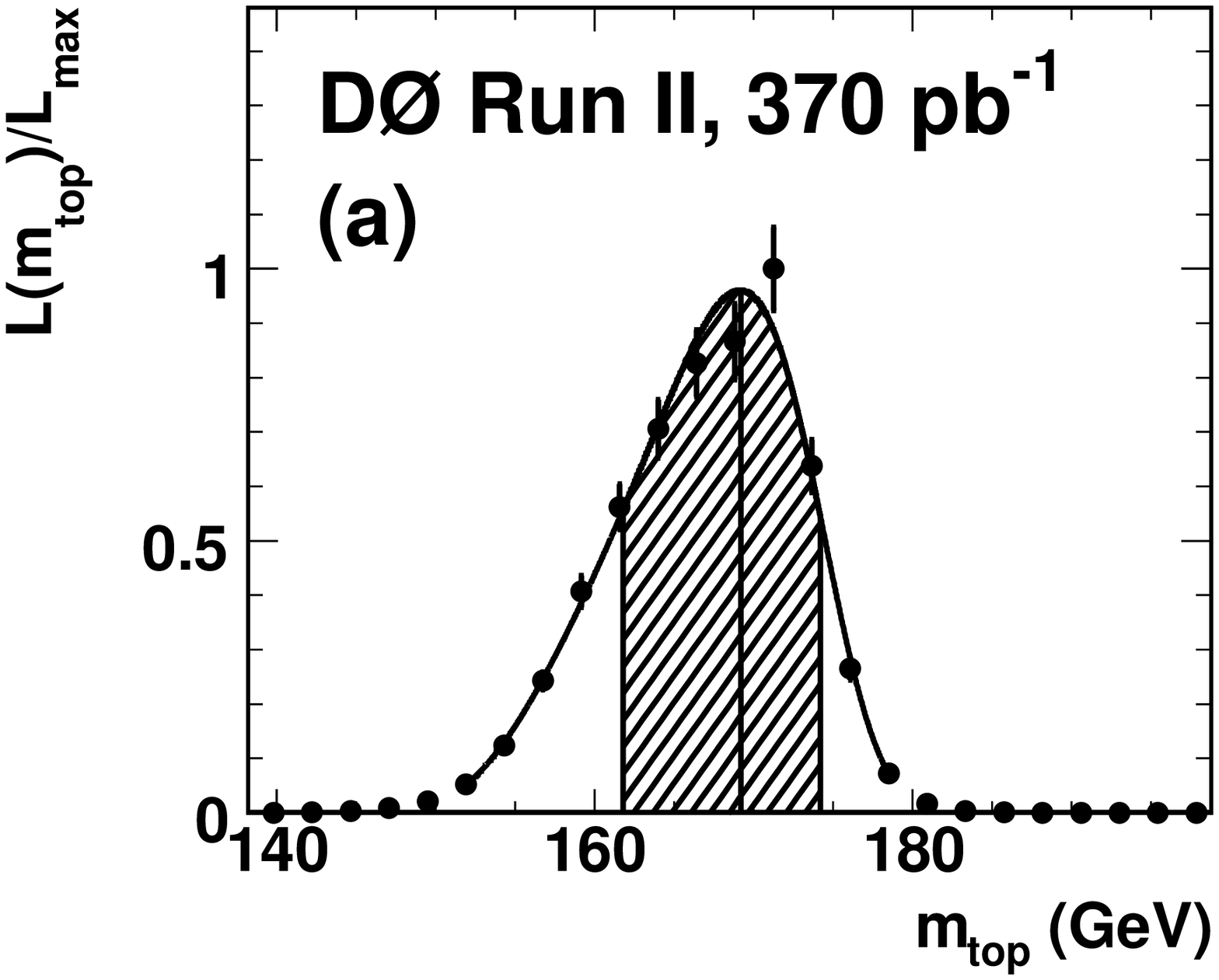}
\includegraphics[width=0.45\textwidth]{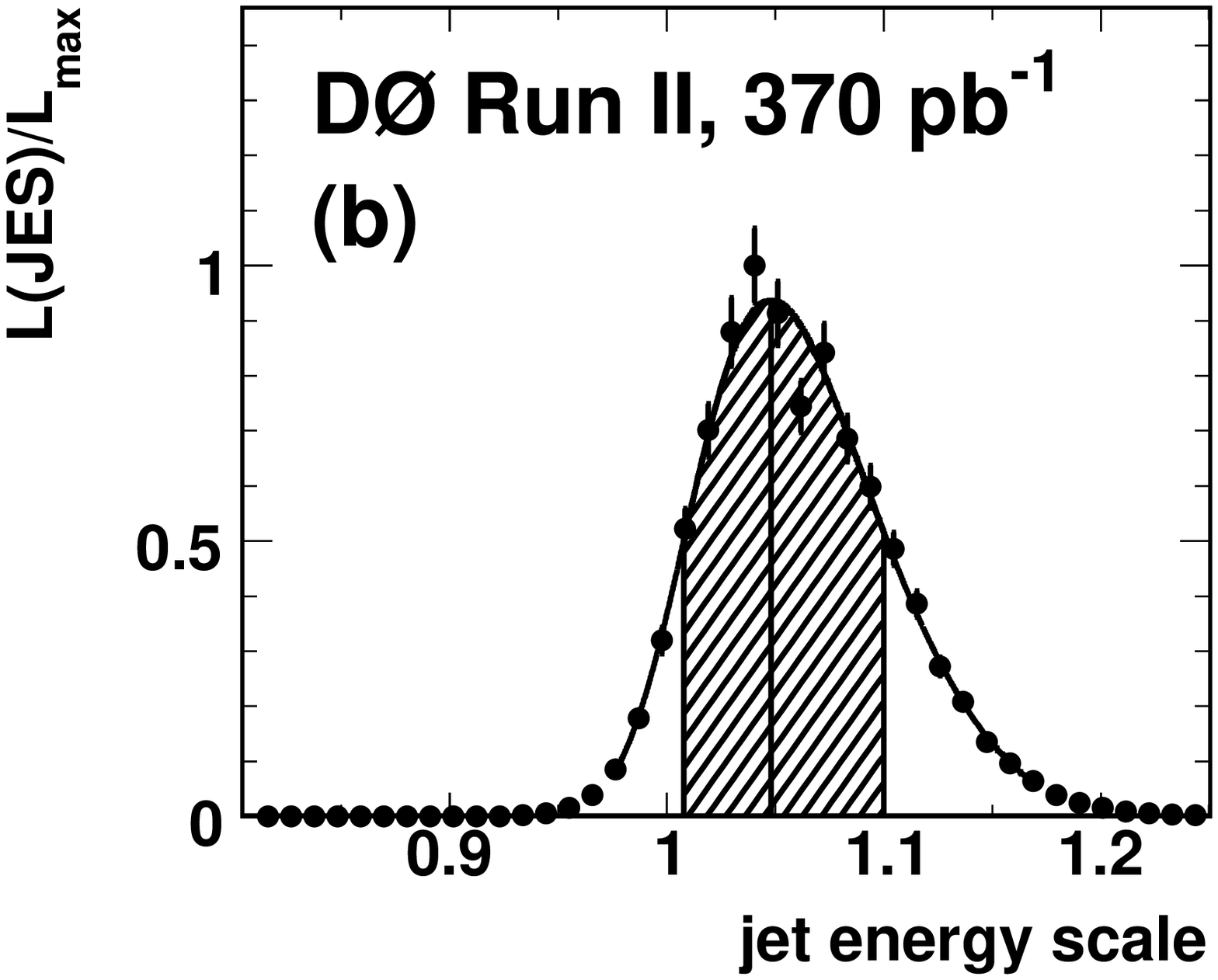}
\caption{Application of the topological Matrix Element method to the data.
The \mtop and $JES$ axes correspond to the calibrated
values. Plot (a) shows the probability as a function of 
assumed top quark mass.  The correlation with the jet energy scale
is taken into account.
The fitted curve is shown, as well as the 
most likely value and the 68\% confidence level region.
The corresponding plot for the $JES$ parameter is shown in (b).}
\label{fig:MEresult}
\end{center}
\end{figure}

\begin{figure}
\begin{center}
\includegraphics[width=0.45\textwidth]{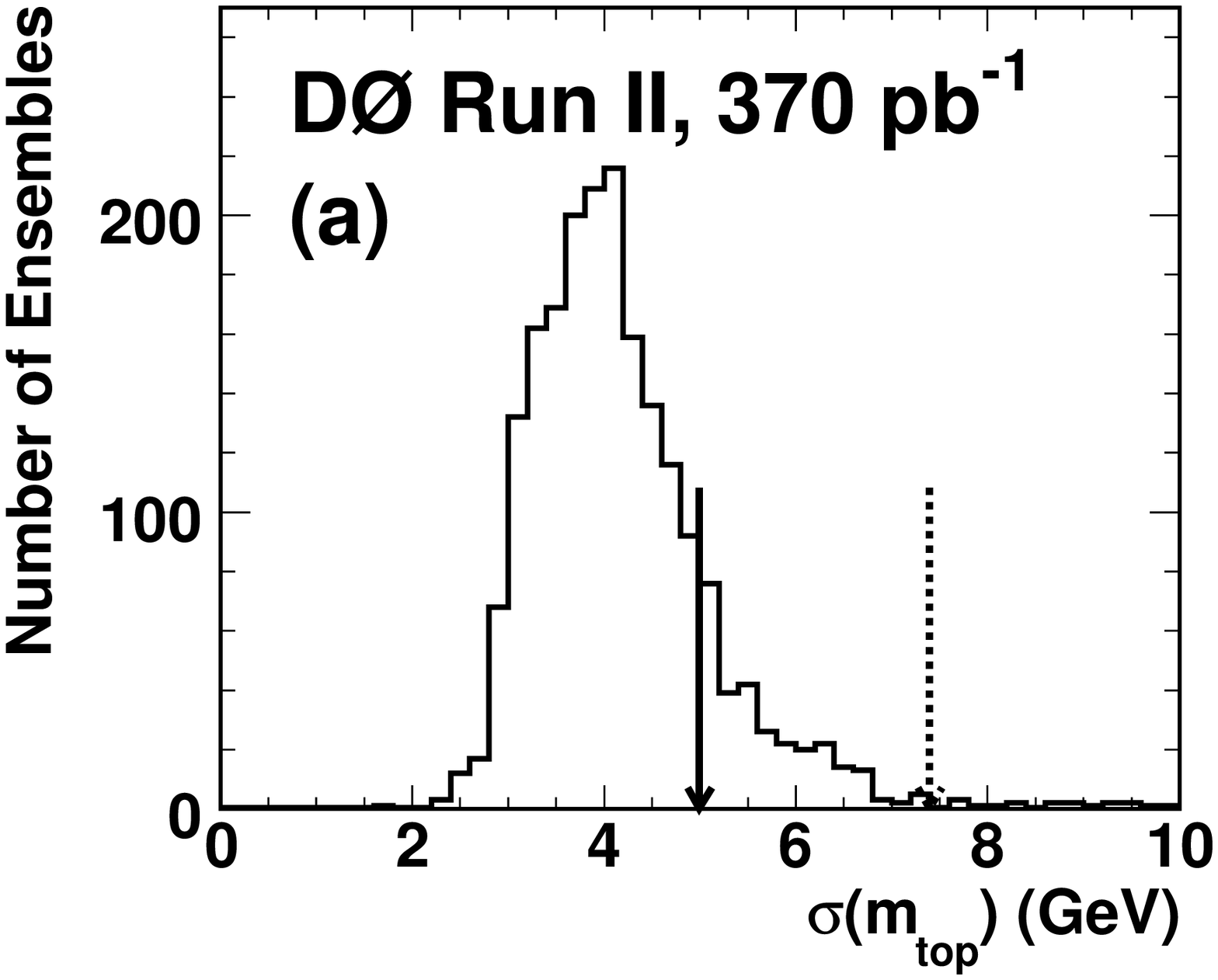}
\includegraphics[width=0.45\textwidth]{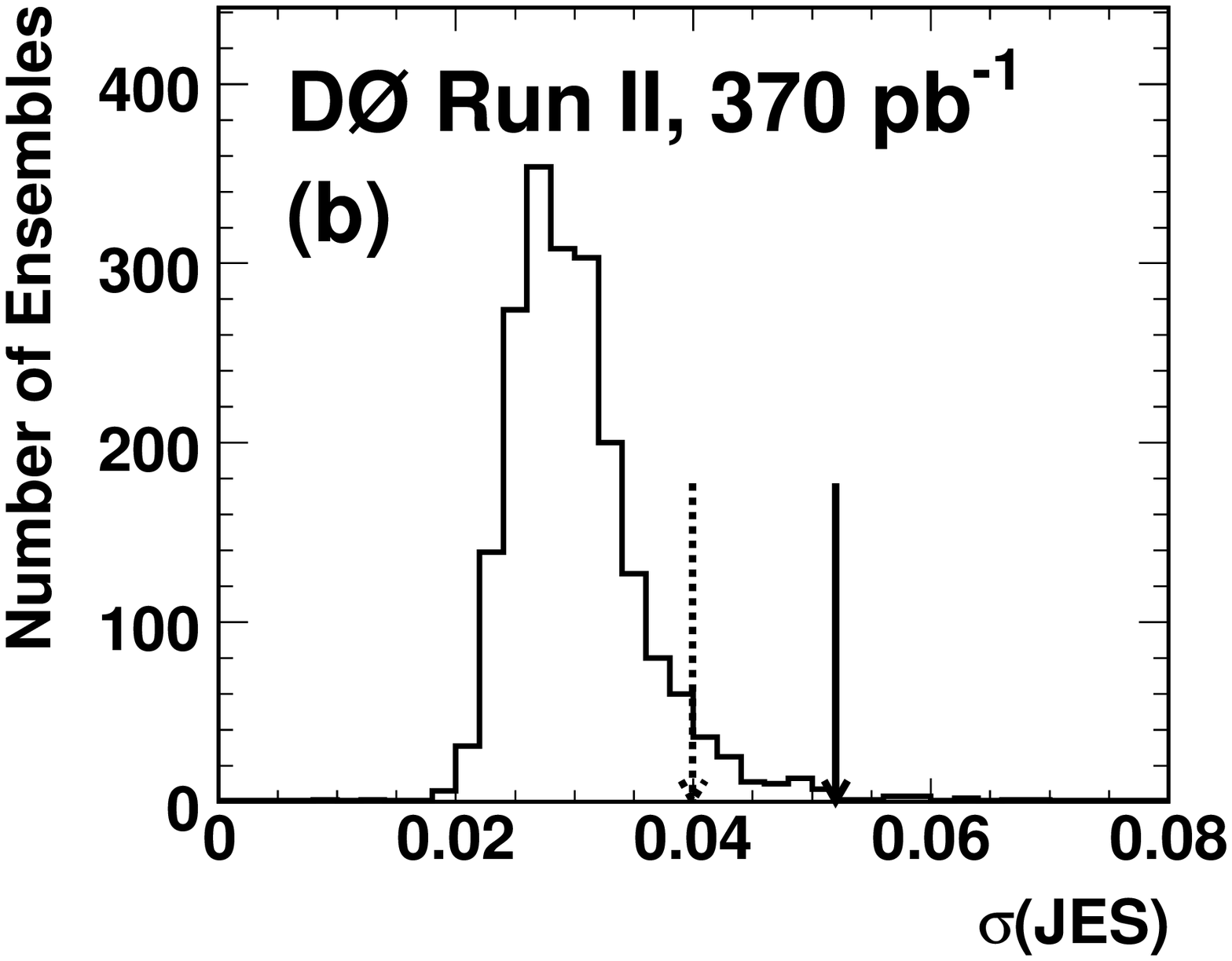}
\caption{Uncertainties on \mtop (a) and $JES$ (b) in the topological
analysis.  The distributions of fitted uncertainties obtained 
from ensemble tests are shown by the histograms.  Both upper and 
lower uncertainties are shown; their distributions are very similar.
The upper (lower) uncertainty in the data is indicated by the solid 
(dashed) arrow.
The probability for a lower uncertainty on \mtop with a magnitude
larger than that observed in the data is 2\%.}
\label{fig:MEresult-error}
\end{center}
\end{figure}

The Matrix Element method is applied to the $370\,\ipb$ lepton+jets data set.
The calibrations for \mtop derived in the previous section are taken
into account.
The calibrated fit result for the combined lepton+jets sample is shown in
Fig.~\ref{fig:MEresult}.
In this figure, the probability as a function of assumed top mass
is shown together with the fitted curve (the polynomial
fitted to the $-\ln L$ values as described in
Section~\ref{sec:MEfit-fit} has been transformed accordingly), 
and the central value and 68\% confidence level interval are
indicated.
The probability as a function of assumed $JES$ parameter is 
also shown.
The top quark mass is measured to be
\begin{eqnarray}
	\mtop^{\ejets}  & = & \resultemstat\,\GeVcc \nonumber \\
	\mtop^{\mujets} & = & \resultmustat\,\GeVcc \nonumber \\
	\mtop^{\ljets}  & = & \resultstatenspace\,\GeVcc \ .
\end{eqnarray}
The statistical uncertainties are consistent with the expectation.
A comparison of the fitted uncertainties on \mtop and $JES$ with the 
expectations from ensemble tests is given in Fig.~\ref{fig:MEresult-error}.
The fit yields a signal fraction \ftop of $\resultftopstat$,
in good agreement with the result of the topological likelihood fit.
The fitted jet energy scale is $JES=\resultjesstat$ and indicates that the
data is consistent with the reference scale.

For a fixed jet energy scale, the statistical uncertainty of the fit is
$\MEerrstatnojes\,\GeVcc$;
thus the component from the jet energy scale uncertainty is
$\MEerrjes\,\GeVcc$.
Systematic uncertainties are discussed in Section~\ref{sec:systuncs}.

\begin{figure}
\begin{center}
\includegraphics[width=0.45\textwidth]{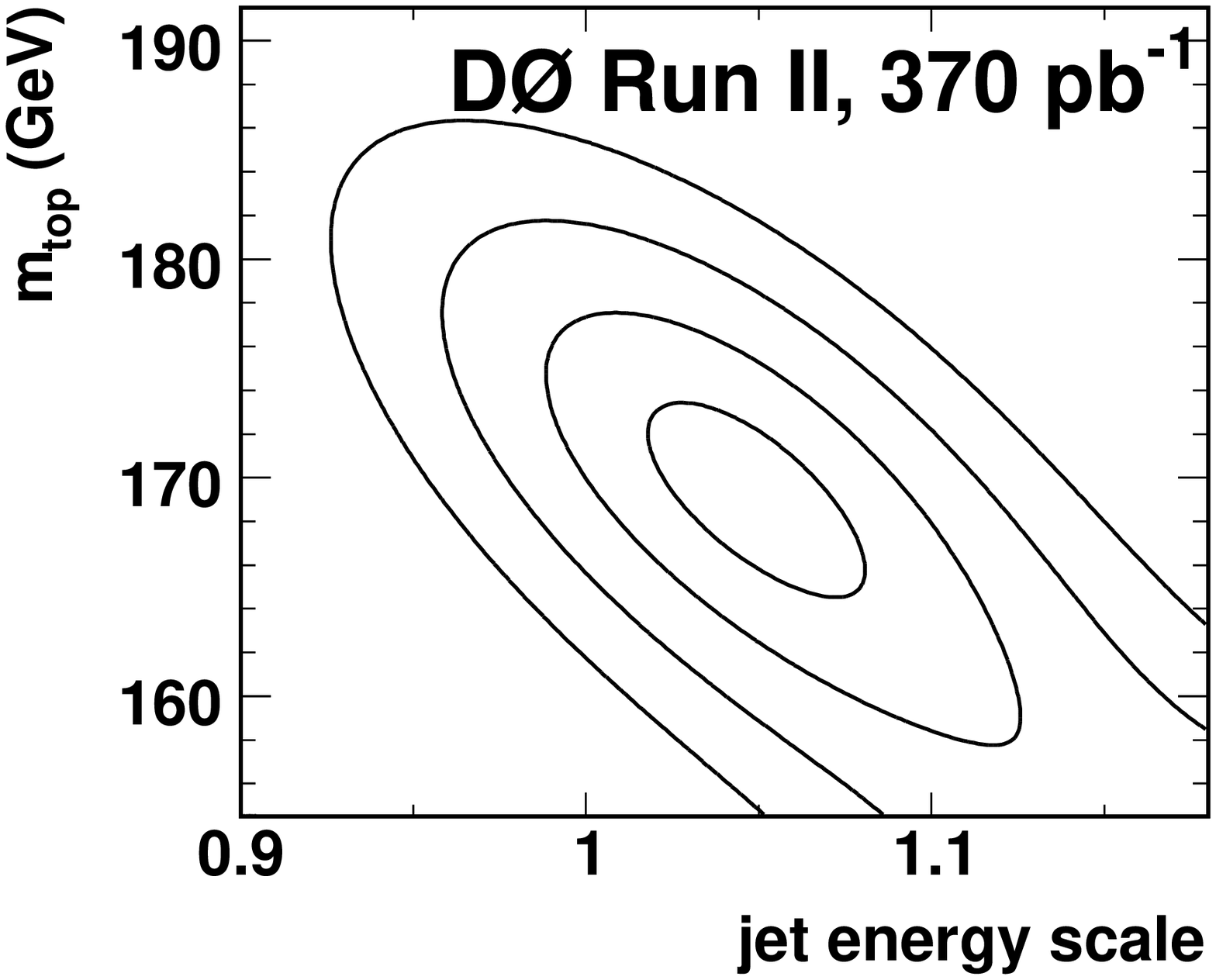}
\caption{Application of the topological Matrix Element method to the data.
Fit of a two-dimensional fourth-order polynomial to the 
$-\ln L$ values as a function of both \mtop and $JES$.  Shown are
the contours corresponding to $\Delta\ln L = 0.5$, $2.0$, $4.5$,
and $8.0$ relative to the minimum.}
\label{fig:lhood2d_topo}
\end{center}
\end{figure}

To show the likelihood as a function of both \mtop and $JES$
simultaneously, the $-\ln L$ values have been fitted with a
two-dimensional fourth-degree polynomial with its minimum fixed to the 
measurements mentioned above.
The resulting contours corresponding to $\Delta\ln L = 0.5$, $2.0$, $4.5$,
and $8.0$ relative to the minimum are shown in
Fig.~\ref{fig:lhood2d_topo}.
Note that the statistical measurement uncertainties quoted on $\mtop$ and 
$JES$ are obtained from the one-dimensional projections as discussed above; 
Fig.~\ref{fig:lhood2d_topo} therefore serves only illustrative purposes.
Because of non-Gaussian tails, the projections of the $\Delta\ln L = 0.5$
contour shown in Fig.~\ref{fig:lhood2d_topo} onto the $\mtop$ and
$JES$ axes do not exactly correspond to these quoted statistical
uncertainties.

\section{Top Quark Mass Measurement Using \boldmath{$b$} Jet Identification}
\label{sec:MEfitbtag}

\subsection{Top Quark Mass Fit}
\label{sec:MEfitbtag-fit}
The incorporation of $b$-tagging information introduces two
significant modifications to the Matrix Element mass fitting
technique.
First, $b$-tagging information is used to determine the relative
weights $w_i$ of the different jet-parton assignments in the signal
probability calculation.
The $w_i$ are parameterized as a function of the jet transverse energy
$\et$ and pseudorapidity $\eta$ and the assumed flavor $\alpha_k$ of
the parton corresponding to the jet, as described in
Section~\ref{sec:tf-btag}.
The signal probability is then computed according to 
Equation~(\ref{eq:MEpsgn}).
The background probability is identical to that used in the topological
analysis according to Equation~(\ref{eq:MEpbkg}).

The second modification is to classify events into three categories
according to the number of $b$-tagged jets.
Each of these categories will have different signal fractions and 
background compositions due to the
relative suppression of \wjets events with dominantly light quark 
and gluon jets.
The event categories are exclusive and correspond to i) no $b$-tagged
jet, ii) exactly one tagged jet, and iii) two or more tagged jets.

When the analysis is separately performed in each $\Ntag$ category, the 
signal fractions $\ftopb$ are determined independently for each
category, and $\pevtb$ is calculated as
\eqMEpevtbtag
To combine the three categories into one analysis the three
purities $\ftopb$ have to be related to one inclusive 
signal purity $\ftop$.
The purity of the \Ntag sample is given by
\begin{equation}
\label{b1}
    \ftopb
  = 
    \frac{N^{\Ntag}_{\rm sig}}{N^{\Ntag}_{\rm sig}+N^{\Ntag}_{\rm bkg}}
  \ .
\end{equation}
The numbers of signal and background events after $b$ tagging,
$N^{\Ntag}_{\rm sig}$ and $N^{\Ntag}_{\rm bkg}$, can be related to the 
corresponding numbers for the inclusive sample, $N_{\rm sig}$ and 
$N_{\rm bkg}$, by
\begin{equation}
N^{\Ntag}_{\rm sig} = N_{\rm sig} \; \epsilon^{\Ntag}_{\rm sig}
\;\;\; \text{and} \;\;\;\; 
N^{\Ntag}_{\rm bkg} = N_{\rm bkg} \; \epsilon^{\Ntag}_{\rm bkg}
\ ,
\label{eq:Ntaggs}
\end{equation}
where $\epsilon^{\Ntag}_{\rm sig}$ and $\epsilon^{\Ntag}_{\rm bkg}$
are the average tagging efficiencies for signal and background,
respectively.
They are defined as 
\eqDATAepssgnbkg
with relative fractions $f_\Phi$ of the different flavor
contributions $\Phi$ to the \wjets background as given in
Table~\ref{tab:HFfract}.
The jets in selected QCD multijet background events have 
kinematic characteristics similar to those of jets in selected \wjets 
background events.
Concerning the event $b$-tagging probabilities, we 
therefore do not distinguish between QCD multijet and \wjets
backgrounds.
The difference between multijet and \wjets kinematics is treated as a
systematic uncertainty. 
The relation between $\ftopb$ and the inclusive signal purity
$\ftop=N_{\rm sig}/(N_{\rm sig}+N_{\rm bkg})$ is then
\begin{equation}
    \ftopb 
  = 
    \frac{\ftop r^{\Ntag}}{\ftop (r^{\Ntag}-1) + 1}
  \ ,
\label{b3}
\end{equation}
where
\begin{equation}
r^{\Ntag} = \frac{\epsilon^{\Ntag}_{\rm sig}}{\epsilon^{\Ntag}_{\rm bkg}} \ .
\label{b4}
\end{equation}
Equation~(\ref{b3}) needs to be corrected for the fact that the fraction
of \ttbar events that are jet-parton matched is different 
in each tag-multiplicity sample.
Thus, a correction factor $c^{\Ntag}$ defined as the ratio of fitted
\ttbar fraction over expected \ttbar fraction is introduced as an 
intercalibration of the \ftopb values. 
The top fraction for a given tag-multiplicity sample is then
\begin{equation}
    f^{\Ntag}_{\rm top} 
  = 
    c^{\Ntag} \frac{f_{\rm top} r^{\Ntag}}{f_{\rm top} (r^{\Ntag}-1) + 1}
  \ .
  \label{b3-corrected.eqn}
\end{equation}
The correction factors $c^{\Ntag}$ are different for \ejets and
\mujets events, and are given in Table~\ref{table:cn}.
\begin{table}
\begin{center}
\begin{tabular}{c@{\ \ \ \ }c@{\ \ \ }c@{\ \ \ }c}
\hline
\hline
                        & \multicolumn{3}{c}{Subsample} \\
Channel                 & 0-tag         & 1-tag   & $\ge2$-tag  \\ 
\hline
\ejets                  & 0.68  & 0.86  & 0.94  \\ 
\mujets                 & 0.77  & 0.84  & 0.90  \\ 
\hline
\hline
\end{tabular}
\caption{\label{table:cn} Signal purity correction factors $c^{\Ntag}$.}
\end{center}
\end{table}

Equation~(\ref{b3-corrected.eqn}) defines the dependence of the signal purity on tagging 
multiplicity, as a function of the ratio of event-tagging efficiencies and 
signal purity before tagging.
In order to extract the top quark mass from the total 
sample of selected events, the likelihoods in the 
individual event categories are then combined as
\eqMElhoodfncbtag
where $N^{\Ntag}$ is the number of events in each of the three tag-categories.
As in the topological analysis, we determine the value
$\ftopbest\left(\mtop,\, JES\right)$ that maximizes the likelihood
$L$ in Equation~(\ref{eq:MElhood-fncb}) for each pair of assumed values of 
\mtop and $JES$.
The top quark mass and jet energy scale are then obtained by
maximizing
\eqMElhoodnoftopbtag
as described in Section~\ref{sec:MEfit-fit}
(for the 0-tag sample, the fit range is restricted to $\Delta\ln L<1$).

\subsection{Calibration of the Method}
\label{sec:MEcalibrationb}

The calibration is obtained following a similar procedure as described 
in Section~\ref{sec:MEcalibration}. 
The number of events in each tag-multiplicity class
is calculated by multipliying the expected number 
of selected events (before tagging) by the average event-tagging
probability in each process:
\begin{eqnarray}
  \nonumber
    N^{\Ntag}_{\ttbar} 
  & = & 
    N^{\rm sel}_{\ttbar} \epsilon^{\Ntag}_{\ttbar}
  \ {\rm and}
\\
  \label{eq:ETests12}
    N^{\Ntag}_{\rm bkg} 
  & = &
    N^{\rm sel}_{\rm bkg} \epsilon^{\Ntag}_{\wjets}
  \ .
\end{eqnarray}
The \wjets background is classified into two categories according
to the differences in event kinematics: \wlj (\W + 4 jets without
heavy flavor) and \whj (\W + 4 jets including heavy flavor). 
Thus, the background composition for the ensemble tests after tagging
is given by
\begin{eqnarray}
  \nonumber
    N^{\Ntag}_{\wlj} 
  & = &
    N^{\rm sel}_{\rm bkg} \ f_{\wlj} \ \epsilon^{\Ntag}_{\wlj}
  \ {\rm and}
  \\
    N^{\Ntag}_{\whj} 
  & = &
    N^{\rm sel}_{\rm bkg} 
    \sum_{\Phi\ne\wlj} f_{\Phi} \ \epsilon^{\Ntag}_{\Phi}
  \ ,
\end{eqnarray}
where $f_{Wjjjj}$ is the fraction of $W+{\rm light\ jets}$,
$\Phi$ denotes one of the five $W+$heavy flavor subprocesses
(see Table~\ref{tab:HFfract}), and 
$f_{\Phi}$ is the corresponding fraction of these subprocesses.
Table~\ref{tableC} shows the \wjets composition in the 0, 1, and
$\ge2$ tag samples used in the ensemble tests.
The average number of events in each tag-category and for each sample
are fluctuated according to a Poisson distribution.
For a given tag-category $\Ntag$, the decision of which jets are tagged is
made by randomly selecting $\Ntag$ jets as tagged jets, taking into 
account the $\et$ and $\eta$ dependence of the tagging efficiencies.
\begin{table}
\begin{center}
\begin{tabular}{cccc}
\hline
\hline
                        & \multicolumn{3}{c}{Subsample} \\
Contribution            & 0-tag         & 1-tag   & $\ge2$-tag  \\ 
\hline
\wlj                    & 90.9\%        & 19.4\%  &  \enspace\enspace0.0\%\\ 
\whj                    & \enspace9.1\% & 80.6\%  &  100.0\%\\ 
\hline
\hline
\end{tabular}
\caption{Background composition used in the ensemble tests for the 
$b$-tagging analysis.  The contribution from \wjets events without
heavy flavor is given in the first line, the contribution from events
with heavy flavor jets in the second line.}
\label{tableC}
\end{center}
\end{table}

\begin{figure}
\begin{center}
\includegraphics[width=\figMEcalibwidth\textwidth]{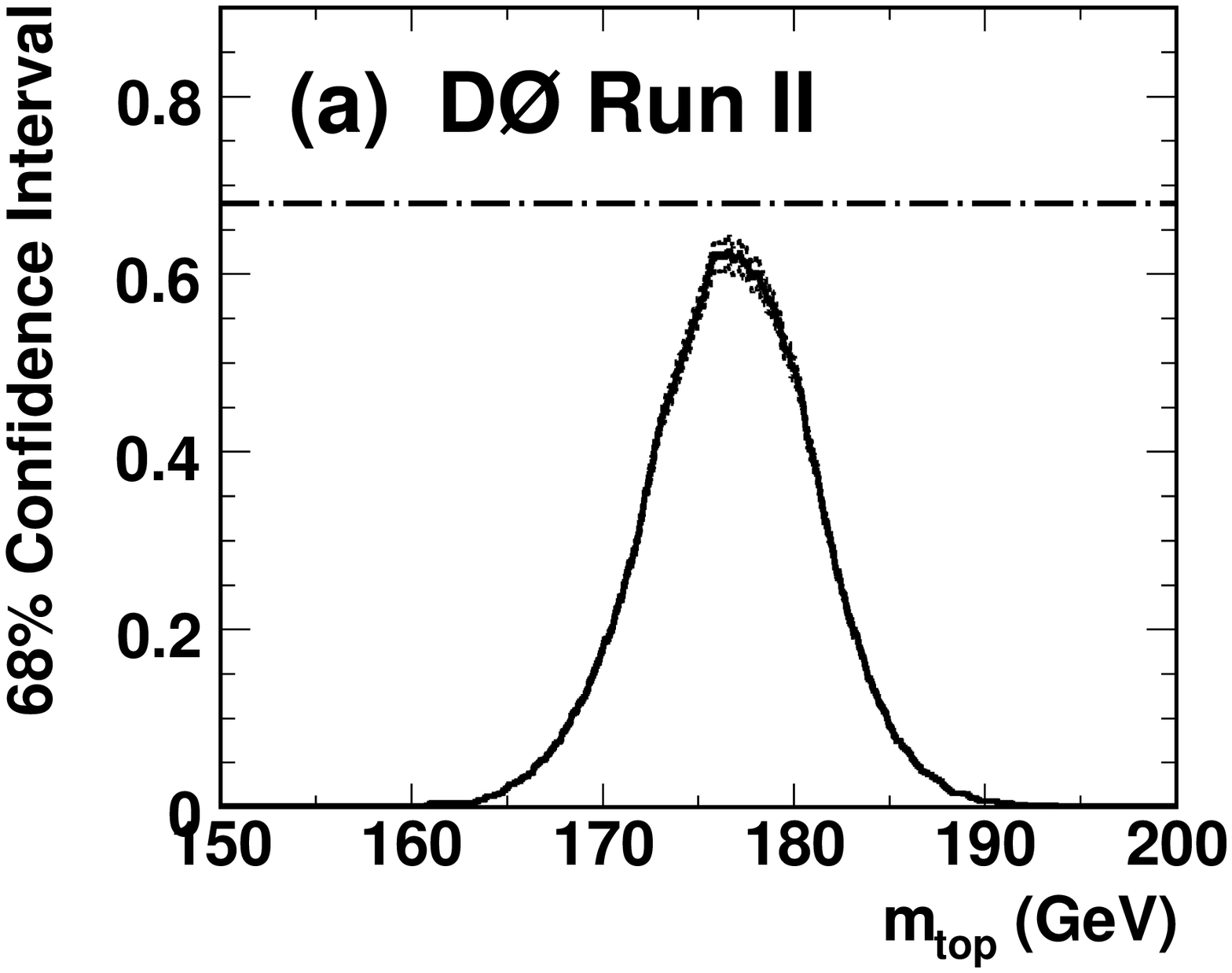}
\hspace{-0.03\textwidth}
\includegraphics[width=\figMEcalibwidth\textwidth]{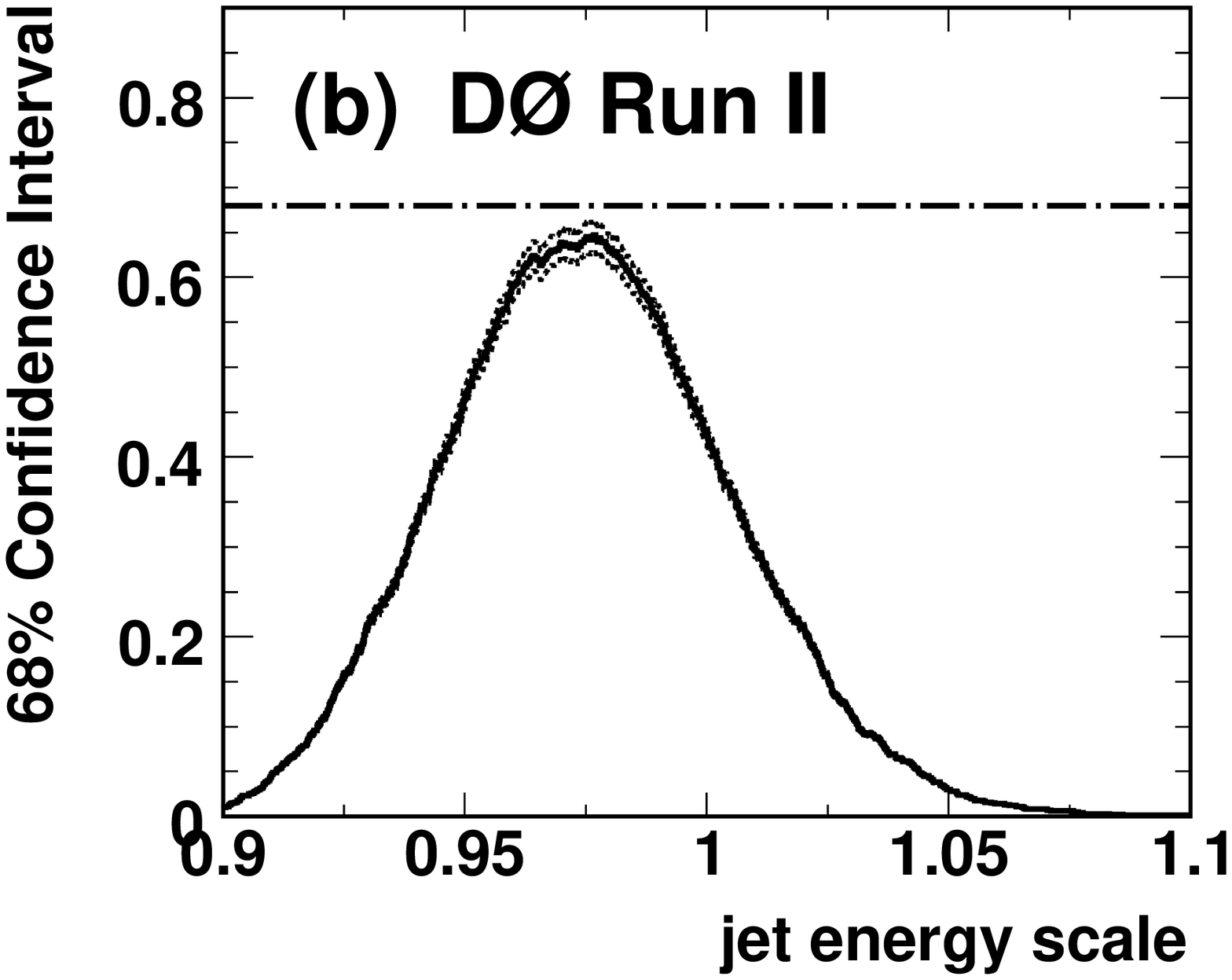}
\caption{Calibration of the Matrix Element mass fitting 
procedure for the $b$-tagging analysis, using ensembles with a
top quark mass of 175\,\GeV and $JES=1.0$.  The 68\% confidence interval
distributions for (a) the measured top quark mass and (b) the jet energy
scale is given by the solid, the upper and lower error
bands by the dashed histograms.  A value of 0.68 as indicated by
the dash-dotted line would mean that the corresponding \mtop ($JES$) value 
is included in the fitted $68\%$ confidence interval in 68\% of the
ensembles.}
\label{fig:CL-btag}
\end{center}
\end{figure}

\begin{figure}
\begin{center}
\includegraphics[width=\figMEcalibwidth\textwidth]{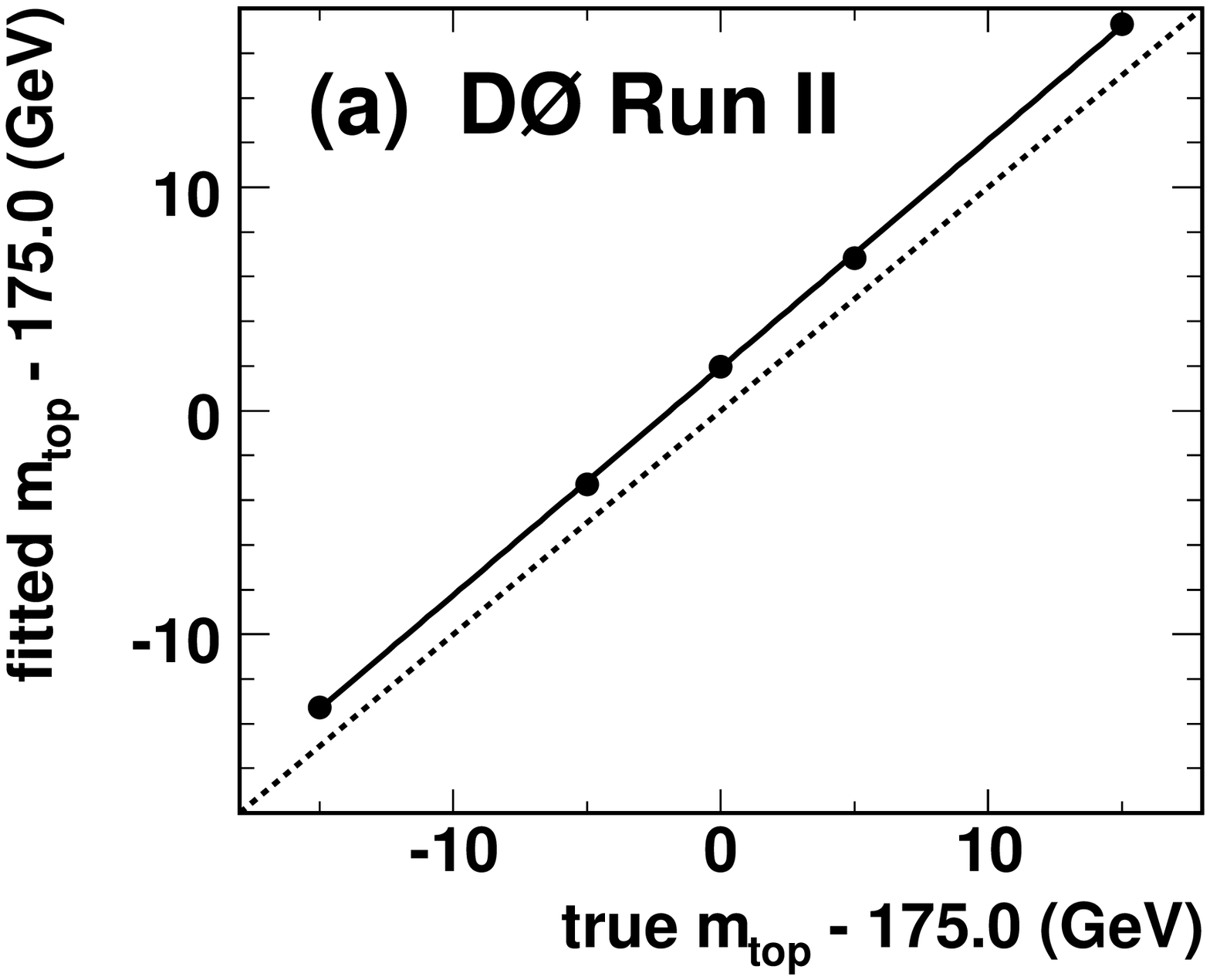}
\hspace{-0.03\textwidth}
\includegraphics[width=\figMEcalibwidth\textwidth]{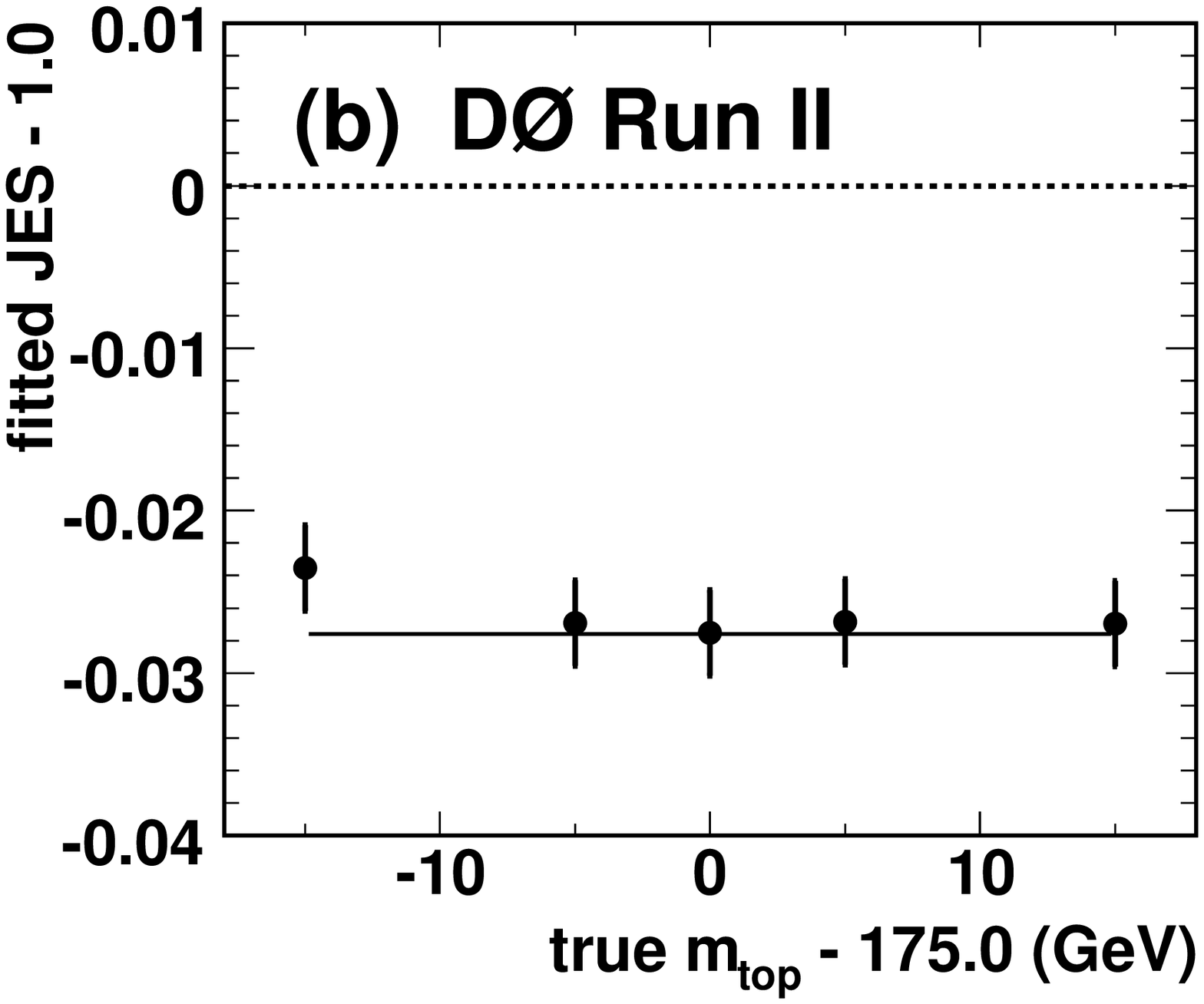}
\includegraphics[width=\figMEcalibwidth\textwidth]{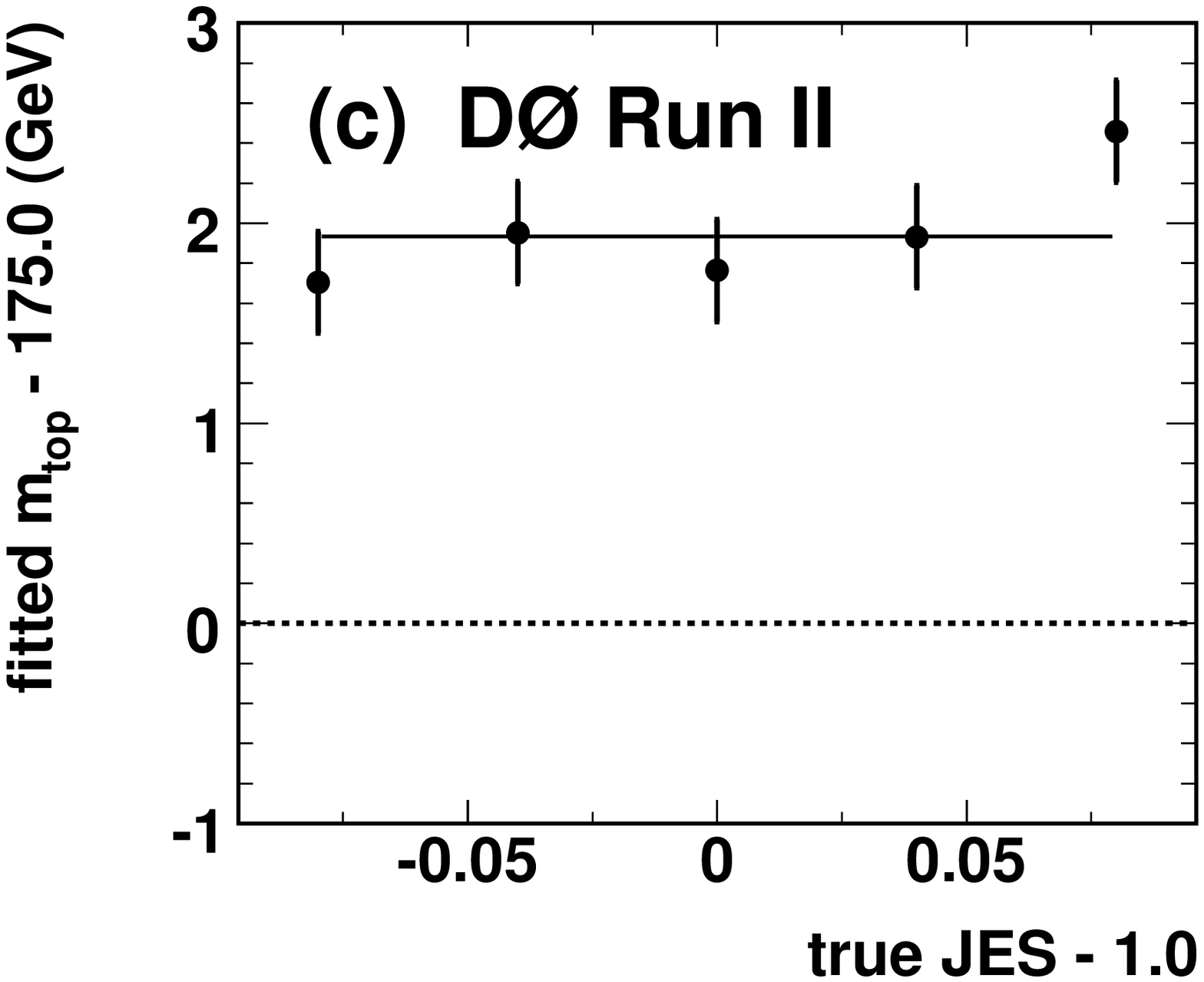}
\hspace{-0.03\textwidth}
\includegraphics[width=\figMEcalibwidth\textwidth]{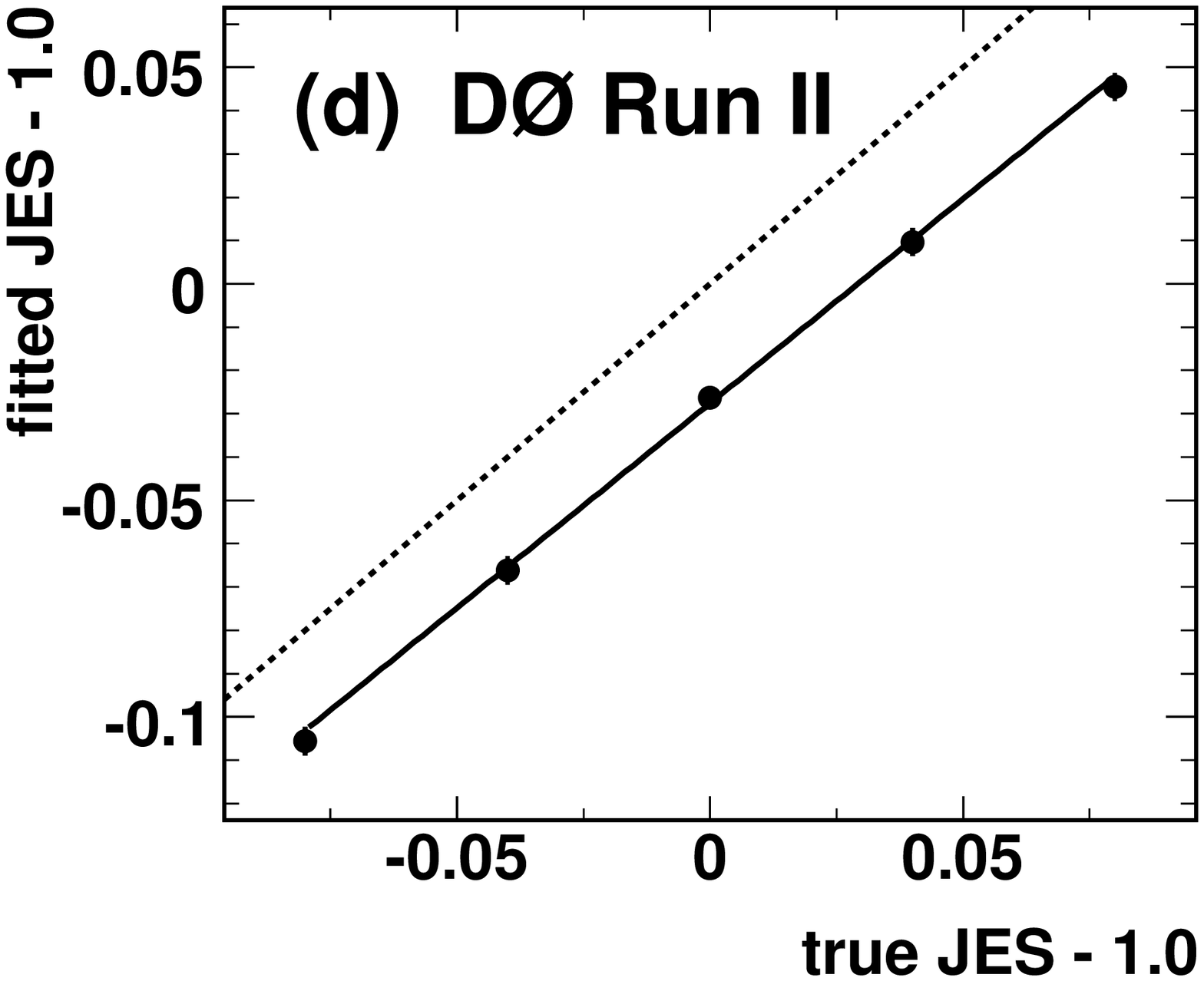}
\caption{Calibration of the Matrix Element mass fitting 
procedure for the $b$-tagging analysis.  The upper plots show the 
reconstructed top mass (a) and the measured jet energy scale
(b) as a function of the input top mass.
The two lower plots show the reconstructed top mass (c) and the 
measured jet energy scale (d) as a function of the
input jet energy scale.
The solid lines show the results of linear fits to the points,
which are used to calibrate the measurement technique.
The dashed lines would be obtained for equal 
fitted and true values of \mtop and $JES$.}
\label{fig:MEcalib-btag}
\end{center}
\end{figure}

\begin{figure}
\begin{center}
\includegraphics[width=\figMEcalibwidth\textwidth]{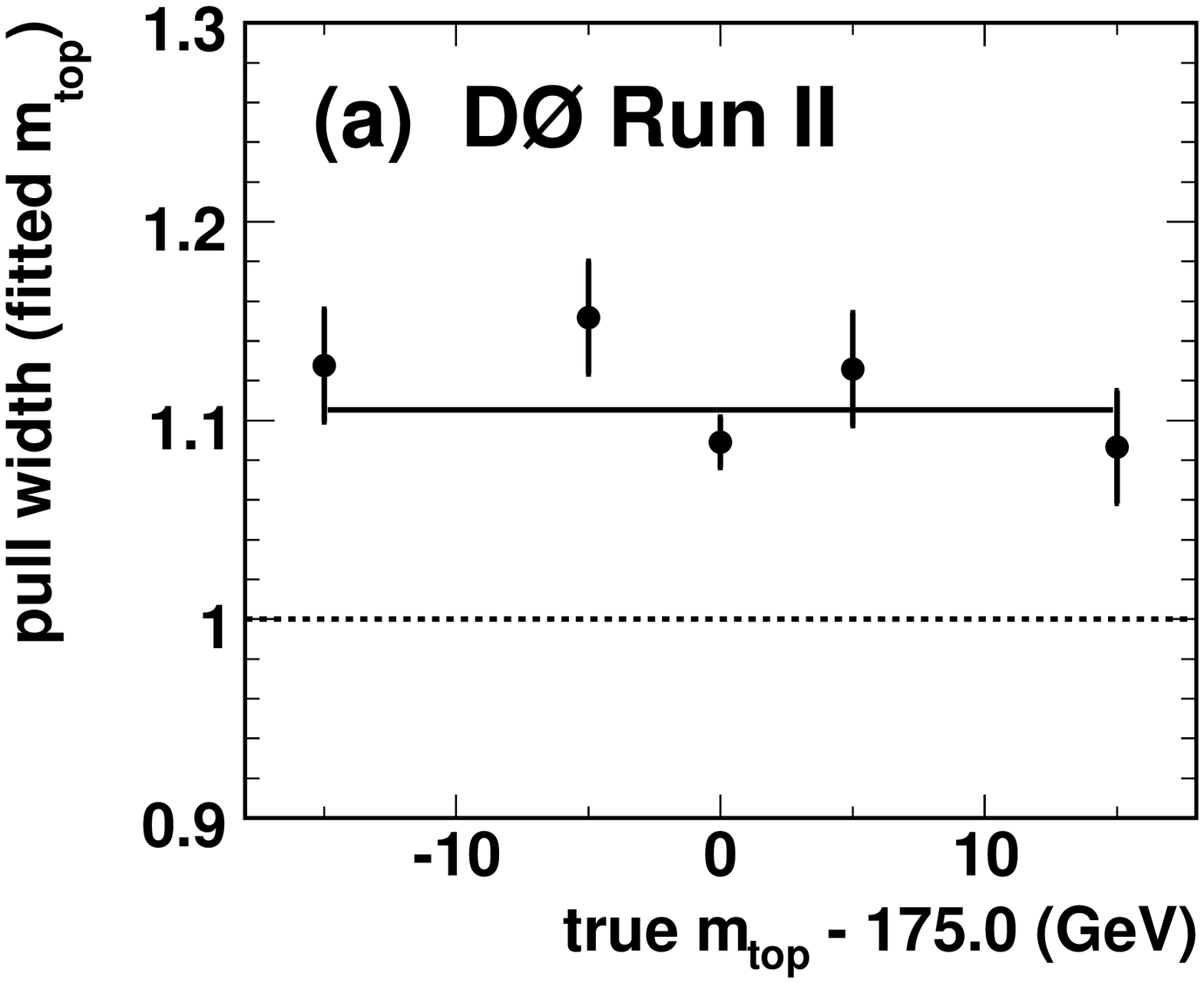}
\hspace{-0.03\textwidth}
\includegraphics[width=\figMEcalibwidth\textwidth]{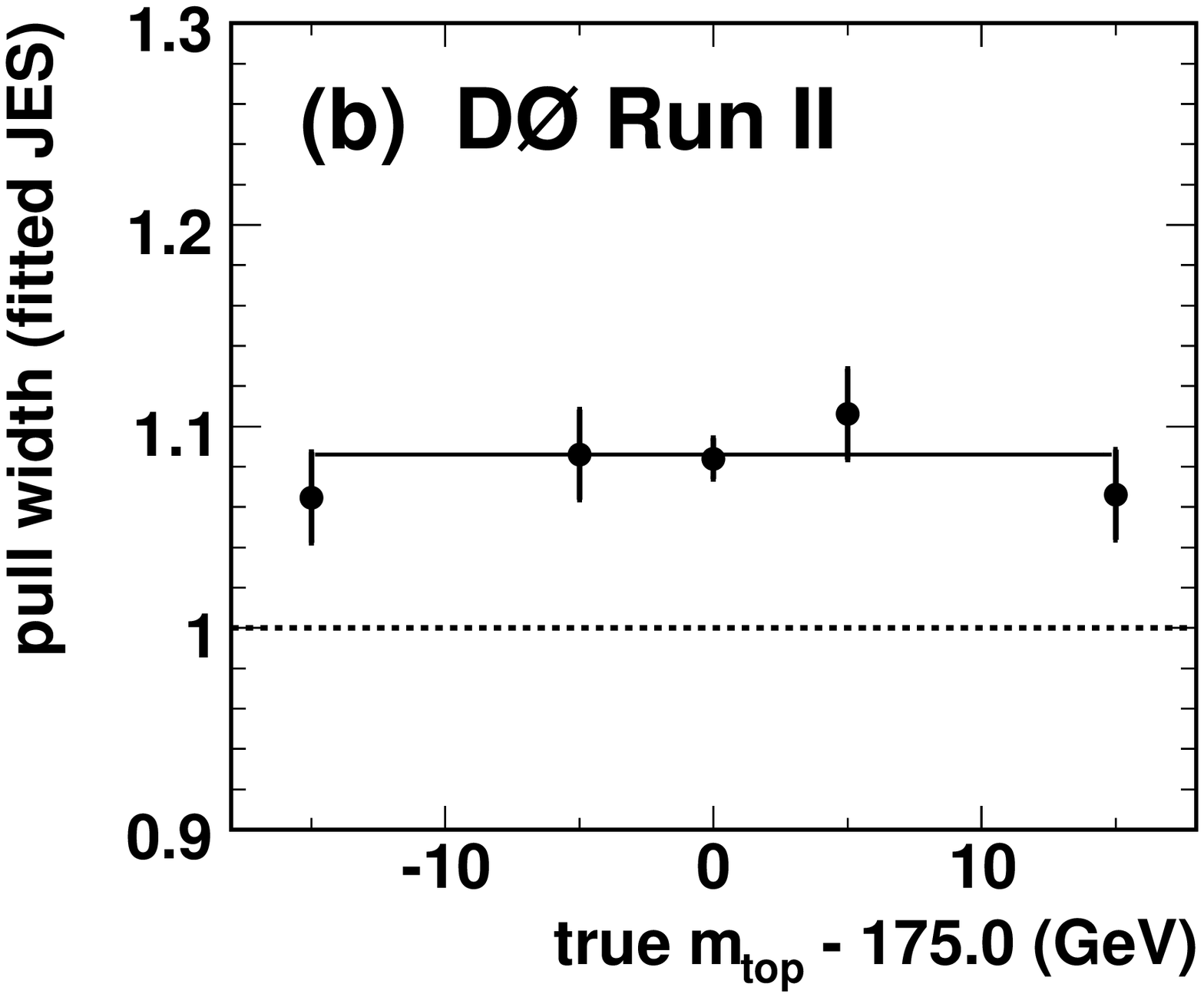}
\includegraphics[width=\figMEcalibwidth\textwidth]{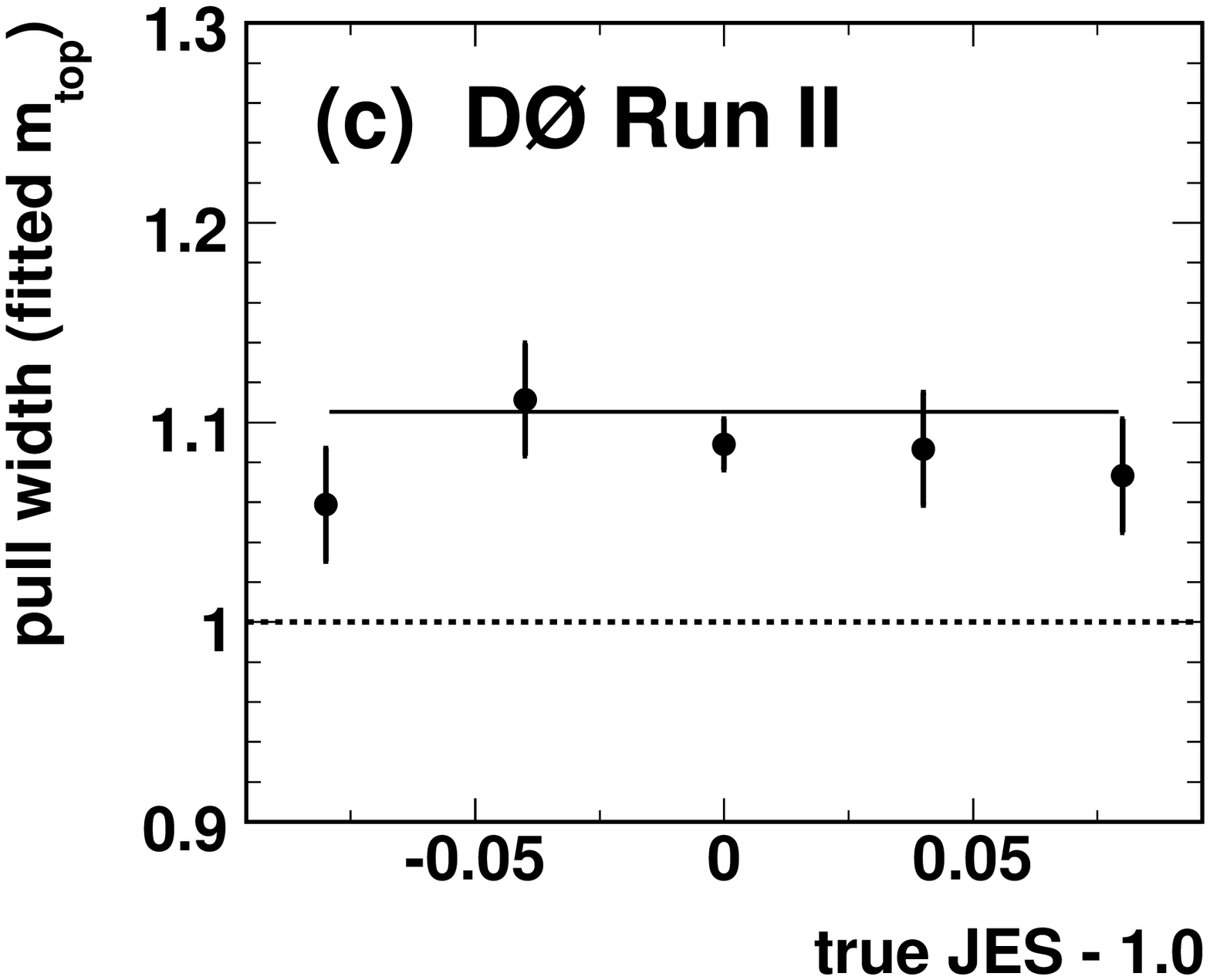}
\hspace{-0.03\textwidth}
\includegraphics[width=\figMEcalibwidth\textwidth]{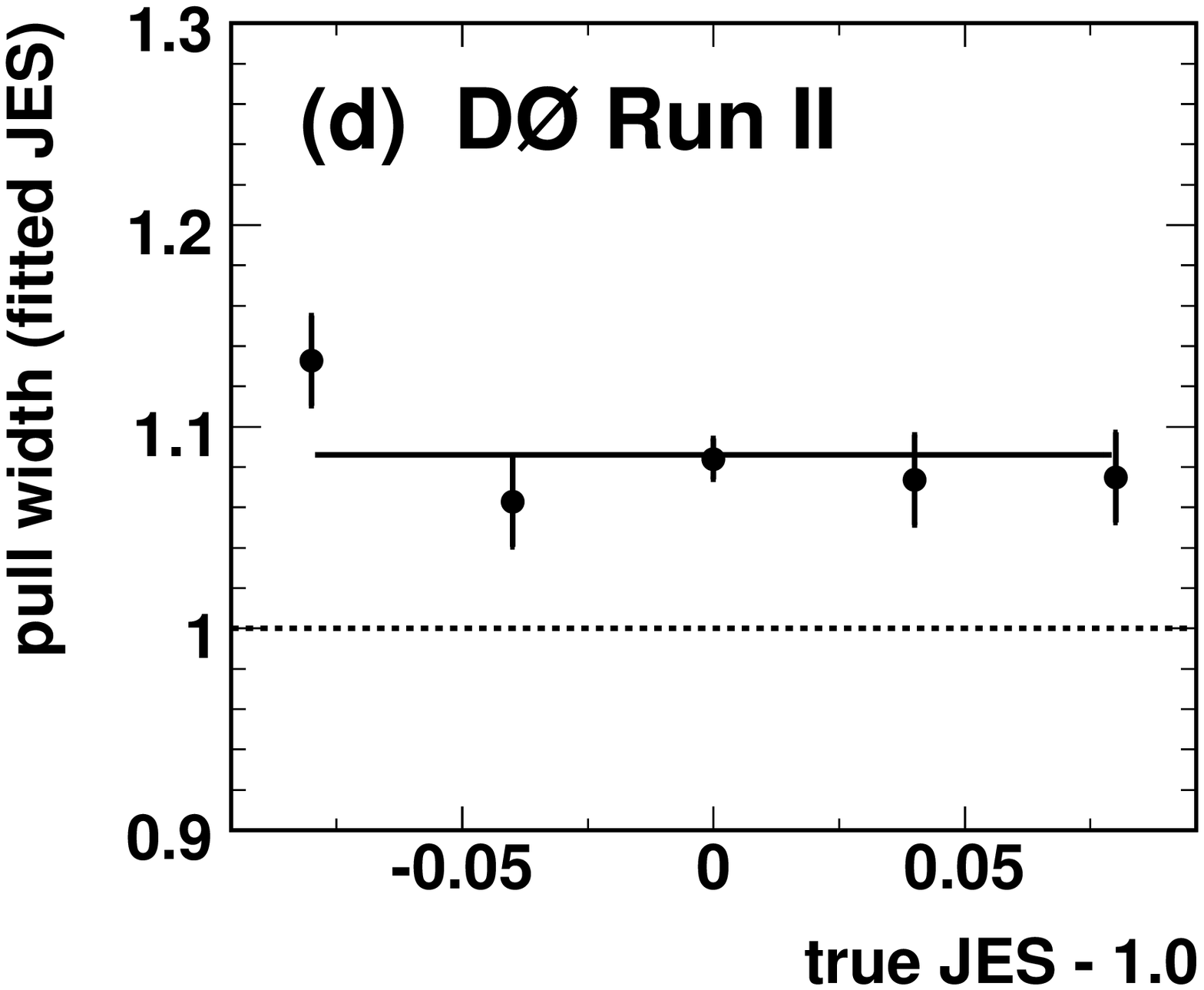}
\caption{Calibration of the Matrix Element mass fitting 
procedure for the  $b$-tagging analysis.  The upper plots show the 
widths of the pull distributions for the top mass
(a) and jet energy scale (b)
as a function of the input top mass.
The two lower plots show the widths of the pull distributions for the top mass
(c) and jet energy scale (d) as a function of the
input jet energy scale.
The solid lines show the mean pull width, while
the dashed lines indicate a pull width of 1.0.}
\label{fig:MEpullcalib-btag}
\end{center}
\end{figure}

In Fig.~\ref{fig:CL-btag}, $68\%$ confidence interval distributions 
are shown for ensembles with $\mtop=175\,\GeV$ and $JES=1.0$.
The calibration results for the combined fit 
are shown in
Figs.~\ref{fig:MEcalib-btag} and~\ref{fig:MEpullcalib-btag}.
The final fit results are corrected for the biases and for the 
deviation of the pull width from 1.0 given in Table~\ref{tab:calib-btag}.

\begin{table}
\begin{center}
\begin{tabular}{c@{\quad}r@{\quad}c@{\quad}c}
\hline
\hline
  &
    \multicolumn{1}{c}{offset $o$}
  &
    slope $s$
  &
    pull width $w$
\\
\hline
    \mtop
  &
    \offsetmtopbtag GeV
  &
    \slopemtopbtag
  &
    \pullwidthmtopbtag
\\
    $JES$
  &
    \offsetjesbtag \phantom{GeV}
  &
    \slopejesbtag
  &
    \pullwidthjesbtag
\\
\hline
\hline
\end{tabular}
\caption{Calibration of the Matrix Element mass fitting 
procedure for the $b$-tagging analysis.  The offsets are 
quoted for a true top quark mass of \offsetatmtopbtag~GeV
and a true jet energy scale of \offsetatjesbtag, respectively.
Only statistical uncertainties are quoted in this table.}
\label{tab:calib-btag}
\end{center}
\end{table}

\subsection{Result}
\label{sec:MEresultb}

\begin{figure}
\begin{center}
\includegraphics[width=\figbtagwidthsmall\textwidth]{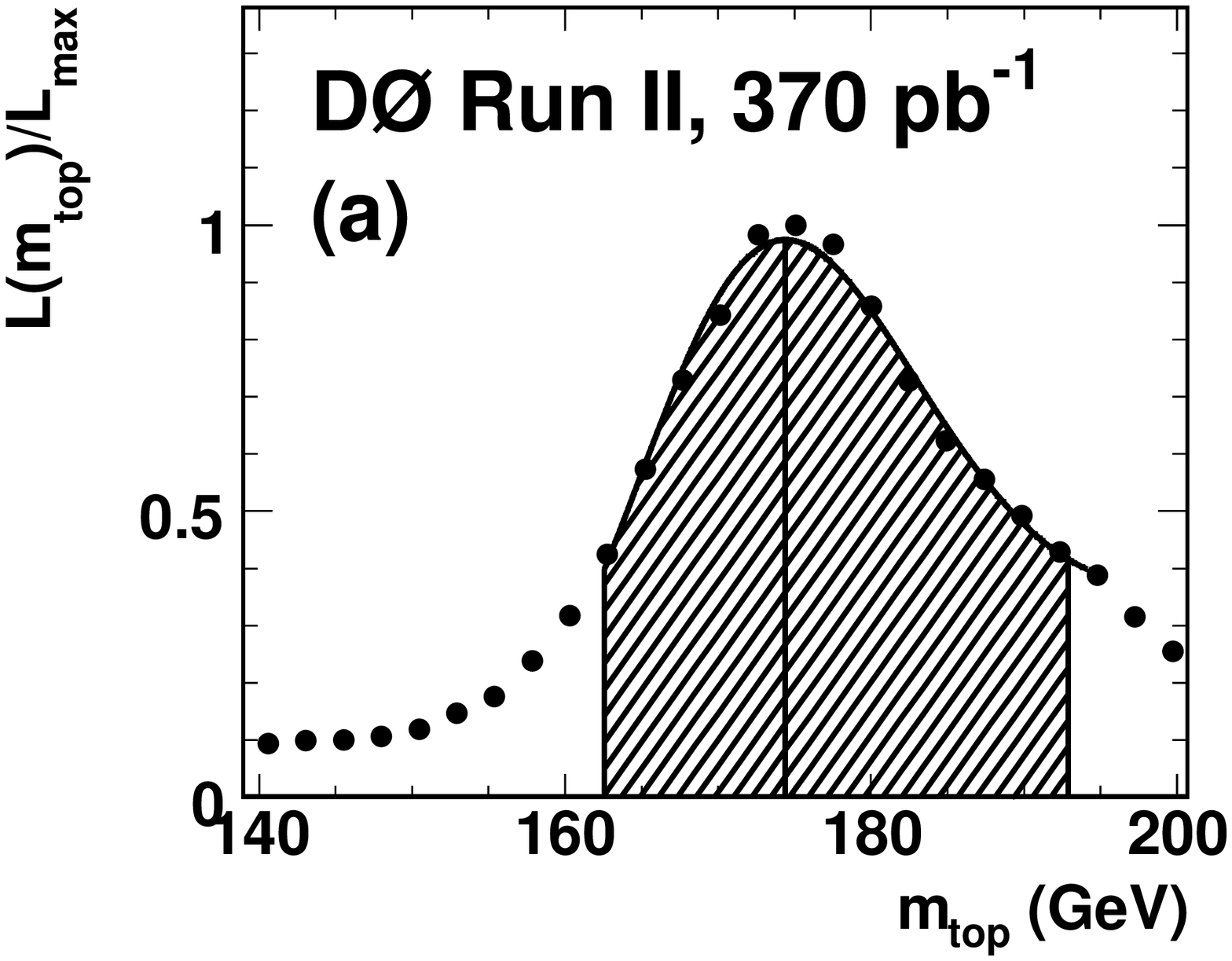}
\hspace{-0.03\textwidth}
\includegraphics[width=\figbtagwidthsmall\textwidth]{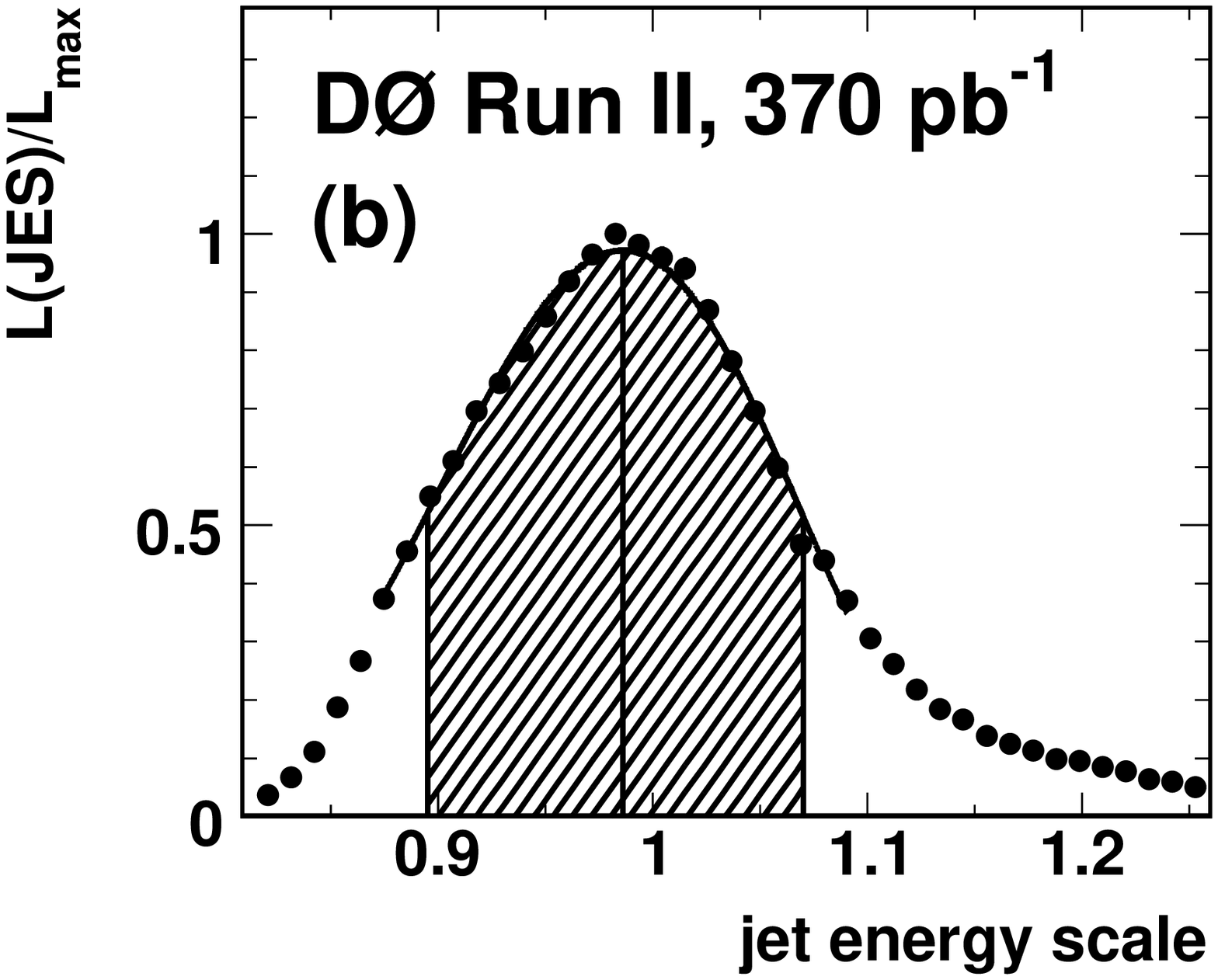}
\includegraphics[width=\figbtagwidthsmall\textwidth]{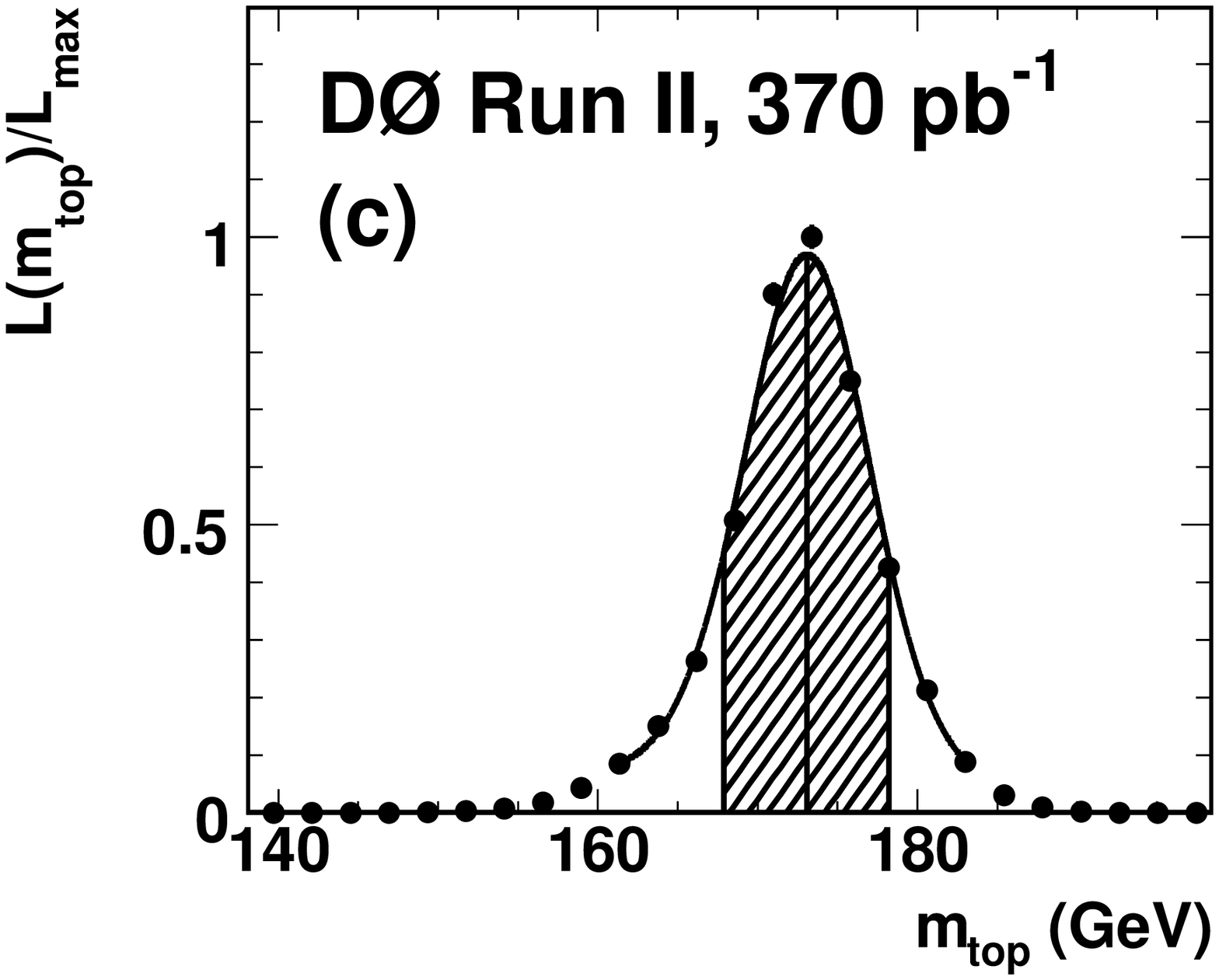}
\hspace{-0.03\textwidth}
\includegraphics[width=\figbtagwidthsmall\textwidth]{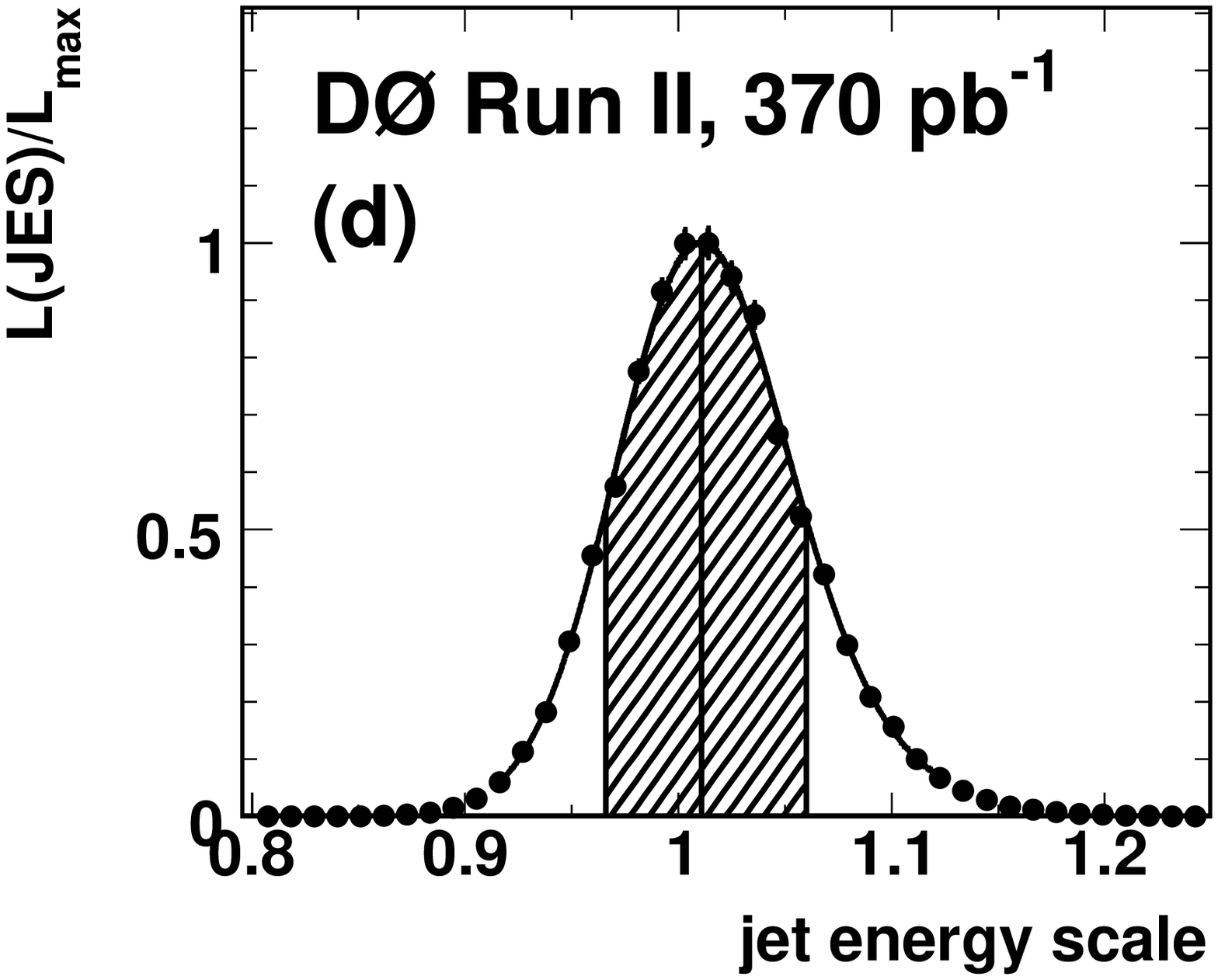}
\includegraphics[width=\figbtagwidthsmall\textwidth]{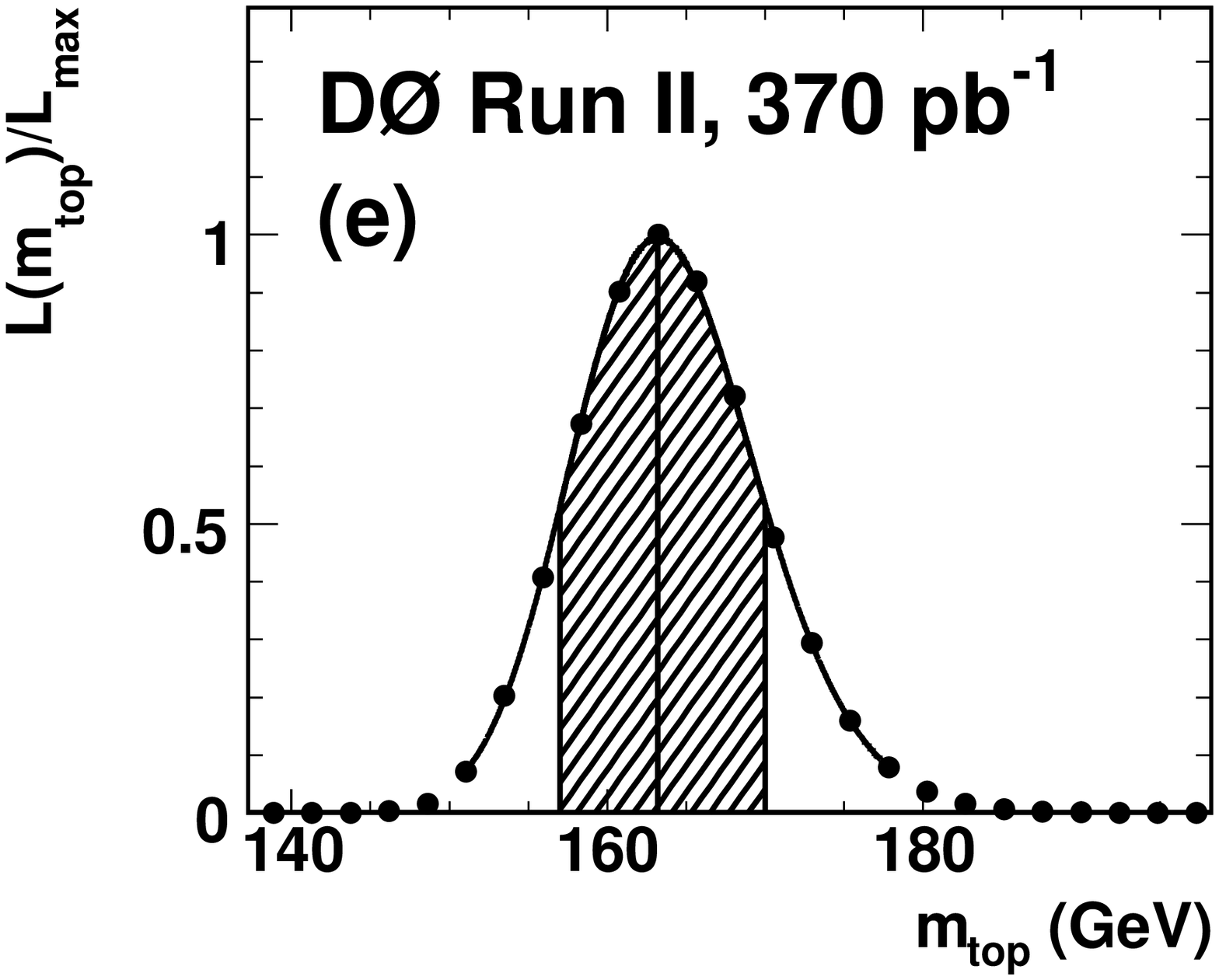}
\hspace{-0.03\textwidth}
\includegraphics[width=\figbtagwidthsmall\textwidth]{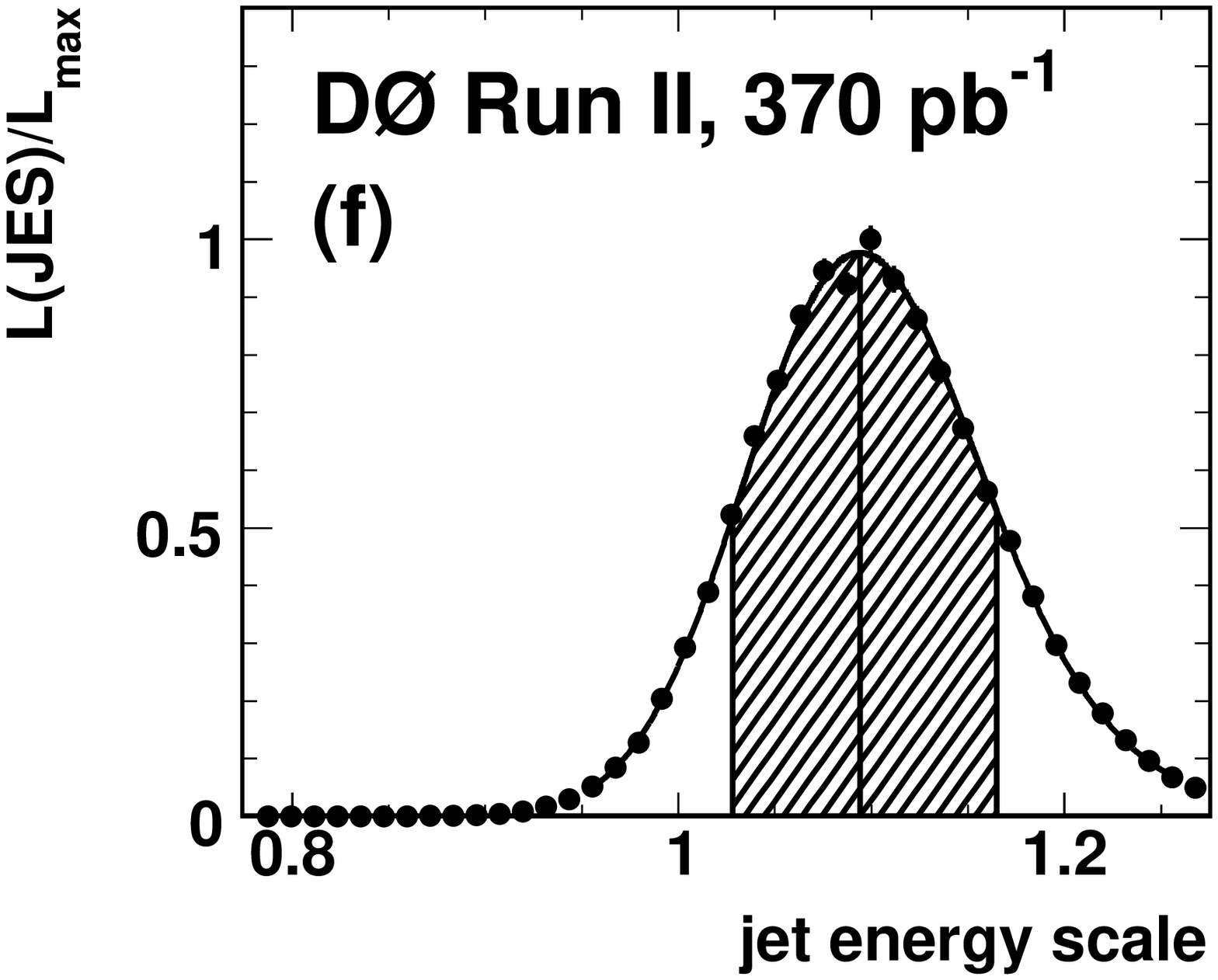}
\caption{Application of the Matrix Element $b$-tagging method to the data.
The fitted \mtop and $JES$ likelihoods for each of the 3
tag-categories: 0-tag ((a) and (b)), 1-tag ((c) and (d)), and
$\ge2$-tag ((e) and (f)).
The 68\% confidence-level interval around the most likely value is shown by the hatched
region under the fitted curve.}
\label{fig:MEresult-Lfitb}
\end{center}
\end{figure}

The Matrix Element $b$-tagging method is applied to the same event sample as in
Section~\ref{sec:MEresult} with a calibration according to the
results from Section~\ref{sec:MEcalibrationb}.
The probability is shown in Fig.~\ref{fig:MEresult-Lfitb} as a function of
\mtop and $JES$ hypothesis for each of the three tag categories.  
The central values and the 68\% confidence level intervals 
are indicated in the figures.
The individual results for the top quark mass are
\begin{eqnarray}
   \mtop^{\ljets}({\rm 0{\text -}tag}) & = & \resultbstatzerotag\,\GeVcc \nonumber \\
   \mtop^{\ljets}({\rm 1{\text -}tag}) & = & \resultbstatonetag \,\GeVcc \nonumber \\
   \mtop^{\ljets}({\rm 2{\text -}tag}) & = & \resultbstattwotag \,\GeVcc 
\end{eqnarray}
in the 0-tag, 1-tag, and $\ge2$-tag categories.
The corresponding results for the jet energy scale are
$JES(0$-${\rm tag}) = \resultbjesstatzerotag$,
$JES(1$-${\rm tag}) = \resultbjesstatonetag$, and
$JES(2$-${\rm tag}) = \resultbjesstattwotag$,
respectively.

\begin{figure}
\begin{center}
\includegraphics[width=\figbtagwidth\textwidth]{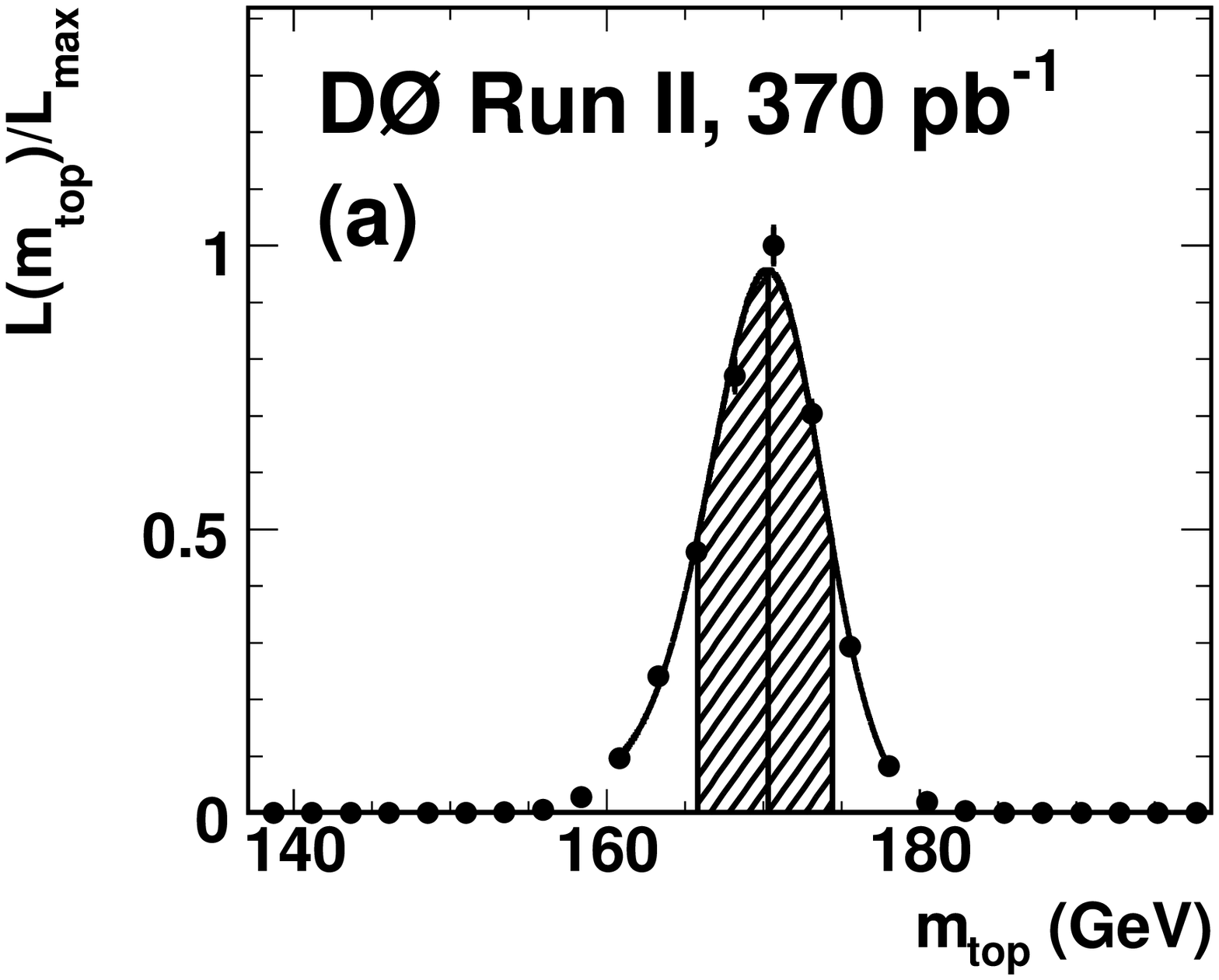}
\includegraphics[width=\figbtagwidth\textwidth]{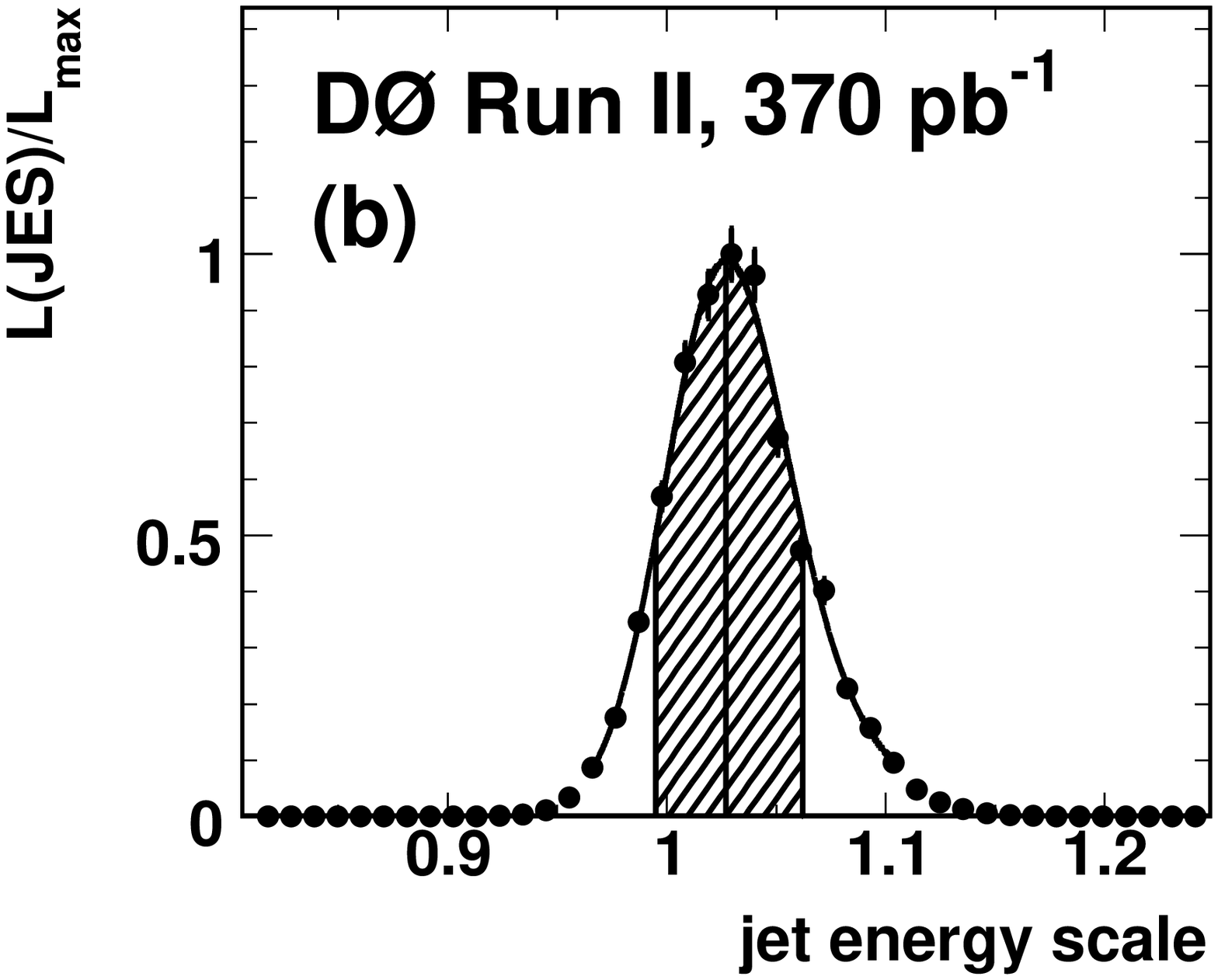}
\caption{Application of the Matrix Element $b$-tagging method to the data.
The final results of the fitted \mtop (a) and $JES$ (b) 
likelihoods for the combined event sample are shown.
The 68\% confidence-level interval around the most likely value is
indicated by the hatched
region under the fitted curve.}
\label{fig:MEresult-btag-final}
\end{center}
\end{figure}

\begin{figure}                                                                  
\begin{center}                                                                  
\includegraphics[width=\figbtagwidth\textwidth]{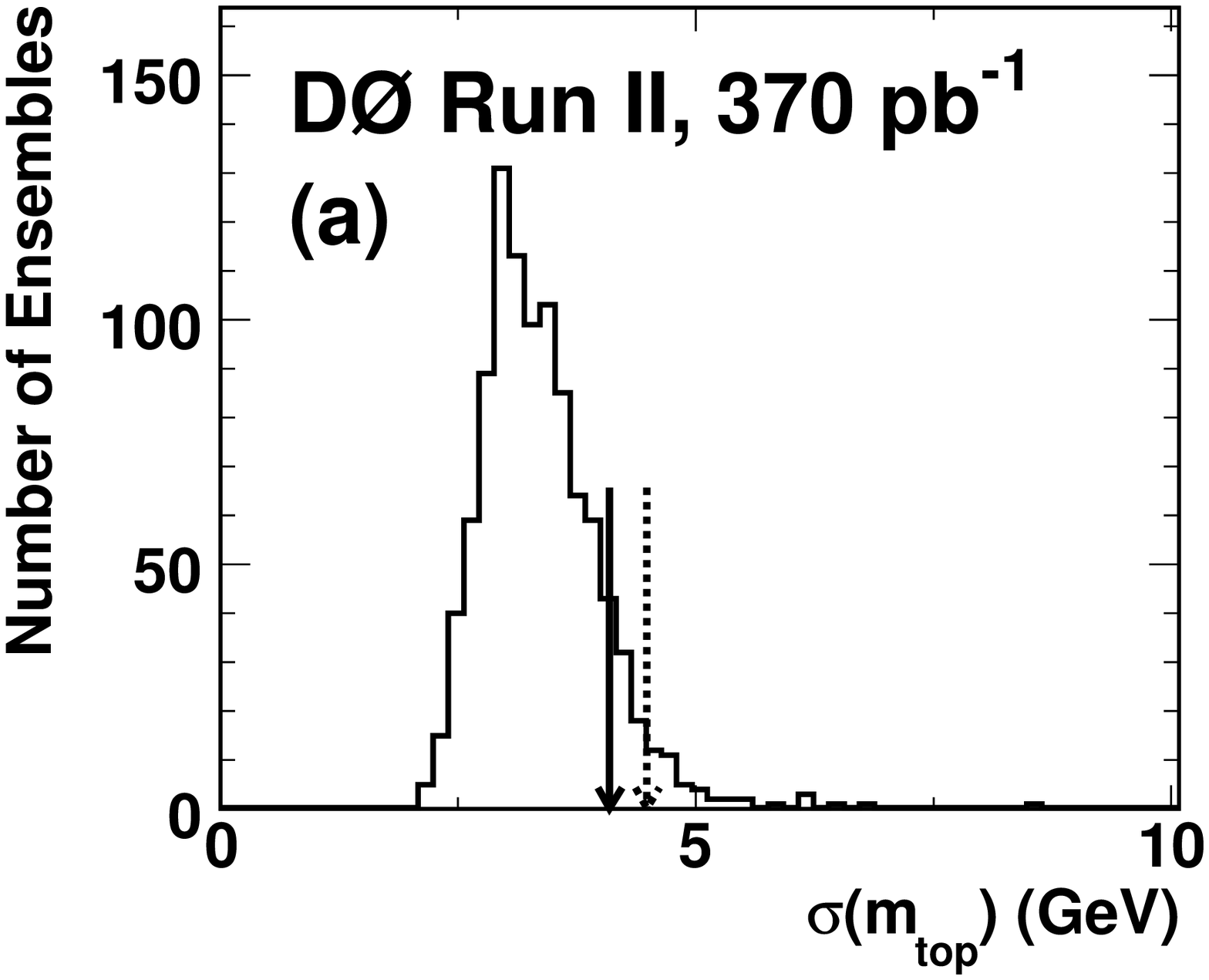}
\includegraphics[width=\figbtagwidth\textwidth]{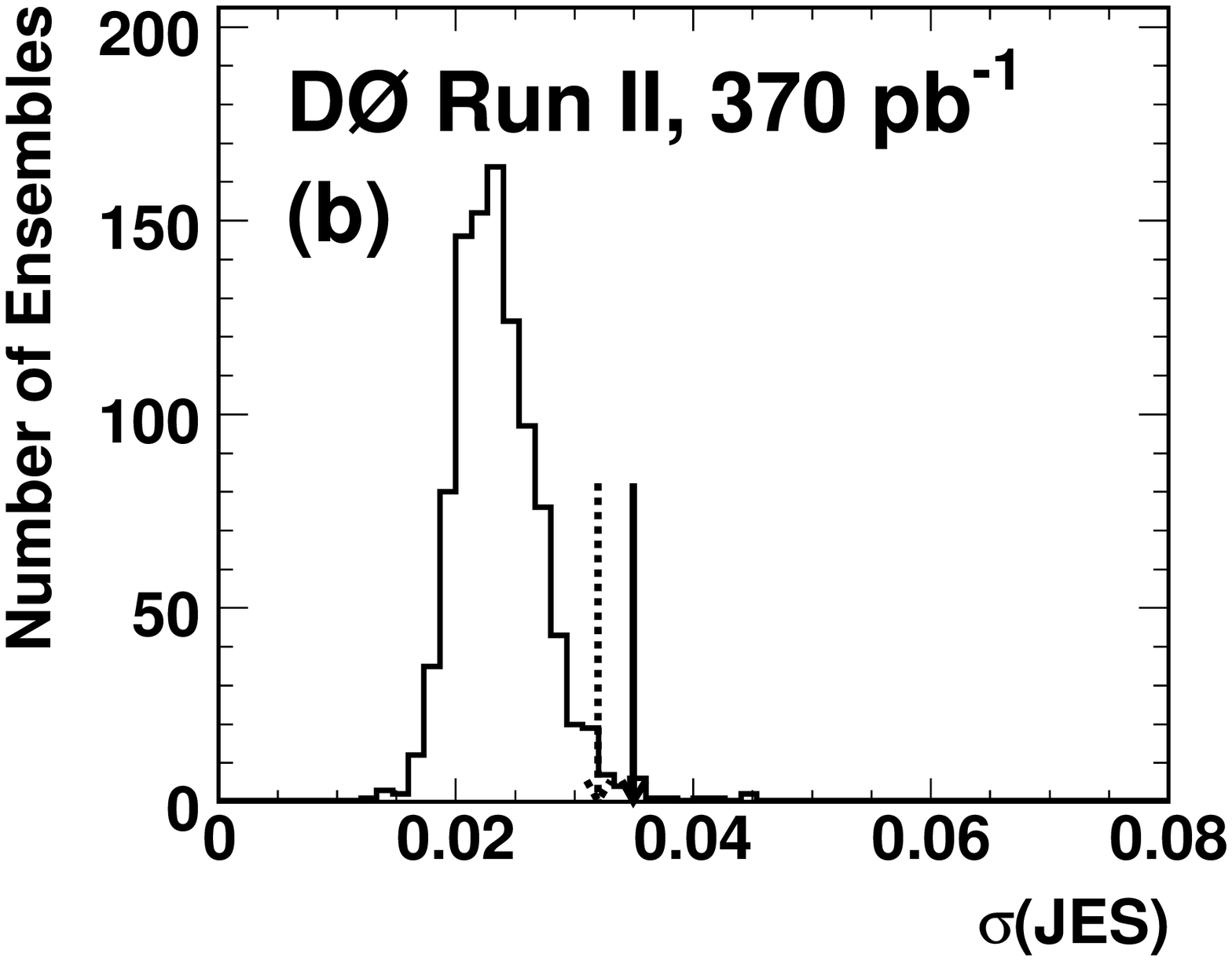}
\caption{Uncertainties on \mtop (a) 
and $JES$ (b) obtained in the $b$-tagging analysis with the combined sample.  
The distributions of fitted uncertainties obtained          
from ensemble tests are shown by the histograms.  Both upper and 
lower uncertainties are shown; their distributions are very similar.
The upper (lower) uncertainty in the data is indicated by the solid 
(dashed) arrow.}
\label{fig:MEresult-btag-error}                                                      
\end{center}                                                                    
\end{figure}

The result for the combined event sample is 
\begin{eqnarray}
	\mtop^{\ejets}  & = & \resultbemstat\,\GeVcc \nonumber \\
	\mtop^{\mujets} & = & \resultbmustat\,\GeVcc \nonumber \\
	\mtop^{\ljets}  & = & \resultbstat\,\GeVcc ;
\end{eqnarray}
the \ljets measurement is shown in Fig.~\ref{fig:MEresult-btag-final}.  
The statistical uncertainties are consistent with the expectation.
Figure~\ref{fig:MEresult-btag-error} shows the distributions of the 
expected \mtop uncertainty compared to the observed result.
The fit yields a signal fraction \ftop of $\resultbftopstat$,
in good agreement with the result of the topological likelihood fit.
The fitted jet energy scale is $JES=\resultbjesstat$ and indicates that the
data is consistent with the reference scale.

For a fixed jet energy scale, the statistical uncertainty of the fit is
$\MEberrstatnojes\,\GeVcc$;
thus the component from the jet energy scale uncertainty is
$\MEberrjes\,\GeVcc$.
Systematic uncertainties are discussed in Section~\ref{sec:systuncs}.

\begin{figure}
\begin{center}
\includegraphics[width=0.45\textwidth]{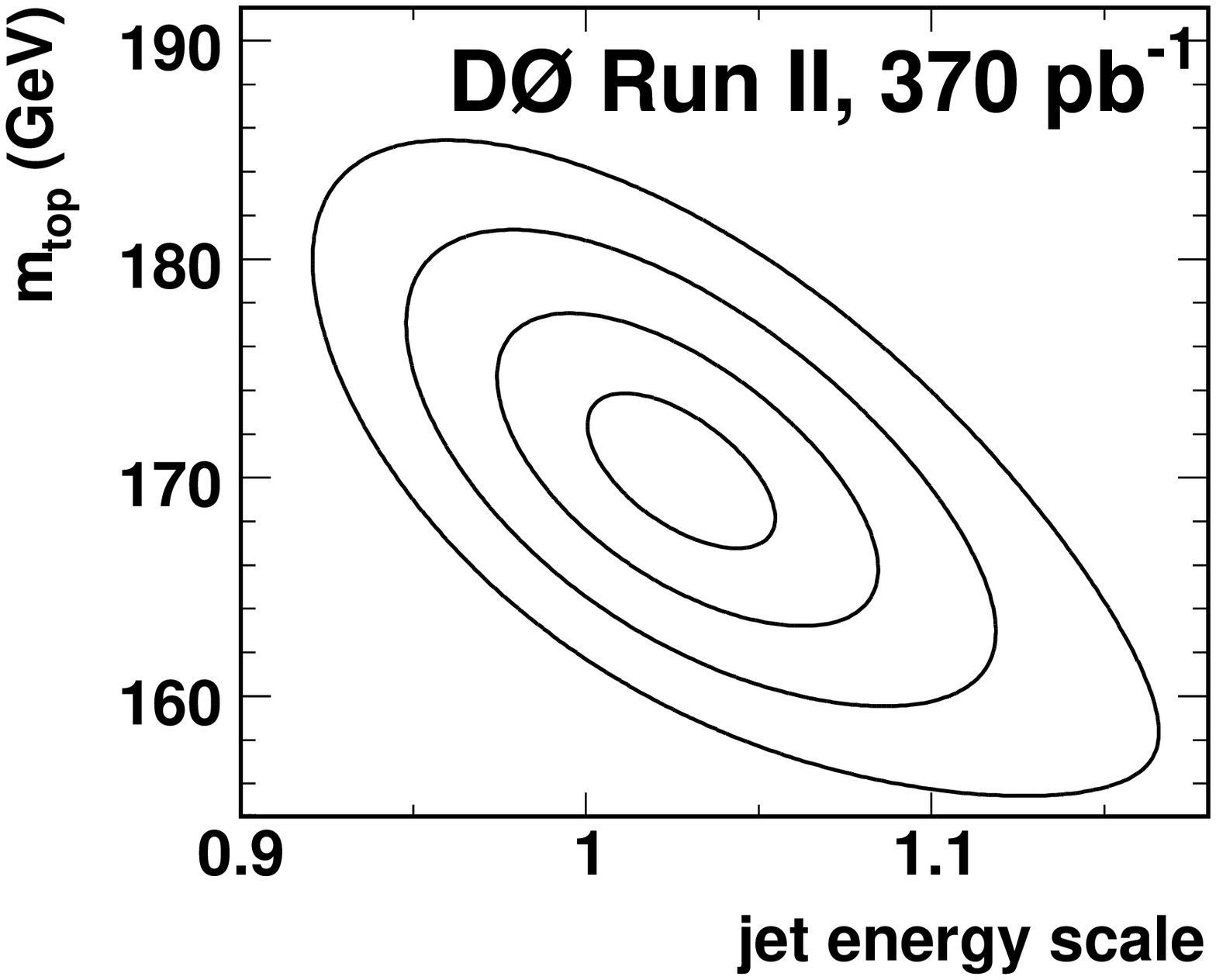}
\caption{Application of the Matrix Element $b$-tagging method to the data.
Fit of a two-dimensional fourth-order polynomial to the 
$-\ln L$ values as a function of both \mtop and $JES$.  Shown are
the contours corresponding to $\Delta\ln L = 0.5$, $2.0$, $4.5$,
and $8.0$ relative to the minimum.}
\label{fig:lhood2d_btag}
\end{center}
\end{figure}

To show the likelihood as a function of both \mtop and $JES$
simultaneously, the $-\ln L$ values have been fitted with a
two-dimensional fourth-degree polynomial with its minimum fixed to the 
measurements mentioned above.
The resulting contours corresponding to $\Delta\ln L = 0.5$, $2.0$, $4.5$,
and $8.0$ relative to the minimum are shown in
Fig.~\ref{fig:lhood2d_btag}.
Note that because of non-Gaussian tails, the projections of the
$\Delta\ln L = 0.5$ contour onto the $\mtop$ and $JES$ axes do not
exactly correspond to the 68\% confidence intervals around the most
likely values; Fig.~\ref{fig:lhood2d_btag} therefore serves only
illustrative purposes. 

\section{Systematic Uncertainties}
\label{sec:systuncs}
Systematic uncertainties arise from three sources: modeling of the physics
processes for \ttbar production and background, modeling of the 
detector performance, and uncertainties in the methods themselves.
Table~\ref{tab:systuncs} lists all uncertainties.
The jet energy scale uncertainty is included in the statistical uncertainty.
The total systematic uncertainty on the top mass measurement is
obtained by adding all contributions in quadrature.
In general, to evaluate systematic uncertainties, the simulation of
events used to calibrate the measurement has been varied, while
the measurement method itself has been kept unchanged.

\begin{table}
\begin{center}
\begin{tabular}{l@{\ \ \ \ \ }c@{\ \ \ \ \ }c}
\hline
\hline
  Source of Uncertainty
& 
  \begin{tabular}{@{}c@{}}Topological\\ Analysis\end{tabular} 
&
  \begin{tabular}{@{}c@{}}$b$-Tagging\\ Analysis\end{tabular} 
\\
\hline
\begin{tabular}{@{}l@{}}
\phantom{.}\vspace{-1ex}\\
Statistical uncertainty\vspace{-0.2ex}\\
and jet energy scale \bigskip
\end{tabular}                      & \MEerrstatlong  & \MEberrstatlong  \\
{\em Physics modeling:}            &                 &                 \\
\quad Signal modeling              & \MEerrsgnmod    & \MEberrsgnmod    \\
\quad Background modeling          & \MEerrbkgmod    & \MEberrbkgmod    \\
\quad PDF uncertainty              & \MEerrpdf       & \MEberrpdf       \\
\quad $b$ fragmentation            & \MEerrbjes      & \MEberrbjes      \\
\quad $b$/$c$ semileptonic decays  & \MEerrbclepbr   & \MEberrbclepbr   \bigskip\\
{\em Detector modeling:}           &                 &                 \\
\quad $JES$ $\pt$ dependence       & \MEerrjespt     & \MEberrjespt     \\ 
\quad $b$ response (h/e)           & \MEerrbresp     & \MEberrbresp     \\
\quad Trigger                      & \MEerrtrg       & \MEberrtrg       \\
\quad $b$ tagging                  & --              & \MEberrbtagging  \bigskip\\
{\em Method:}                      &                 &                 \\
\quad Signal fraction              & \MEerrftop      & \MEberrftop      \\
\quad QCD contamination            & \MEerrqcd       & \MEberrqcd       \\
\quad MC calibration               & \MEerrmccalib   & \MEberrmccalib   \bigskip\\
Total systematic uncertainty       & \MEerrsystlong  & \MEberrsystlong  \bigskip\\
Total uncertainty                  & \MEerrtotallong & \MEberrtotallong \\
\vspace{-2ex}\\
\hline
\hline
\end{tabular}
\caption{Summary of uncertainties on the top quark mass.  All 
values are quoted in GeV.}
\label{tab:systuncs}
\end{center}
\end{table}

\subsection{Physics Modeling}
\label{sec:physicsmodeling}
\begin{itemize}
\item
  {\bf Signal modeling:} 
  When $\ttbar$ events are produced in
  association with a jet, the additional jet can be misinterpreted as
  a product of the \ttbar decay.  Also, the \ttbar system may then have
  significant transverse momentum, in contrast to the assumption made
  in the calculation of \psgn.
  In spite of the event selection that
  requires exactly four jets, these events can be selected if one of
  the jets from the $\ttbar$ decay is not reconstructed. 
  
  Such events are present in the simulated events used for the 
  calibration of the method.
  To assess the uncertainty in the modeling of these effects, 
  events have been generated using a dedicated simulation of the
  production of $\ttbar$ events together with an additional parton.
  The fraction of such events is estimated to be no larger than 30\% 
  (according to the difference between cross section
  calculations in leading and next-to-leading order).

  Two large ensembles of simulated events are composed according to
  the sample composition in the data, one using only events with an
  additional parton for the signal, and the second with the default simulation.
  The result obtained with the default calibration is quoted as central
  value.
  A systematic uncertainty of $30\%$ of the difference in top mass results 
  between these two ensembles is quoted.

  In addition, simulated $gg\to\ttbar$ and $\qqbar\to\ttbar$ events
  have been compared.  The top mass calibration has been rederived
  using only $gg\to\ttbar$ or $\qqbar\to\ttbar$ events to simulate
  the signal, and no significant difference has been found.  Thus 
  no additional uncertainty on the result is assigned.
\item
  {\bf Background modeling:} 
  In order to study the sensitivity of the
  measurement to the choice of background model, the standard \wjets
  Monte Carlo sample is replaced by an alternative sample with the
  default factorization scale of $Q^2 = m_{\W}^{2} + \sum_{j}
  \ptj^{2}$ replaced by $Q'^2 = \left<\ptj\right>^{2}$. 
  One large ensemble of events is composed using both the default 
  and the alternative background model.
  The difference of results obtained with these ensembles is
  symmetrized and is assigned as a systematic uncertainty.
\item
  {\bf PDF uncertainty:} Leading-order matrix elements are used to
  calculate both \psgn and \pbkg. Consequently, both calculations
  use a leading order parton distribution function (PDF): 
  CTEQ5L~\cite{bib:CTEQ5L}.
  To study the systematic uncertainty on \mtop due to this choice, the
  variations provided with the next-to-leading-order PDF set 
  CTEQ6M~\cite{bib:CTEQ6Mvar}
  are used, and
  the result obtained with each of these variations is compared with
  the result using the default CTEQ6M parametrization. 
  The difference between the results obtained with the CTEQ5L and MRST
  leading order PDF sets is taken as another uncertainty.
  Finally, the effect of a variation of $\alpha_s$ is evaluated.
  In all cases, a large ensemble has been composed of events simulated 
  with CTEQ5L, and these have been reweighted such
  that distributions according to the desired PDF set are obtained.
  The individual systematic uncertainties are added in quadrature.
  The systematic uncertainty is dominated by that from the variation
  of CTEQ6M parameters.
\item
  {\bf {\boldmath$b$} fragmentation:} 
  While the
  overall jet energy scale uncertainty is included in the statistical
  uncertainty from the fit, differences in the $b$/light jet energy
  scale ratio between data and simulation may still affect the
  measurement. Possible effects from such differences are studied
  using simulated $\ttbar$ events with different fragmentation models
  for $b$ jets.  The default Bowler~\cite{bib:BOWLER}
  scheme with $r_b=1.0$ is replaced with $r_b=0.69$ or with 
  Peterson~\cite{bib:PETERSON} fragmentation with
  $\epsilon_b=0.00191$.
  Simulation studies show that the variation of $r_b$ results in a
  change of the mean scaled energy $\langle x_B\rangle$ of $b$ hadrons
  that is larger than the uncertainties reported 
  in~\cite{bib:bfragLEP}, while
  the uncertainty on the shape of the $x_B$ distribution is taken
  into account by the comparison of the Bowler and Peterson schemes.
  One large ensemble is built using events from each of the three
  simulations.  The absolute values of the deviations of top mass
  results from the standard sample are added
  in quadrature and symmetrized.
\item
  {\bf {\boldmath$b/c$} semileptonic decays:}
  The reconstructed energy of $b$ jets containing a semileptonic
  bottom or charm decay is in general lower than that of jets 
  containing only hadronic decays.  This can only be taken into account for
  jets in which a soft muon is reconstructed.  Thus, the fitted top
  quark mass still depends on the semileptonic $b$ and $c$ decay
  branching ratios.  They have been varied by reweighting events
  in one large ensemble of simulated events within the bounds given
  in~\cite{bib:PDG}.
\end{itemize}

\subsection{Detector Modeling}
\label{sec:detectormodeling}
\begin{itemize}
\item
  {\bf\boldmath $JES$ {\boldmath$\pt$} and {\boldmath$|\eta|$} dependence:} 
  The relative difference 
  between the jet energy scales in data and Monte Carlo is fitted 
  with a global scale factor, and the corresponding uncertainty is 
  included in the quoted (stat.\,+\,JES) uncertainty.
  Any discrepancy between data and simulation other than a global 
  scale difference may lead to an additional uncertainty on the top 
  quark mass.  
  To estimate this uncertainty, the energies of jets in the events
  of one large ensemble have been scaled by a factor of 
  $(1+0.02\frac{E_{\rm jet}}{100\,{\rm GeV}})$
  where $E_{\rm jet}$ is the default jet energy.  
  The value $0.02$ is suggested by studies of $\gamma$+jets events.
  The top mass result from the modified ensemble has been compared to 
  the default number, and the symmetrized difference 
  is taken as a systematic uncertainty.

  Similarly, to estimate the effect of a possible $|\eta|$ dependence
  of the jet energy scale ratio between data and simulation, the jet
  energies have also been scaled by a factor $(1-0.01|\eta|)$ as
  suggested by $\gamma$+jets events.  No significant effect on the 
  top quark mass has been observed and thus no additional 
  systematic uncertainty is assigned.
\item
  {\bf Relative {\boldmath$b$}/light jet energy scale:} 
  Variations of the h/e calorimeter
  response lead to differences in the $b$/light jet energy
  scale ratio between data and simulation in addition to the
  variations of the $b$ fragmentation function considered in
  Section~\ref{sec:physicsmodeling}.  
  This uncertainty has been evaluated by scaling the energies of $b$
  jets in one large ensemble and studying the effect on the top quark
  mass.
\item
  {\bf Trigger:} The trigger efficiencies used in composing ensembles
  for the calibration of the measurement
  are varied by their uncertainties, and the uncertainties
  from all variations are summed in quadrature.
\item
  {\bf\boldmath $b$ tagging:} The $b$-tagging efficiencies are varied within
  the uncertainties as determined from the data, and the variations
  are propagated to the final result.           
\end{itemize}

Note that no systematic uncertainty is quoted due to multiple
interactions/uranium noise as opposed to the \runi measurement.
The effect is much smaller in \runii as a consequence of the reduced
integration time in the calorimeter readout.
It is moreover covered by the jet energy scale uncertainty, as the
offset correction is computed seperately for data and Monte Carlo in
\runii, accounting for effects arising from electronic noise and
pileup.

\subsection{Method}
\label{sec:methods}
\begin{itemize}
\item
  {\bf Signal fraction:} The normalization procedure of the background 
  probability described in Section~\ref{sec:MEpbkg} is chosen such
  that the signal fraction $\ftop$ as 
  measured with the topological likelihood fit
  and given in Table~\ref{tab:MEcomposition} is reproduced.  However,
  the signal fraction is slightly
  overestimated for low true signal fractions, which leads to a small
  bias in the resulting top mass. The signal fraction in ensemble
  tests used for the calibration
  is varied within the uncertainties determined from the
  topological likelihood fit, and the resulting variation of the top
  quark mass is taken as a systematic uncertainty.  
\item
  {\bf QCD background:} The \wjets simulation is used to model the
  small QCD background in the selected event sample in the analysis.
  The systematic uncertainty from this assumption is computed by
  selecting a dedicated QCD-enriched sample of events from data by
  inverting the lepton isolation cut in the event selection. The
  calibration of the method is repeated with ensembles formed where
  these events are used to model the QCD background events whose
  fraction is given in Table~\ref{tab:MEcomposition}. The
  resulting change is assigned as a systematic uncertainty.
\item
  {\bf MC calibration:} The calibration of the top mass measurement
  is varied according to the statistical uncertainty of the calibration
  curves shown in Figs.~\ref{fig:MEcalib-topo}
  and~\ref{fig:MEcalib-btag}.
\end{itemize}

\section{Summary}
\label{sec:conclusions}
A measurement of the top quark mass using lepton+jets \ttbar events 
in 370\,\ipb of data collected
with the \dzero detector at \runii of the Fermilab Tevatron Collider
has been presented.
The events are analysed with the Matrix Element method, which is
designed to make maximal use of the kinematic information in the 
selected events.
To avoid a large systematic uncertainty, an overall scale factor $JES$ for the
energy of calorimeter jets is determined simultaneously with the top
quark mass.
This in-situ calibration of the jet energy scale helps reduce the
overall uncertainty on the top quark mass when combining with other
measurements.

The resulting top quark mass is
\begin{equation}
  \mtop = \result\,\GeV
\end{equation}
for an analysis that uses only topological information, and
\begin{equation}
  \mtop = \resultb\,\GeV
\end{equation}
when $b$-tagging information is included.
The jet energy scale is $JES = \resultjesstat$ in the topological
analysis and $JES = \resultbjesstat$ when $b$ tagging is included, 
indicating consistency
with the reference scale.
The two results are consistent with each other.
To obtain a value for the top quark mass in combination with other
measurements, the second, more precise value should be used.

\section*{Acknowledgements}
%
We thank the staffs at Fermilab and collaborating institutions, 
and acknowledge support from the 
DOE and NSF (USA);
CEA and CNRS/IN2P3 (France);
FASI, Rosatom and RFBR (Russia);
CAPES, CNPq, FAPERJ, FAPESP and FUNDUNESP (Brazil);
DAE and DST (India);
Colciencias (Colombia);
CONACyT (Mexico);
KRF and KOSEF (Korea);
CONICET and UBACyT (Argentina);
FOM (The Netherlands);
PPARC (United Kingdom);
MSMT (Czech Republic);
CRC Program, CFI, NSERC and WestGrid Project (Canada);
BMBF and DFG (Germany);
SFI (Ireland);
The Swedish Research Council (Sweden);
Research Corporation;
Alexander von Humboldt Foundation;
and the Marie Curie Program.
%



\begin{thebibliography}{99}
\bibitem{bib:Zbible}
  The ALEPH, DELPHI, L3, OPAL, and SLD Collaborations, the 
  LEP Electroweak Working Group, SLD Electroweak Group, 
  and SLD Heavy Flavour Group,
  Phys.\ Rept.\  {\bf 427}, 257 (2006).
%
\bibitem{bib:topdiscovery}
  F.~Abe {\it et al.},
  Phys.\ Rev.\ Lett.\  {\bf 74}, 2626 (1995);\\
  S.~Abachi {\it et al.},
  Phys.\ Rev.\ Lett.\  {\bf 74}, 2632 (1995).
%
\bibitem{bib:topmassruni}
  F.~Abe {\it et al.},
  Phys.\ Rev.\ Lett.\  {\bf 80}, 2767 (1998);\\
  F.~Abe {\it et al.},
  Phys.\ Rev.\ Lett.\  {\bf 80}, 2779 (1998);\\
  F.~Abe {\it et al.},
  Phys.\ Rev.\ Lett.\  {\bf 82}, 271 (1999);\\
  F.~Affolder {\it et al.},
  Phys.\ Rev.\ D {\bf 63}, 032003 (2001);\\
%
  S.~Abachi {\it et al.}, 
  Phys.\ Rev.\ Lett.\  {\bf 79}, 1197 (1997);\\
  B.~Abbott {\it et al.}, 
  Phys.\ Rev.\ Lett.\  {\bf 80}, 2063 (1998);\\
  B.~Abbott {\it et al.},
  Phys.\ Rev.\ D {\bf 58}, 052001 (1998);\\
  B.~Abbott {\it et al.},
  Phys.\ Rev.\ D {\bf 60}, 052001 (1999);\\
  V.~M.~Abazov {\it et~al.},
  Nature {\bf 429}, 638 (2004);\\
  V.~M.~Abazov {\it et al.}
  Phys.\ Lett.\ B {\bf 606}, 25 (2005).
%
\bibitem{bib:topmasscdfrunii}
  A.~Abulencia {\it et al.},
  Phys.\ Rev.\ D {\bf 73}, 032003 (2006);\\
  A.~Abulencia {\it et al.},
  Phys.\ Rev.\ Lett.\  {\bf 96}, 022004 (2006).
%
\bibitem{bib:nature}
  V.~M.~Abazov {\it et~al.}, 
  Nature {\bf 429}, 638 (2004).
%
\bibitem{bib:philipp}
  P.~Schieferdecker,
  FERMILAB-THESIS {\bf 2005-46}.
%
\bibitem{run2det} 
  V.~M.~Abazov {\it et al.}, 
  Nucl. Instrum. Methods A {\bf 565}, 463 (2006).
%
\bibitem{run1det} 
  S. Abachi {\it et al.},
  Nucl. Instrum. Methods Phys. Res. A {\bf 338}, 185 (1994).
%
\bibitem{run2muon} 
  V.~M.~Abazov {\it et al.}, 
  physics/0503151 (2005).
%
\bibitem{bib:btagxsect} 
  V.~M.~Abazov {\it et al.}, 
  Phys.\ Lett.\ B {\bf 626}, 35 (2005).
%
\bibitem{bib:alpgen} 
  M.~L.~Mangano {\it et al.}, 
  JHEP {\bf 307}, 1 (2003).
%
\bibitem{bib:pythia}
  T.~Sj\"ostrand, P.~Ed\'en, C.~Friberg, L.~L\"onnblad, G.~Miu, S.~Mrenna,
  and E.~Norrbin,
  Computer Phys. Commun. {\bf 135}, 238 (2001).
%
\bibitem{bib:topoxsect} 
  V.~M.~Abazov {\it et al.}, 
  Phys.\ Lett.\ B {\bf 626}, 45 (2005).
%
\bibitem{bib:MAHLON}
  G.~Mahlon and S.~J.~Parke,
  Phys.\ Lett.\ B {\bf 411}, 173 (1997).
%
\bibitem{bib:PDG}
  W.-M. Yao {\it et al.}
  J.~Phys.~G {\bf 33}, 1 (2006).
%
\bibitem{bib:VEGAS1}
  G.~P.~Lepage, 
  Journal of Computational Physics {\bf 27}, 192 (1978).
%
\bibitem{bib:VEGAS2}
  G.~P.~Lepage, 
  Cornell preprint CLNS:80-447, (1980).
%
\bibitem{bib:CTEQ5L}
  H.~L.~Lai {\it et al.},
  Eur.\ Phys.\ J.\ C {\bf 12}, 375 (2000).
%
\bibitem{bib:VECBOS}
  F.~A.~Berends, H.~Kuijf, B.~Tausk, and W.~T.~Giele, 
  Nucl.~Phys.\ {\bf B357}, 32 (1991).
%
\bibitem{bib:MADGRAPH} 
  F.~Maltoni, T.~Stelzer, 
  JHEP {\bf 302}, 27 (2003).
%
\bibitem{bib:CTEQ6Mvar}
  J.~Pumplin {\it et al.}, 
  JHEP~{\bf 0207}, 012 (2002).
%
\bibitem{bib:BOWLER}
  M.~G.~Bowler, 
  Z.~Phys C {\bf 11}, 169 (1981).
%
\bibitem{bib:PETERSON} 
  C.~Peterson {\it et. al}, 
  Phys.~Rev.~D {\bf 27}, 105 (1983).
%
\bibitem{bib:bfragLEP}
  K.~Abe {\it et. al}, 
  Phys.\ Rev.\ Lett.\ {\bf 84}, 4300 (2000);\\
  A.~Heister {\it et. al}, 
  Phys.\ Lett.\ B {\bf 512}, 30 (2001);\\
  G.~Abbiendi {\it et. al}, 
  Eur.\ Phys.\ J.\ C {\bf 29}, 463 (2003).

\bibitem[*]{kurca}
On leave from IEP SAS Ko{\v s}ice, Slovakia.
\bibitem[\dag]{voutilainen}
Visitor from Helsinki Institute of Physics, Helsinki, Finland.
%
\vskip 0.25cm

\end{thebibliography}
\end{document}